\pdfoutput=1
\documentclass[reprint,
superscriptaddress,
 amsmath,amssymb,
 aps,floatfix,
prl
]{revtex4-2}

\usepackage{graphicx}
\usepackage{placeins}
\usepackage{multirow}
\usepackage{amsthm}
\usepackage{mathrsfs}
\usepackage[dvipsnames]{xcolor}
\usepackage{textcomp}
\usepackage[normalem]{ulem}
\usepackage{booktabs}
\usepackage[caption=false]{subfig}
\usepackage{algorithm2e}
\usepackage{listings}
\usepackage{enumitem}
\usepackage{cases}
\usepackage[useregional]{datetime2}
\usepackage[colorlinks=true, linkcolor=darkBlue, citecolor=darkBlue, urlcolor=darkBlue]{hyperref}
\usepackage{cleveref}

\usepackage{physics}
\usepackage{braket}
\usepackage{url}
\usepackage{dcolumn}
\usepackage{bm}
\usepackage{comment}
\usepackage{orcidlink}
\usepackage{tikz}
\usepackage{microtype}

\allowdisplaybreaks

\definecolor{darkBlue}{RGB}{0,0,200}
\definecolor{darkGreen}{RGB}{13, 59, 2}

\newcommand{\ii}{\mathrm{i}}
\newcommand{\ee}{\mathrm{e}}
\newcommand{\QDE}{Q_{\mathrm{DE}}}

\newcommand{\roruni}[2]{\href{https://ror.org/#2}{#1}}
\newcommand{\Rig}{\mathcal{R}}
\newcommand{\Spread}{\mathcal{S}}
\newcommand{\AGP}{\mathcal{A}}

\newcommand{\Np}{N_{\mathrm p}}

\begin{document}

 \title{Nonequilibrium Spin-Glass Order in a Long-Range Quantum Ising Glass and Its Sensitivity to Noise}

\author{Giuseppe D'Auria \!\orcidlink{0000-0001-7818-870X}}
\affiliation{International School for Advanced Studies (\roruni{SISSA}{004fze387}), Via Bonomea 265, 34136 Trieste, Italy}
\author{Alessandro Silva \!\orcidlink{0000-0002-5668-7134}}
\affiliation{International School for Advanced Studies (\roruni{SISSA}{004fze387}), Via Bonomea 265, 34136 Trieste, Italy}

\date{\today}

\begin{abstract}
We study spin-glass order after a transverse-field quench in a long-range quantum Ising glass. The diagonal-ensemble Edwards–Anderson order coincides at the no-quench and zero-field endpoints, yet in between cold shells overshoot it and hot shells undershoot it. This stems from a competition in correlation space between the rigidity of populated eigenstates and the spread of their pair patterns. Under weak longitudinal dephasing the zero-field endpoint is exactly pinned while the characteristic finite-quench response fades, giving a self-referenced noise probe for quantum spin simulators.
\end{abstract}

\maketitle

The nonequilibrium dynamics of quantum many-body systems with quenched disorder has become a central frontier at the intersection of  condensed matter physics and quantum computing~\cite{Polkovnikov2011,DAlessio2016,Abanin2019}.  The interplay of randomness, frustration, and quantum fluctuations can stabilize exotic phases that are forbidden in thermal equilibrium, challenging the standard paradigms of ergodicity and thermalization~\cite{NandkishoreHuse2015,Abanin2019,Else2020}. 
Quantum spin glasses provide a particularly rich setting, combining random interactions and coherent quantum dynamics to produce frozen, spatially irregular order~\cite{BinderYoung1986,MillerHuse1993}. Their frustration and rugged energy landscapes also make them demanding benchmarks for programmable quantum processors~\cite{Harrigan2021,Pelofske2024}. Superconducting quantum annealers have realized programmable transverse-field spin glasses~\cite{Harris2018}, coherent dynamics across quantum phase transitions~\cite{King2022,King2023}, and spin-glass quench dynamics claimed to be beyond the reach  of classical simulation~\cite{King2025}, while trapped-ion~\cite{islam_2011,Zhang2017,MonroeRMP2021} and Rydberg-atom platforms~\cite{Ebadi2021,Ebadi2022} now enable controlled preparation, driving, and measurement of such glassy many-body systems.

A paradigmatic model bridging classical glass physics and quantum simulation is the one-dimensional long-range Ising spin glass of Kotliar, Anderson, and Stein (KAS)~\cite{KAS83}. Power-law couplings $J_{ij}\sim\xi_{ij}/|i-j|^{\alpha}$ with $1/2<\alpha<1$ effectively raise the dimensionality and support a finite-temperature spin-glass phase already in one dimension, with mean-field behavior for $\alpha<2/3$ and non-mean-field behavior for $2/3<\alpha<1$~\cite{KAS83,Leuzzi08,Moore10,Monthus14,DefenuRMP23}. Long-range power-law Ising interactions are native to trapped-ion simulators, where they have been combined with programmable random local fields~\cite{MonroeRMP2021,smith2016}; engineering arbitrary coupling matrices, including random-sign ones of the KAS type, has been proposed via multimode driving with individual addressing~\cite{Korenblit2012}, while programmable random couplings are already routinely realized on quantum annealers~\cite{Harris2018,King2023}. A transverse field adds quantum fluctuations that compete with glassiness and can drive a transition to a paramagnet. On the static side, attention has focused on the eigenstate structure of this and related quantum spin glasses~\cite{Laumann2014,Smelyanskiy2020}. In many-body localized (MBL) or nonergodic extended regimes, finite-energy-density eigenstates can retain Edwards--Anderson (EA) order~\cite{Javanmard2019,Baldwin2017,Mossi17,rademaker_abanin_2020}. Out of equilibrium, by contrast, studies of quantum spin glasses have largely relied on controlled semiclassical or mean-field descriptions~\cite{Correale23,Pappalardi20,Thomson20}. A spin-$1/2$ qubit simulator realizes neither limit, and the quantum-to-classical crossover is known to be delicate~\cite{hosseinabadi24}. The fully quantum, non-mean-field dynamics of the KAS glass, which is the regime programmable quantum hardware actually accesses, therefore remains largely unexplored.

In this Letter, we study the stationary spin-glass order reached after a transverse-field quench in the spin-$1/2$ transverse-field KAS (TFKAS) model in the non-mean-field regime $2/3<\alpha<1$, using exact diagonalization. We characterize the stationary state by the EA order built from equal-time $\sigma^z\sigma^z$ correlations in the diagonal ensemble. This quantity takes exactly the same value without a quench and at the zero-field endpoint, so any departure between these two anchors is a genuinely nonequilibrium effect. Resolving it by the energy density of the initial state, we find a pronounced nonmonotonic response. Low-energy shells overshoot the no-quench EA order, whereas sufficiently excited shells undershoot it. We trace this behavior to a geometric competition in correlation space between the growing rigidity of the populated final eigenstates and the spread of their mutually incompatible frozen patterns. We also show that the cold-shell overshoot is captured by an isolated-resonance theory, detailed in the Supplemental Material. Finally, we show that under weak local longitudinal dephasing the zero-field endpoint remains exactly protected, while the finite-quench response is progressively washed out. Since both are accessible from projective $\sigma^z$ snapshots without tomography, the protocol probes nonequilibrium spin-glass order and at the same time provides a self-referenced witness of weak dephasing in programmable quantum simulators.

\begin{figure}[t]
\centering
\includegraphics[width=0.85\columnwidth]{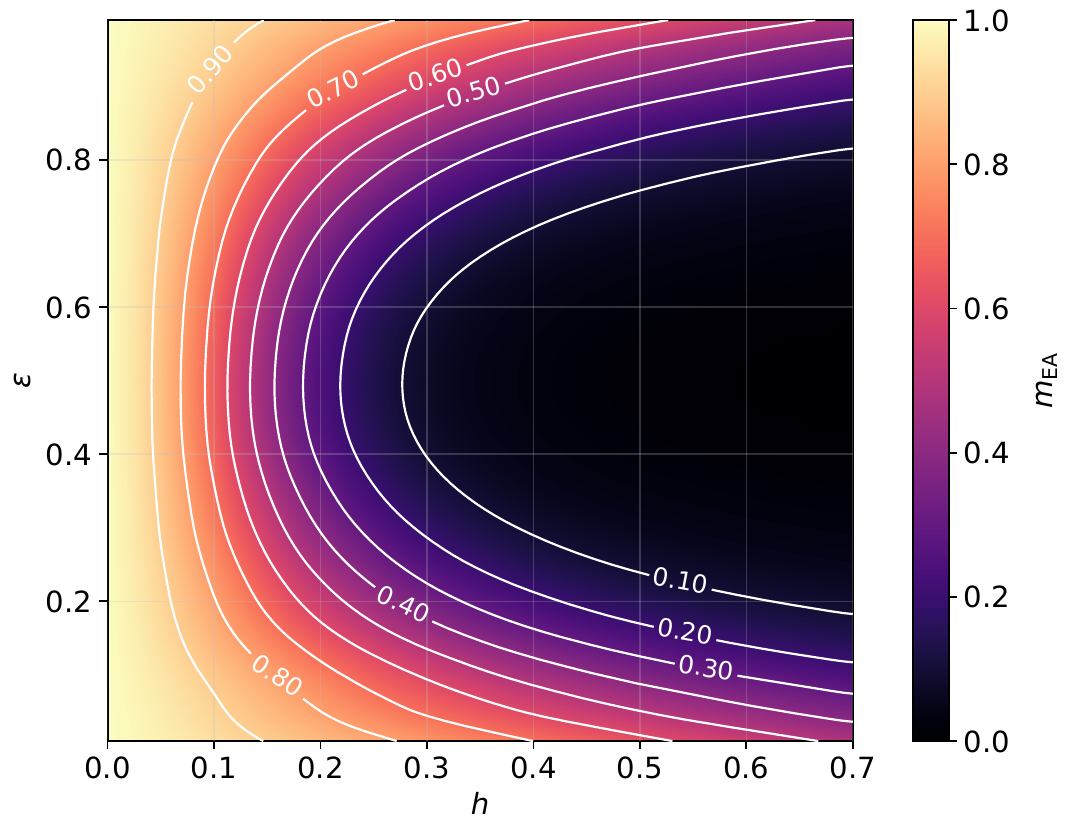}
\caption{Finite-size static eigenstate landscape for $L=13$. The color scale is the shell- and disorder-averaged Edwards--Anderson order $m_{\rm EA}(\epsilon,h;L)$ of Eq.~\eqref{eq:mEA}, with shell width $\Delta\epsilon=0.05$ and $2\times10^3$ disorder realizations. Static glass order decreases monotonically with $h$ and is lost first at mid-spectrum, whereas the post-quench response (Fig.~\ref{fig:geometry}) is not monotonic.}
\label{fig:static}
\end{figure}

We consider the TFKAS model, a periodic chain of $L$ spin-$1/2$ degrees of freedom,
\begin{equation}
H(h)=V-hX,\quad
V=\sum_{i<j}\frac{\xi_{ij}}{r_{ij}^{\alpha}}\sigma_i^z\sigma_j^z,
\quad X=\sum_i \sigma_i^x,
\label{eq:H}
\end{equation}
where $r_{ij}=\min(|i-j|,L-|i-j|)$, the $\xi_{ij}$ are independent zero-mean, unit-variance Gaussian variables, and $1/2<\alpha<1$ ensures that local fields have finite variance without Kac rescaling. In the numerics we set $\alpha=0.85$, inside the non-mean-field window. $H(h)$ commutes with the global spin flip $P=\prod_i\sigma^x_i$, and its disorder ensemble is symmetric under $E\to-E$ (via $\xi_{ij}\to-\xi_{ij}$ and $\prod_i\sigma^z_i$), so we restrict to the lower half of the spectrum. For an eigenstate $H(h)\ket{n}=E_n\ket{n}$ we use the rescaled energy density $\epsilon_n=(E_n-E_{\min})/(E_{\max}-E_{\min})$ and the finite-size eigenstate EA order parameter
\begin{equation}
m_{\rm EA}^{(n)}=\frac{1}{\Np}\sum_{i<j}\braket{n|\sigma_i^z\sigma_j^z|n}^2,\qquad \Np=\tbinom{L}{2},
\label{eq:mEA}
\end{equation}
where $\Np$ is the number of spin pairs; this normalization matches that of the post-quench order parameter introduced below. Its shell- and disorder-averaged map $m_{\rm EA}(\epsilon,h;L)$, shown in Fig.~\ref{fig:static}, serves only as a static reference. As $h$ increases, EA order is lost first at the center of the spectrum, $\epsilon\simeq1/2$, while the spectral edges remain glassy up to substantially larger fields. 

For a fixed disorder realization we drive the system out of equilibrium by a sudden quench of the transverse field, $h_I\to h_F$, parametrized by $\lambda:=h_F/h_I$. We prepare an eigenstate $\ket{n_I}$ of $H_I\equiv H(h_I)$ and evolve it with $H_F\equiv H(h_F)$. Since the spectrum is nondegenerate within each parity sector, long-time averages of observables are described by the diagonal ensemble $\rho_{\rm DE}^{(n)}=\sum_m p_m^{(n)}\ketbra{m_F}{m_F}$, with $p_m^{(n)}=|\braket{m_F|n_I}|^2$~\cite{RigolNature08,Polkovnikov2011}. We then define
\begin{equation}
q_{\rm DE}^{(n)}:=\frac{1}{\Np}\sum_{i<j}
\left[\Tr\!\left(\rho_{\rm DE}^{(n)}\sigma_i^z\sigma_j^z\right)\right]^2,
\label{eq:qDE}
\end{equation}
the EA functional of the dephased stationary state. Equation~\eqref{eq:qDE} squares the time-averaged correlator, $\overline{C_{ij}}^{\,2}$, rather than averaging the squared one, $\overline{C_{ij}^{2}}\geq\overline{C_{ij}}^{\,2}$. The latter also contains temporal fluctuations. Thus $q_{\rm DE}$ isolates the frozen pair pattern that survives dephasing, which is the appropriate order parameter for the glassiness of the post-quench state (see Appendix~\ref{app:derivation} in the End Matter). It also reduces the infinite-time problem to eigenstate quantities, avoiding long real-time evolutions that are demanding both numerically and experimentally. Averaging Eq.~\eqref{eq:qDE} over disorder and over initial states in an energy-density shell of $H_I$ of width $\Delta\epsilon=0.05$ centered at $\epsilon$ defines $\QDE(\lambda,\epsilon)$.

Two exact anchors constrain the response. At $\lambda=1$ there is no quench and $\rho_{\rm DE}=\ketbra{n_I}{n_I}$. At $\lambda=0$, $H_F=V$ and every
$\sigma_i^z\sigma_j^z$ commutes with $H_F$, so each pair correlator is conserved
by the post-quench dynamics and by dephasing alike. Hence
\begin{equation}
\lim_{\lambda\to0}\QDE(\lambda,\epsilon)
=\QDE(1,\epsilon)\equiv Q_I(\epsilon).
\label{eq:endpoints}
\end{equation}
\begin{figure*}[t]
\centering
\includegraphics[width=0.7\textwidth]{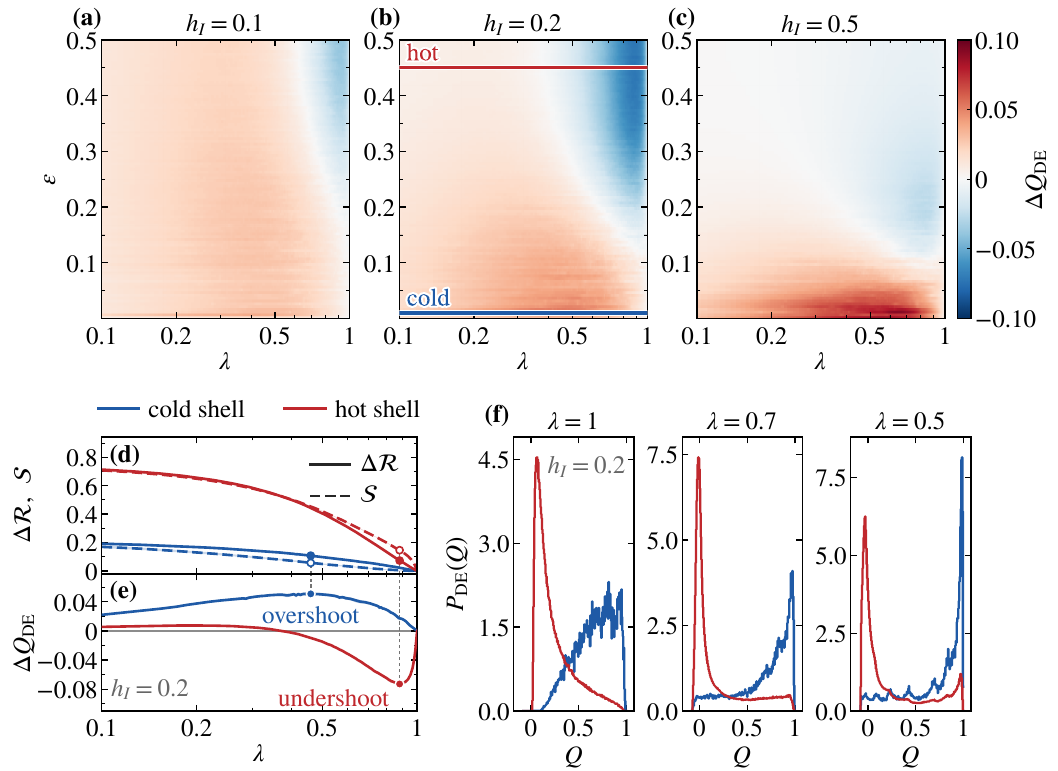}
\caption{Nonequilibrium spin-glass order and its origin in correlation space, for $L=13$, $\alpha=0.85$, and $10^3$ disorder realizations. (a)--(c) Energy-resolved excursion $\Delta\QDE(\lambda,\epsilon)=\QDE(\lambda,\epsilon)-Q_I(\epsilon)$ for $h_I=0.1,0.2,0.5$; red marks an overshoot and blue an undershoot of the common endpoint value. The lines in (b) mark the cold ($\epsilon\in[0.01,0.02]$, blue) and hot ($\epsilon\in[0.45,0.46]$, red) shells used in (d)--(f). (d) Gain in self-overlap $\Delta\Rig$ (solid) and pattern spread $\Spread$ (dashed) for $h_I=0.2$. (e) Their difference $\Delta\QDE=\Delta\Rig-\Spread$ [Eq.~\eqref{eq:mechanism}], with the maximal overshoot and undershoot marked; dotted lines connect them to (d). (f) Overlap distributions $P_{\rm DE}(Q)$ of the populated final eigenstates at $\lambda=1$, $0.7$, and $0.5$, averaged over the initial states of each shell. At $\lambda=1$ the distribution reflects only the spread of self-overlaps among initial states; as $\lambda$ decreases, the overlap weight is redistributed,
and by $\lambda=0.5$ it begins to accumulate near the discrete
classical values $Q=(Lq^2-1)/(L-1)$, $q=(2k+1)/L$,
anticipating the peak structure of the zero-field limit.}
\label{fig:geometry}
\end{figure*}
Although $H(0)=V$ is twofold degenerate under the global flip $P$, Eq.~\eqref{eq:endpoints} is unaffected because $\sigma_i^z\sigma_j^z$ is $P$-even; see Appendix~\ref{app:derivation} in the End Matter. Equation~\eqref{eq:endpoints} is the internal reference of
the protocol: two settings that are prepared and measured independently must
return the same number, so any measured discrepancy is attributable to the
device.
Despite this exact equality, the curve is not flat between its endpoints.  We refer to excursions above and below the common endpoint value $Q_I$ as an \emph{overshoot} and an \emph{undershoot}, respectively.  Figures~\ref{fig:geometry}(a)--\ref{fig:geometry}(c) show a broad overshoot in cold shells and, for sufficiently excited shells, an undershoot near the no-quench point.

The nonmonotonic enhancement of glassy order above the static no-quench value is a systematic feature of the data, most persistent in the cold shells and at weaker initial transverse field, while becoming progressively more fragile in hotter shells and as $h_I$ is increased.
Conversely, a cusp at $\lambda=1$ preceded by a local undershoot for $\lambda = 1^-$ emerges for hotter initial energy-density shells, signaling the glassy-to-paramagnet crossover in the static regime. Interestingly, such an undershoot appears in a regime where the static system would be paramagnetic. These features persist across disorder realizations and system sizes; their finite-size evolution is analyzed in Sec.~\ref{sec:fss_nonmonotonic} of the Supplemental Material~\cite{SM}.

The origin of this nonmonotonic response becomes transparent in the language of spin-glass overlaps. We associate with each final eigenstate its pair-correlation pattern
\begin{equation}
\bm c_m:=\bigl(\langle m_F|\sigma_i^z\sigma_j^z|m_F\rangle\bigr)_{i<j}
\label{eq:cm}
\end{equation}
and define the pair overlap between two final eigenstates as $Q_{mm'}:=\Np^{-1}\,\bm c_m\cdot\bm c_{m'}$. For classical configurations $s,s'$ this reduces to $Q_{ss'}=(L\,q_{ss'}^2-1)/(L-1)$, with $q_{ss'}=L^{-1}\sum_i s_is'_i$ the usual spin overlap. Hence $Q_{mm'}$ is a flip-invariant eigenstate generalization of the squared overlap, and $Q_{mm}=m_{\rm EA}^{(m)}$ is the self-overlap of Eq.~\eqref{eq:mEA}. In this language Eq.~\eqref{eq:qDE} reads $q_{\rm DE}=\sum_{m,m'}p_mp_{m'}Q_{mm'}$, the mean of the overlap distribution $P_{\rm DE}(Q)=\sum_{m,m'}p_mp_{m'}\,\delta(Q-Q_{mm'})$ of the populated final eigenstates, in close analogy with the overlap distribution that distinguishes coexisting pure states in a spin glass~\cite{Parisi83,MezardParisiVirasoro1987}. Separating self- from mutual overlaps gives the exact decomposition
\begin{equation}
q_{\rm DE}
=\sum_m p_m\,Q_{mm}
-\frac{1}{2\Np}\sum_{m,m'}p_mp_{m'}\,\|\bm c_m-\bm c_{m'}\|^2 ,
\label{eq:decomp}
\end{equation}
where $\|\cdot\|$ denotes the Euclidean norm in $\mathbb{R}^{N_p}$.
We denote the shell and disorder averages of the two terms by $\Rig$ and $\Spread$, respectively. The first, $\Rig$, is the mean self-overlap, i.e., the eigenstate EA order of the populated final eigenstates. It measures how glassy each of them is individually. The second, $\Spread\geq0$, is half the mean squared distance between their patterns. It vanishes if all populated eigenstates share the same frozen pattern and is of order one if they belong to incompatible valleys. It thus plays the role of the broadening of $P_{\rm DE}(Q)$ away from the self-overlap. Equivalently, $\Spread$ is the eigenstate-to-eigenstate fluctuation of the pair correlators within the populated energy window, which is exponentially small in $L$ under ETH~\cite{KimIkedaHuse2014} but finite in the glassy regime. Since $\Spread(1)=0$ and $\Rig(1)=Q_I$,
\begin{equation}
\Delta\QDE=\Delta\Rig-\Spread,\qquad
\Delta\Rig:=\Rig(\lambda)-\Rig(1),
\label{eq:mechanism}
\end{equation}
with $\Delta\QDE:=\QDE(\lambda)-Q_I$. Both terms grow toward $1-Q_I$ as $\lambda\to0$, where the eigenstates of $V$ are classical configurations with $Q_{mm}=1$. The endpoint identity~\eqref{eq:endpoints} is therefore an exact cancellation, and an overshoot or undershoot reflects the sign of their imbalance at intermediate $\lambda$.

Hot-shell eigenstates can thus individually become more glassy while their average pattern becomes less ordered. Lowering the field makes each populated eigenstate more glassy, as anticipated by the static map of Fig.~\ref{fig:static}, so that $\Delta\Rig>0$. At the same time, the quench distributes the initial weight over several final eigenstates, at the cost $\Spread$. In cold shells the weight remains concentrated on a few of them, with inverse participation ratio $P_2=\sum_m p_m^2\gtrsim1/2$, as for an isolated two-level resonance, whereas in hot shells it fragments over many, with $P_2\ll1/2$. Figures~\ref{fig:geometry}(d) and \ref{fig:geometry}(e) show that in the cold shell the rigidity gain wins, $\Delta\Rig>\Spread$, producing the overshoot. In the hot shell, by contrast, the incompatibility of the populated patterns dominates, $\Delta\Rig<\Spread$, over a broad window below $\lambda=1$, producing the undershoot, which is deepest near $\lambda\approx0.9$.

Figure~\ref{fig:geometry}(f) makes this mechanism visible through the overlap distributions, whose mean is $\QDE$. In the cold shell the quench moves the weight toward $Q\simeq1$. A mutual overlap close to unity requires both that each populated eigenstate be strongly glassy and that the states share the same frozen pattern. The few populated eigenstates thus become more rigid without becoming incompatible, and the mean of $P_{\rm DE}(Q)$ rises above $Q_I$. In the hot shell, instead, the mutual overlaps of the many populated eigenstates concentrate in a sharp peak at $Q\simeq0$. Their nearly orthogonal patterns cancel in the mean pattern $\bar{\bm c}=\sum_m p_m\bm c_m$, and the mean of $P_{\rm DE}(Q)$ drops below $Q_I$. Since $\Delta\Rig>0$ at the same time [Fig.~\ref{fig:geometry}(d)], this is the microscopic content of the statement above: the undershoot reflects the mutual incompatibility of increasingly glassy eigenstates rather than a loss of glassiness. The same qualitative behavior is observed for $L=16$, the largest system size investigated, as shown in Fig.~\ref{fig:L16} of the Supplemental Material~\cite{SM}, further supporting the physical picture discussed above.

The origin of the spread $\Spread$ lies in the variation of the eigenbasis with $\lambda$, generated by the adiabatic gauge potential (AGP), whose off-diagonal elements in the parallel gauge are $\AGP_{m\ell}(\lambda)=\ii\braket{m(\lambda)|\partial_\lambda\ell(\lambda)}=-\ii h_I\bra{m(\lambda)}X\ket{\ell(\lambda)}/[E_\ell(\lambda)-E_m(\lambda)]$. The AGP flow determines both $\Rig(\lambda)$ and $\Spread(\lambda)$ from the initial state alone (Sec.~\ref{sec:AGP} of the Supplemental Material~\cite{SM}). Close to the no-quench point, and away from exact degeneracies, the spread of a single initial state $\ket{n}$, whose average is $\Spread$, reads
\begin{equation}
\Spread^{(n)}(1+\delta\lambda) = \frac{\delta\lambda^2}{\Np}\sum_{m\neq n} \bigl|\AGP_{mn}(1)\bigr|^2 \,d_{mn}^{2} + O\!\left(|\delta \lambda|^3\right),
\label{eq:pattern_absorption}
\end{equation}
where $d_{mn}^{2}=\|\bm c_m(1)-\bm c_n(1)\|^2$ is the distance between the unperturbed patterns of $\ket{n}$ and $\ket{m}$. Thus $\Spread$ is a \emph{pattern-weighted fidelity susceptibility}: it is large only when the quench hybridizes the initial state with nearby levels carrying very different frozen patterns. The self-overlap of the adiabatically connected pattern, instead, changes already at linear order in $\delta\lambda$, so that immediately below $\lambda=1$ the rigidity gain always dominates. The width of this perturbative window is set by the smallest relevant gaps, and it closes much sooner in hot shells, which controls the cold/hot dichotomy. In \textit{hot shells} the level spacings are smallest, and many overlapping avoided crossings of $H_F$ are encountered as soon as $\lambda$ decreases below unity. Because avoided crossings that share a level do not commute, the initial state fragments across multiple spectral branches, driving $P_2$ well below $1/2$ (see Fig.~\ref{fig:agp_microscopic} of the Supplemental Material~\cite{SM}), and $\Spread$ rapidly overtakes $\Delta\Rig$, producing the cusp at $\lambda=1$ and the undershoot. In \textit{cold shells} avoided crossings are dilute and well separated, so that the weight is shared between only two eigenstates at a time, $P_2\gtrsim1/2$; the spread then remains bounded and the rigidity gain wins, producing the overshoot. This behavior is captured quantitatively by an isolated-resonance theory, in which each resonance contributes constructively to $\QDE$ (Sec.~\ref{app:cold_shell_theory} of the Supplemental Material~\cite{SM}); dynamical and spectral diagnostics supporting this picture are given in Secs.~\ref{sec:resonant_network} and~\ref{sec:spectral}.

\begin{figure*}[ht]
\centering
\includegraphics[width=0.8\linewidth]{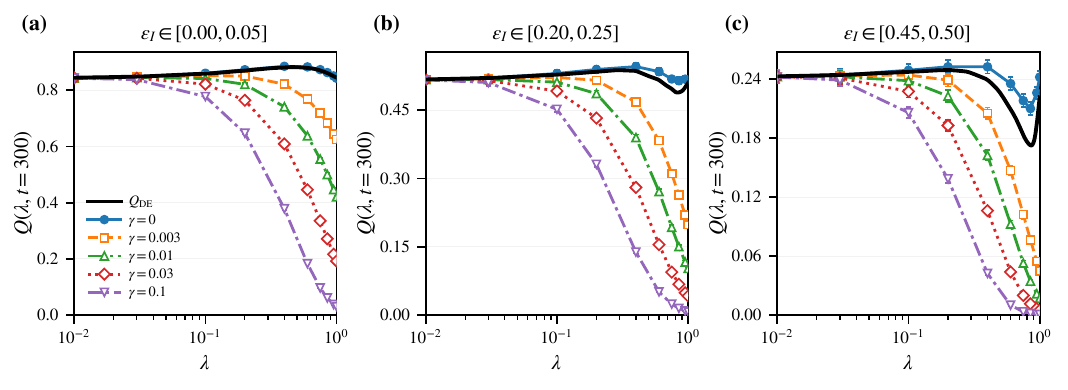}
\caption{
Weak longitudinal dephasing of the post-quench response. Disorder-averaged $Q(\lambda,\gamma;t)$ at $t=300$ for $L=13$, $\alpha=0.85$, and $h_I=0.2$, using $10^3$ Monte Carlo trajectories per disorder realization, averaged over $10^3$ disorder realizations. Panels correspond to $\epsilon_I\in[0,0.05]$, $[0.20,0.25]$, and $[0.45,0.50]$ (left to right). Colored curves show the GKSL dynamics for the indicated $\gamma$, while the black curve denotes the closed-system diagonal ensemble $\QDE(\lambda)$. The endpoint $\lambda\to0$ remains exactly pinned at $Q_I$ [Eq.~\eqref{eq:protected}], whereas dephasing progressively suppresses the
finite-$\lambda$ response and removes the cold-shell overshoot by $\gamma\simeq10^{-2}$. The residual offset of the $\gamma=0$ data from $\QDE$ near $\lambda=1$ reflects coherences with $|E_m-E_n|\lesssim t^{-1}$.
}
\label{fig:noise}
\end{figure*}

In trapped-ion qubits decoherence is dominated by pure dephasing of the qubit~\cite{Bruzewicz2019}, which acts as longitudinal or transverse noise depending on whether the Ising axis coincides with the qubit axis, as in Penning-trap simulators~\cite{Britton2012,Garttner2017}, or lies in its transverse plane, as in M{\o}lmer--S{\o}rensen-based Paul-trap simulators~\cite{MonroeRMP2021}. Let us therefore  ask how much of this structure survives when the same quench is executed on a device subject to weak noise, starting with the longitudinal case. 
The system is prepared in an eigenstate $\ket{n_I}$ of $H_I$ belonging to a fixed energy-density shell, the transverse field is quenched to $h_F=\lambda h_I$, and the state is propagated for a fixed time $t$ under the full Gorini--Kossakowski--Sudarshan--Lindblad (GKSL) generator
\begin{equation}
\frac{d\rho}{dt}
=-\ii\,[H_F,\rho]
+\gamma\sum_{i=1}^{L}\bigl(\sigma_i^z\rho\,\sigma_i^z-\rho\bigr),
\qquad
\rho(0)=\ketbra{n_I}{n_I},
\label{eq:lindblad}
\end{equation}
with $\gamma$ the local dephasing rate.  We then evaluate
\begin{equation}
Q(\lambda,\gamma;t)=
\frac{1}{\Np}\sum_{i<j}
\bigl[\Tr\!\left(\rho(t)\,\sigma_i^z\sigma_j^z\right)\bigr]^{2},
\label{eq:Qopen}
\end{equation}
and perform the disorder averages only afterwards.

The longitudinal pair operators are dark under the dissipator itself. Their decay at finite transverse field is instead generated by the coherent dynamics, since $[X,\sigma_i^z\sigma_j^z]
=
-2\ii
\left(
\sigma_i^y\sigma_j^z
+
\sigma_i^z\sigma_j^y
\right)$,
which rotates them into operators damped by the dephasing channel. At $\lambda=0$, however, $H_F$ contains only longitudinal Ising terms and commutes with every $\sigma_i^z\sigma_j^z$, yielding the exact protected endpoint
\begin{equation}
Q(0,\gamma;t)=Q_I,
\qquad
\forall\,\gamma\geq0,\quad t\geq0.
\label{eq:protected}
\end{equation}
For any fixed $\lambda>0$ and $\gamma>0$, by contrast, the finite system ultimately relaxes to the maximally mixed state and $Q\to0$. The finite-$\lambda$ response is therefore noise sensitive, while $\lambda=0$ provides an exact internal reference.

Figure~\ref{fig:noise} shows the shell-dependent suppression and reshaping of the finite-$\lambda$ response under increasing longitudinal dephasing at $t=300$. Increasing $\gamma$ progressively suppresses the cold-shell overshoot, whereas in hotter shells it modifies the recovery toward $Q_I$ near $\lambda=1$. The effect is thus shell dependent rather than a uniform attenuation of $Q$. The $\gamma=0$ data closely follow the diagonal-ensemble result $\QDE(\lambda)$ over most of the interval, while the residual deviations near $\lambda=1$ arise from slowly dephasing coherences with $|E_m-E_n|\lesssim t^{-1}$.

The protection of  $\lambda=0$ is specific to noise that commutes with the Ising interaction, but the  endpoint remains exactly solvable for general local Pauli noise. Replacing the dissipator in Eq.~\eqref{eq:lindblad} by $\sum_{i}\sum_{a=x,y,z}\gamma_a\bigl(\sigma_i^a\rho\,\sigma_i^a-\rho\bigr)$ and using $[V,\sigma_i^z\sigma_j^z]=0$ and $\sigma_i^{a}\sigma_i^z\sigma_i^{a}=-\sigma_i^z$ for $a=x,y$, each pair correlator obeys $d\langle\sigma_i^z\sigma_j^z\rangle/dt=-4(\gamma_x+\gamma_y)\langle\sigma_i^z\sigma_j^z\rangle$, so that
\begin{equation}
Q(0;t)=Q_I\,e^{-8(\gamma_x+\gamma_y)t},
\label{eq:pauli_endpoint}
\end{equation}
independently of the initial state, energy shell, and disorder realization. The longitudinal rate $\gamma_z$ drops out, and Eq.~\eqref{eq:protected} is recovered for $\gamma_x=\gamma_y=0$. The $\lambda=0$ point thus directly measures the spin-flip component of the noise, $\gamma_x+\gamma_y=-(8t)^{-1}\ln[Q(0;t)/Q_I]$, whereas the finite-$\lambda$ response is also sensitive to longitudinal dephasing, so that the same protocol separates the two classes of noise. In the weak-noise kinetic limit, transverse noise indeed suppresses both the zero-field reference and the finite-quench response, in a shell-dependent way (see Fig.~\ref{fig:transverse_noise_slow} of the Supplemental Material~\cite{SM}).

In conclusion, we have shown that nonequilibrium spin-glass order after a transverse-field quench results from an exact competition, naturally expressed in terms of overlaps, between the self-overlap of the populated final eigenstates and the spread of their frozen pair patterns. In cold shells the weight remains on a few compatible eigenstates, and the rigidity gain produces an overshoot that is captured by an isolated-resonance theory. In hot shells it fragments over many eigenstates with mutually incompatible patterns, which cancel and produce an undershoot. Post-quench glass order is thus controlled not only by the EA order of individual eigenstates, but also by the compatibility of their frozen patterns. Under local longitudinal dephasing the zero-field endpoint remains exactly protected while the finite-$\lambda$ response is progressively suppressed. For general local Pauli noise the same endpoint decays as $e^{-8(\gamma_x+\gamma_y)t}$, so that the protocol separates spin-flip noise from longitudinal dephasing. Since all the required information is contained in equal-time $\sigma^z\sigma^z$ correlations, it is accessible from projective snapshots without state tomography. This suggests frustrated many-body dynamics as a self-referenced probe of decoherence in programmable quantum simulators.

\begin{acknowledgments}
    
All numerical simulations were coded and run by G.D.; A.S. and G.D. acknowledge support from the PNRR MUR program under D.M. 629/2024, Mission 4, Component 1, Investment 4.1.
The authors thank Lara Faoro, Mikhail Feigel'man, Lev Ioffe, Anatoli Polkovnikov, Marco Schirò, and  Bojan \v{Z}unkovi\v{c}  for insightful discussions regarding the critical aspects of the model.
\end{acknowledgments}

\bibliography{bibliography.bib}

\begin{thebibliography}{49}%
\makeatletter
\providecommand \@ifxundefined [1]{%
 \@ifx{#1\undefined}
}%
\providecommand \@ifnum [1]{%
 \ifnum #1\expandafter \@firstoftwo
 \else \expandafter \@secondoftwo
 \fi
}%
\providecommand \@ifx [1]{%
 \ifx #1\expandafter \@firstoftwo
 \else \expandafter \@secondoftwo
 \fi
}%
\providecommand \natexlab [1]{#1}%
\providecommand \enquote  [1]{``#1''}%
\providecommand \bibnamefont  [1]{#1}%
\providecommand \bibfnamefont [1]{#1}%
\providecommand \citenamefont [1]{#1}%
\providecommand \href@noop [0]{\@secondoftwo}%
\providecommand \href [0]{\begingroup \@sanitize@url \@href}%
\providecommand \@href[1]{\@@startlink{#1}\@@href}%
\providecommand \@@href[1]{\endgroup#1\@@endlink}%
\providecommand \@sanitize@url [0]{\catcode `\\12\catcode `\$12\catcode `\&12\catcode `\#12\catcode `\^12\catcode `\_12\catcode `\%12\relax}%
\providecommand \@@startlink[1]{}%
\providecommand \@@endlink[0]{}%
\providecommand \url  [0]{\begingroup\@sanitize@url \@url }%
\providecommand \@url [1]{\endgroup\@href {#1}{\urlprefix }}%
\providecommand \urlprefix  [0]{URL }%
\providecommand \Eprint [0]{\href }%
\providecommand \doibase [0]{https://doi.org/}%
\providecommand \selectlanguage [0]{\@gobble}%
\providecommand \bibinfo  [0]{\@secondoftwo}%
\providecommand \bibfield  [0]{\@secondoftwo}%
\providecommand \translation [1]{[#1]}%
\providecommand \BibitemOpen [0]{}%
\providecommand \bibitemStop [0]{}%
\providecommand \bibitemNoStop [0]{.\EOS\space}%
\providecommand \EOS [0]{\spacefactor3000\relax}%
\providecommand \BibitemShut  [1]{\csname bibitem#1\endcsname}%
\let\auto@bib@innerbib\@empty
%</preamble>
\bibitem [{\citenamefont {Polkovnikov}\ \emph {et~al.}(2011)\citenamefont {Polkovnikov}, \citenamefont {Sengupta}, \citenamefont {Silva},\ and\ \citenamefont {Vengalattore}}]{Polkovnikov2011}%
  \BibitemOpen
  \bibfield  {author} {\bibinfo {author} {\bibfnamefont {A.}~\bibnamefont {Polkovnikov}}, \bibinfo {author} {\bibfnamefont {K.}~\bibnamefont {Sengupta}}, \bibinfo {author} {\bibfnamefont {A.}~\bibnamefont {Silva}},\ and\ \bibinfo {author} {\bibfnamefont {M.}~\bibnamefont {Vengalattore}},\ }\bibfield  {title} {\bibinfo {title} {Colloquium: Nonequilibrium dynamics of closed interacting quantum systems},\ }\href {https://doi.org/10.1103/RevModPhys.83.863} {\bibfield  {journal} {\bibinfo  {journal} {Rev. Mod. Phys.}\ }\textbf {\bibinfo {volume} {83}},\ \bibinfo {pages} {863} (\bibinfo {year} {2011})}\BibitemShut {NoStop}%
\bibitem [{\citenamefont {D'Alessio}\ \emph {et~al.}(2016)\citenamefont {D'Alessio}, \citenamefont {Kafri}, \citenamefont {Polkovnikov},\ and\ \citenamefont {Rigol}}]{DAlessio2016}%
  \BibitemOpen
  \bibfield  {author} {\bibinfo {author} {\bibfnamefont {L.}~\bibnamefont {D'Alessio}}, \bibinfo {author} {\bibfnamefont {Y.}~\bibnamefont {Kafri}}, \bibinfo {author} {\bibfnamefont {A.}~\bibnamefont {Polkovnikov}},\ and\ \bibinfo {author} {\bibfnamefont {M.}~\bibnamefont {Rigol}},\ }\bibfield  {title} {\bibinfo {title} {From quantum chaos and eigenstate thermalization to statistical mechanics and thermodynamics},\ }\href {https://doi.org/10.1080/00018732.2016.1198134} {\bibfield  {journal} {\bibinfo  {journal} {Adv. Phys.}\ }\textbf {\bibinfo {volume} {65}},\ \bibinfo {pages} {239} (\bibinfo {year} {2016})}\BibitemShut {NoStop}%
\bibitem [{\citenamefont {Abanin}\ \emph {et~al.}(2019)\citenamefont {Abanin}, \citenamefont {Altman}, \citenamefont {Bloch},\ and\ \citenamefont {Serbyn}}]{Abanin2019}%
  \BibitemOpen
  \bibfield  {author} {\bibinfo {author} {\bibfnamefont {D.~A.}\ \bibnamefont {Abanin}}, \bibinfo {author} {\bibfnamefont {E.}~\bibnamefont {Altman}}, \bibinfo {author} {\bibfnamefont {I.}~\bibnamefont {Bloch}},\ and\ \bibinfo {author} {\bibfnamefont {M.}~\bibnamefont {Serbyn}},\ }\bibfield  {title} {\bibinfo {title} {Colloquium: Many-body localization, thermalization, and entanglement},\ }\href {https://doi.org/10.1103/RevModPhys.91.021001} {\bibfield  {journal} {\bibinfo  {journal} {Rev. Mod. Phys.}\ }\textbf {\bibinfo {volume} {91}},\ \bibinfo {pages} {021001} (\bibinfo {year} {2019})}\BibitemShut {NoStop}%
\bibitem [{\citenamefont {Nandkishore}\ and\ \citenamefont {Huse}(2015)}]{NandkishoreHuse2015}%
  \BibitemOpen
  \bibfield  {author} {\bibinfo {author} {\bibfnamefont {R.}~\bibnamefont {Nandkishore}}\ and\ \bibinfo {author} {\bibfnamefont {D.~A.}\ \bibnamefont {Huse}},\ }\bibfield  {title} {\bibinfo {title} {Many-body localization and thermalization in quantum statistical mechanics},\ }\href {https://doi.org/10.1146/annurev-conmatphys-031214-014726} {\bibfield  {journal} {\bibinfo  {journal} {Annu. Rev. Condens. Matter Phys.}\ }\textbf {\bibinfo {volume} {6}},\ \bibinfo {pages} {15} (\bibinfo {year} {2015})}\BibitemShut {NoStop}%
\bibitem [{\citenamefont {Else}\ \emph {et~al.}(2020)\citenamefont {Else}, \citenamefont {Monroe}, \citenamefont {Nayak},\ and\ \citenamefont {Yao}}]{Else2020}%
  \BibitemOpen
  \bibfield  {author} {\bibinfo {author} {\bibfnamefont {D.~V.}\ \bibnamefont {Else}}, \bibinfo {author} {\bibfnamefont {C.}~\bibnamefont {Monroe}}, \bibinfo {author} {\bibfnamefont {C.}~\bibnamefont {Nayak}},\ and\ \bibinfo {author} {\bibfnamefont {N.~Y.}\ \bibnamefont {Yao}},\ }\bibfield  {title} {\bibinfo {title} {Discrete time crystals},\ }\href {https://doi.org/10.1146/annurev-conmatphys-031119-050658} {\bibfield  {journal} {\bibinfo  {journal} {Annu. Rev. Condens. Matter Phys.}\ }\textbf {\bibinfo {volume} {11}},\ \bibinfo {pages} {467} (\bibinfo {year} {2020})}\BibitemShut {NoStop}%
\bibitem [{\citenamefont {Binder}\ and\ \citenamefont {Young}(1986)}]{BinderYoung1986}%
  \BibitemOpen
  \bibfield  {author} {\bibinfo {author} {\bibfnamefont {K.}~\bibnamefont {Binder}}\ and\ \bibinfo {author} {\bibfnamefont {A.~P.}\ \bibnamefont {Young}},\ }\bibfield  {title} {\bibinfo {title} {Spin glasses: Experimental facts, theoretical concepts, and open questions},\ }\href {https://doi.org/10.1103/RevModPhys.58.801} {\bibfield  {journal} {\bibinfo  {journal} {Rev. Mod. Phys.}\ }\textbf {\bibinfo {volume} {58}},\ \bibinfo {pages} {801} (\bibinfo {year} {1986})}\BibitemShut {NoStop}%
\bibitem [{\citenamefont {Miller}\ and\ \citenamefont {Huse}(1993)}]{MillerHuse1993}%
  \BibitemOpen
  \bibfield  {author} {\bibinfo {author} {\bibfnamefont {J.}~\bibnamefont {Miller}}\ and\ \bibinfo {author} {\bibfnamefont {D.~A.}\ \bibnamefont {Huse}},\ }\bibfield  {title} {\bibinfo {title} {Zero-temperature critical behavior of the infinite-range quantum {Ising} spin glass},\ }\href {https://doi.org/10.1103/PhysRevLett.70.3147} {\bibfield  {journal} {\bibinfo  {journal} {Phys. Rev. Lett.}\ }\textbf {\bibinfo {volume} {70}},\ \bibinfo {pages} {3147} (\bibinfo {year} {1993})}\BibitemShut {NoStop}%
\bibitem [{\citenamefont {Harrigan}\ \emph {et~al.}(2021)\citenamefont {Harrigan} \emph {et~al.}}]{Harrigan2021}%
  \BibitemOpen
  \bibfield  {author} {\bibinfo {author} {\bibfnamefont {M.~P.}\ \bibnamefont {Harrigan}} \emph {et~al.},\ }\bibfield  {title} {\bibinfo {title} {Quantum approximate optimization of non-planar graph problems on a planar superconducting processor},\ }\href {https://doi.org/10.1038/s41567-020-01105-y} {\bibfield  {journal} {\bibinfo  {journal} {Nat. Phys.}\ }\textbf {\bibinfo {volume} {17}},\ \bibinfo {pages} {332} (\bibinfo {year} {2021})}\BibitemShut {NoStop}%
\bibitem [{\citenamefont {Pelofske}\ \emph {et~al.}(2024)\citenamefont {Pelofske}, \citenamefont {B{\"a}rtschi}, \citenamefont {Cincio}, \citenamefont {Golden},\ and\ \citenamefont {Eidenbenz}}]{Pelofske2024}%
  \BibitemOpen
  \bibfield  {author} {\bibinfo {author} {\bibfnamefont {E.}~\bibnamefont {Pelofske}}, \bibinfo {author} {\bibfnamefont {A.}~\bibnamefont {B{\"a}rtschi}}, \bibinfo {author} {\bibfnamefont {L.}~\bibnamefont {Cincio}}, \bibinfo {author} {\bibfnamefont {J.}~\bibnamefont {Golden}},\ and\ \bibinfo {author} {\bibfnamefont {S.}~\bibnamefont {Eidenbenz}},\ }\bibfield  {title} {\bibinfo {title} {Scaling whole-chip {QAOA} for higher-order {Ising} spin glass models on heavy-hex graphs},\ }\href {https://doi.org/10.1038/s41534-024-00906-w} {\bibfield  {journal} {\bibinfo  {journal} {npj Quantum Inf.}\ }\textbf {\bibinfo {volume} {10}},\ \bibinfo {pages} {109} (\bibinfo {year} {2024})}\BibitemShut {NoStop}%
\bibitem [{\citenamefont {Harris}\ \emph {et~al.}(2018)\citenamefont {Harris}, \citenamefont {Sato}, \citenamefont {Berkley}, \citenamefont {Reis}, \citenamefont {Altomare}, \citenamefont {Amin}, \citenamefont {Boothby}, \citenamefont {Bunyk}, \citenamefont {Deng}, \citenamefont {Enderud} \emph {et~al.}}]{Harris2018}%
  \BibitemOpen
  \bibfield  {author} {\bibinfo {author} {\bibfnamefont {R.}~\bibnamefont {Harris}}, \bibinfo {author} {\bibfnamefont {Y.}~\bibnamefont {Sato}}, \bibinfo {author} {\bibfnamefont {A.~J.}\ \bibnamefont {Berkley}}, \bibinfo {author} {\bibfnamefont {M.}~\bibnamefont {Reis}}, \bibinfo {author} {\bibfnamefont {F.}~\bibnamefont {Altomare}}, \bibinfo {author} {\bibfnamefont {M.~H.}\ \bibnamefont {Amin}}, \bibinfo {author} {\bibfnamefont {K.}~\bibnamefont {Boothby}}, \bibinfo {author} {\bibfnamefont {P.}~\bibnamefont {Bunyk}}, \bibinfo {author} {\bibfnamefont {C.}~\bibnamefont {Deng}}, \bibinfo {author} {\bibfnamefont {C.}~\bibnamefont {Enderud}}, \emph {et~al.},\ }\bibfield  {title} {\bibinfo {title} {Phase transitions in a programmable quantum spin glass simulator},\ }\href {https://doi.org/10.1126/science.aat2025} {\bibfield  {journal} {\bibinfo  {journal} {Science}\ }\textbf {\bibinfo {volume} {361}},\ \bibinfo {pages} {162} (\bibinfo {year} {2018})}\BibitemShut {NoStop}%
\bibitem [{\citenamefont {King}\ \emph {et~al.}(2022)\citenamefont {King}, \citenamefont {Suzuki}, \citenamefont {Raymond}, \citenamefont {Zucca}, \citenamefont {Lanting}, \citenamefont {Altomare}, \citenamefont {Berkley}, \citenamefont {Ejtemaee}, \citenamefont {Hoskinson}, \citenamefont {Huang} \emph {et~al.}}]{King2022}%
  \BibitemOpen
  \bibfield  {author} {\bibinfo {author} {\bibfnamefont {A.~D.}\ \bibnamefont {King}}, \bibinfo {author} {\bibfnamefont {S.}~\bibnamefont {Suzuki}}, \bibinfo {author} {\bibfnamefont {J.}~\bibnamefont {Raymond}}, \bibinfo {author} {\bibfnamefont {A.}~\bibnamefont {Zucca}}, \bibinfo {author} {\bibfnamefont {T.}~\bibnamefont {Lanting}}, \bibinfo {author} {\bibfnamefont {F.}~\bibnamefont {Altomare}}, \bibinfo {author} {\bibfnamefont {A.~J.}\ \bibnamefont {Berkley}}, \bibinfo {author} {\bibfnamefont {S.}~\bibnamefont {Ejtemaee}}, \bibinfo {author} {\bibfnamefont {E.}~\bibnamefont {Hoskinson}}, \bibinfo {author} {\bibfnamefont {S.}~\bibnamefont {Huang}}, \emph {et~al.},\ }\bibfield  {title} {\bibinfo {title} {Coherent quantum annealing in a programmable 2,000 qubit {Ising} chain},\ }\href {https://doi.org/10.1038/s41567-022-01741-6} {\bibfield  {journal} {\bibinfo  {journal} {Nat. Phys.}\ }\textbf {\bibinfo {volume} {18}},\ \bibinfo {pages} {1324} (\bibinfo {year} {2022})}\BibitemShut {NoStop}%
\bibitem [{\citenamefont {King}\ \emph {et~al.}(2023)\citenamefont {King}, \citenamefont {Raymond}, \citenamefont {Lanting}, \citenamefont {Harris}, \citenamefont {Zucca} \emph {et~al.}}]{King2023}%
  \BibitemOpen
  \bibfield  {author} {\bibinfo {author} {\bibfnamefont {A.~D.}\ \bibnamefont {King}}, \bibinfo {author} {\bibfnamefont {J.}~\bibnamefont {Raymond}}, \bibinfo {author} {\bibfnamefont {T.}~\bibnamefont {Lanting}}, \bibinfo {author} {\bibfnamefont {R.}~\bibnamefont {Harris}}, \bibinfo {author} {\bibfnamefont {A.}~\bibnamefont {Zucca}}, \emph {et~al.},\ }\bibfield  {title} {\bibinfo {title} {Quantum critical dynamics in a 5,000-qubit programmable spin glass},\ }\href {https://doi.org/10.1038/s41586-023-05867-2} {\bibfield  {journal} {\bibinfo  {journal} {Nature}\ }\textbf {\bibinfo {volume} {617}},\ \bibinfo {pages} {61} (\bibinfo {year} {2023})}\BibitemShut {NoStop}%
\bibitem [{\citenamefont {King}\ \emph {et~al.}(2025)\citenamefont {King}, \citenamefont {Nocera}, \citenamefont {Rams}, \citenamefont {Dziarmaga}, \citenamefont {Wiersema}, \citenamefont {Bernoudy}, \citenamefont {Raymond}, \citenamefont {Kaushal}, \citenamefont {Heinsdorf}, \citenamefont {Harris} \emph {et~al.}}]{King2025}%
  \BibitemOpen
  \bibfield  {author} {\bibinfo {author} {\bibfnamefont {A.~D.}\ \bibnamefont {King}}, \bibinfo {author} {\bibfnamefont {A.}~\bibnamefont {Nocera}}, \bibinfo {author} {\bibfnamefont {M.~M.}\ \bibnamefont {Rams}}, \bibinfo {author} {\bibfnamefont {J.}~\bibnamefont {Dziarmaga}}, \bibinfo {author} {\bibfnamefont {R.}~\bibnamefont {Wiersema}}, \bibinfo {author} {\bibfnamefont {W.}~\bibnamefont {Bernoudy}}, \bibinfo {author} {\bibfnamefont {J.}~\bibnamefont {Raymond}}, \bibinfo {author} {\bibfnamefont {N.}~\bibnamefont {Kaushal}}, \bibinfo {author} {\bibfnamefont {N.}~\bibnamefont {Heinsdorf}}, \bibinfo {author} {\bibfnamefont {R.}~\bibnamefont {Harris}}, \emph {et~al.},\ }\bibfield  {title} {\bibinfo {title} {Beyond-classical computation in quantum simulation},\ }\href {https://doi.org/10.1126/science.ado6285} {\bibfield  {journal} {\bibinfo  {journal} {Science}\ }\textbf {\bibinfo {volume} {388}},\ \bibinfo {pages} {199} (\bibinfo {year} {2025})}\BibitemShut {NoStop}%
\bibitem [{\citenamefont {Islam}\ \emph {et~al.}(2011)\citenamefont {Islam}, \citenamefont {Edwards}, \citenamefont {Kim}, \citenamefont {Korenblit}, \citenamefont {Noh}, \citenamefont {Carmichael}, \citenamefont {Lin}, \citenamefont {Duan}, \citenamefont {Joseph~Wang}, \citenamefont {Freericks},\ and\ \citenamefont {Monroe}}]{islam_2011}%
  \BibitemOpen
  \bibfield  {author} {\bibinfo {author} {\bibfnamefont {R.}~\bibnamefont {Islam}}, \bibinfo {author} {\bibfnamefont {E.}~\bibnamefont {Edwards}}, \bibinfo {author} {\bibfnamefont {K.}~\bibnamefont {Kim}}, \bibinfo {author} {\bibfnamefont {S.}~\bibnamefont {Korenblit}}, \bibinfo {author} {\bibfnamefont {C.}~\bibnamefont {Noh}}, \bibinfo {author} {\bibfnamefont {H.}~\bibnamefont {Carmichael}}, \bibinfo {author} {\bibfnamefont {G.-D.}\ \bibnamefont {Lin}}, \bibinfo {author} {\bibfnamefont {L.-M.}\ \bibnamefont {Duan}}, \bibinfo {author} {\bibfnamefont {C.-C.}\ \bibnamefont {Joseph~Wang}}, \bibinfo {author} {\bibfnamefont {J.}~\bibnamefont {Freericks}},\ and\ \bibinfo {author} {\bibfnamefont {C.}~\bibnamefont {Monroe}},\ }\bibfield  {title} {\bibinfo {title} {Onset of a quantum phase transition with a trapped ion quantum simulator},\ }\href {https://doi.org/10.1038/ncomms1374} {\bibfield  {journal} {\bibinfo  {journal} {Nat. Commun.}\ }\textbf {\bibinfo {volume} {2}},\ \bibinfo {pages} {377} (\bibinfo {year}
  {2011})}\BibitemShut {NoStop}%
\bibitem [{\citenamefont {Zhang}\ \emph {et~al.}(2017)\citenamefont {Zhang}, \citenamefont {Pagano}, \citenamefont {Hess}, \citenamefont {Kyprianidis}, \citenamefont {Becker}, \citenamefont {Kaplan}, \citenamefont {Gorshkov}, \citenamefont {Gong},\ and\ \citenamefont {Monroe}}]{Zhang2017}%
  \BibitemOpen
  \bibfield  {author} {\bibinfo {author} {\bibfnamefont {J.}~\bibnamefont {Zhang}}, \bibinfo {author} {\bibfnamefont {G.}~\bibnamefont {Pagano}}, \bibinfo {author} {\bibfnamefont {P.~W.}\ \bibnamefont {Hess}}, \bibinfo {author} {\bibfnamefont {A.}~\bibnamefont {Kyprianidis}}, \bibinfo {author} {\bibfnamefont {P.}~\bibnamefont {Becker}}, \bibinfo {author} {\bibfnamefont {H.}~\bibnamefont {Kaplan}}, \bibinfo {author} {\bibfnamefont {A.~V.}\ \bibnamefont {Gorshkov}}, \bibinfo {author} {\bibfnamefont {Z.-X.}\ \bibnamefont {Gong}},\ and\ \bibinfo {author} {\bibfnamefont {C.}~\bibnamefont {Monroe}},\ }\bibfield  {title} {\bibinfo {title} {Observation of a many-body dynamical phase transition with a 53-qubit quantum simulator},\ }\href {https://doi.org/10.1038/nature24654} {\bibfield  {journal} {\bibinfo  {journal} {Nature}\ }\textbf {\bibinfo {volume} {551}},\ \bibinfo {pages} {601} (\bibinfo {year} {2017})}\BibitemShut {NoStop}%
\bibitem [{\citenamefont {Monroe}\ \emph {et~al.}(2021)\citenamefont {Monroe}, \citenamefont {Campbell}, \citenamefont {Duan}, \citenamefont {Gong}, \citenamefont {Gorshkov}, \citenamefont {Hess}, \citenamefont {Islam}, \citenamefont {Kim}, \citenamefont {Linke}, \citenamefont {Pagano}, \citenamefont {Richerme}, \citenamefont {Senko},\ and\ \citenamefont {Yao}}]{MonroeRMP2021}%
  \BibitemOpen
  \bibfield  {author} {\bibinfo {author} {\bibfnamefont {C.}~\bibnamefont {Monroe}}, \bibinfo {author} {\bibfnamefont {W.}~\bibnamefont {Campbell}}, \bibinfo {author} {\bibfnamefont {L.-M.}\ \bibnamefont {Duan}}, \bibinfo {author} {\bibfnamefont {Z.-X.}\ \bibnamefont {Gong}}, \bibinfo {author} {\bibfnamefont {A.}~\bibnamefont {Gorshkov}}, \bibinfo {author} {\bibfnamefont {P.}~\bibnamefont {Hess}}, \bibinfo {author} {\bibfnamefont {R.}~\bibnamefont {Islam}}, \bibinfo {author} {\bibfnamefont {K.}~\bibnamefont {Kim}}, \bibinfo {author} {\bibfnamefont {N.}~\bibnamefont {Linke}}, \bibinfo {author} {\bibfnamefont {G.}~\bibnamefont {Pagano}}, \bibinfo {author} {\bibfnamefont {P.}~\bibnamefont {Richerme}}, \bibinfo {author} {\bibfnamefont {C.}~\bibnamefont {Senko}},\ and\ \bibinfo {author} {\bibfnamefont {N.}~\bibnamefont {Yao}},\ }\bibfield  {title} {\bibinfo {title} {Programmable quantum simulations of spin systems with trapped ions},\ }\href {https://doi.org/10.1103/revmodphys.93.025001} {\bibfield  {journal}
  {\bibinfo  {journal} {Rev. Mod. Phys.}\ }\textbf {\bibinfo {volume} {93}},\ \bibinfo {pages} {025001} (\bibinfo {year} {2021})}\BibitemShut {NoStop}%
\bibitem [{\citenamefont {Ebadi}\ \emph {et~al.}(2021)\citenamefont {Ebadi}, \citenamefont {Wang}, \citenamefont {Levine}, \citenamefont {Keesling}, \citenamefont {Semeghini}, \citenamefont {Omran}, \citenamefont {Bluvstein}, \citenamefont {Samajdar}, \citenamefont {Pichler}, \citenamefont {Ho}, \citenamefont {Choi}, \citenamefont {Sachdev}, \citenamefont {Greiner}, \citenamefont {Vuleti{\'c}},\ and\ \citenamefont {Lukin}}]{Ebadi2021}%
  \BibitemOpen
  \bibfield  {author} {\bibinfo {author} {\bibfnamefont {S.}~\bibnamefont {Ebadi}}, \bibinfo {author} {\bibfnamefont {T.~T.}\ \bibnamefont {Wang}}, \bibinfo {author} {\bibfnamefont {H.}~\bibnamefont {Levine}}, \bibinfo {author} {\bibfnamefont {A.}~\bibnamefont {Keesling}}, \bibinfo {author} {\bibfnamefont {G.}~\bibnamefont {Semeghini}}, \bibinfo {author} {\bibfnamefont {A.}~\bibnamefont {Omran}}, \bibinfo {author} {\bibfnamefont {D.}~\bibnamefont {Bluvstein}}, \bibinfo {author} {\bibfnamefont {R.}~\bibnamefont {Samajdar}}, \bibinfo {author} {\bibfnamefont {H.}~\bibnamefont {Pichler}}, \bibinfo {author} {\bibfnamefont {W.~W.}\ \bibnamefont {Ho}}, \bibinfo {author} {\bibfnamefont {S.}~\bibnamefont {Choi}}, \bibinfo {author} {\bibfnamefont {S.}~\bibnamefont {Sachdev}}, \bibinfo {author} {\bibfnamefont {M.}~\bibnamefont {Greiner}}, \bibinfo {author} {\bibfnamefont {V.}~\bibnamefont {Vuleti{\'c}}},\ and\ \bibinfo {author} {\bibfnamefont {M.~D.}\ \bibnamefont {Lukin}},\ }\bibfield  {title} {\bibinfo {title}
  {Quantum phases of matter on a 256-atom programmable quantum simulator},\ }\href {https://doi.org/10.1038/s41586-021-03582-4} {\bibfield  {journal} {\bibinfo  {journal} {Nature}\ }\textbf {\bibinfo {volume} {595}},\ \bibinfo {pages} {227} (\bibinfo {year} {2021})}\BibitemShut {NoStop}%
\bibitem [{\citenamefont {Ebadi}\ \emph {et~al.}(2022)\citenamefont {Ebadi}, \citenamefont {Keesling}, \citenamefont {Cain}, \citenamefont {Wang}, \citenamefont {Levine}, \citenamefont {Bluvstein}, \citenamefont {Semeghini}, \citenamefont {Omran}, \citenamefont {Liu}, \citenamefont {Samajdar} \emph {et~al.}}]{Ebadi2022}%
  \BibitemOpen
  \bibfield  {author} {\bibinfo {author} {\bibfnamefont {S.}~\bibnamefont {Ebadi}}, \bibinfo {author} {\bibfnamefont {A.}~\bibnamefont {Keesling}}, \bibinfo {author} {\bibfnamefont {M.}~\bibnamefont {Cain}}, \bibinfo {author} {\bibfnamefont {T.~T.}\ \bibnamefont {Wang}}, \bibinfo {author} {\bibfnamefont {H.}~\bibnamefont {Levine}}, \bibinfo {author} {\bibfnamefont {D.}~\bibnamefont {Bluvstein}}, \bibinfo {author} {\bibfnamefont {G.}~\bibnamefont {Semeghini}}, \bibinfo {author} {\bibfnamefont {A.}~\bibnamefont {Omran}}, \bibinfo {author} {\bibfnamefont {J.-G.}\ \bibnamefont {Liu}}, \bibinfo {author} {\bibfnamefont {R.}~\bibnamefont {Samajdar}}, \emph {et~al.},\ }\bibfield  {title} {\bibinfo {title} {Quantum optimization of maximum independent set using {Rydberg} atom arrays},\ }\href {https://doi.org/10.1126/science.abo6587} {\bibfield  {journal} {\bibinfo  {journal} {Science}\ }\textbf {\bibinfo {volume} {376}},\ \bibinfo {pages} {1209} (\bibinfo {year} {2022})}\BibitemShut {NoStop}%
\bibitem [{\citenamefont {Kotliar}\ \emph {et~al.}(1983)\citenamefont {Kotliar}, \citenamefont {Anderson},\ and\ \citenamefont {Stein}}]{KAS83}%
  \BibitemOpen
  \bibfield  {author} {\bibinfo {author} {\bibfnamefont {G.}~\bibnamefont {Kotliar}}, \bibinfo {author} {\bibfnamefont {P.~W.}\ \bibnamefont {Anderson}},\ and\ \bibinfo {author} {\bibfnamefont {D.~L.}\ \bibnamefont {Stein}},\ }\bibfield  {title} {\bibinfo {title} {One-dimensional spin-glass model with long-range random interactions},\ }\href {https://doi.org/10.1103/PhysRevB.27.602} {\bibfield  {journal} {\bibinfo  {journal} {Phys. Rev. B}\ }\textbf {\bibinfo {volume} {27}},\ \bibinfo {pages} {602} (\bibinfo {year} {1983})}\BibitemShut {NoStop}%
\bibitem [{\citenamefont {Leuzzi}\ \emph {et~al.}(2008)\citenamefont {Leuzzi}, \citenamefont {Parisi}, \citenamefont {Ricci-Tersenghi},\ and\ \citenamefont {Ruiz-Lorenzo}}]{Leuzzi08}%
  \BibitemOpen
  \bibfield  {author} {\bibinfo {author} {\bibfnamefont {L.}~\bibnamefont {Leuzzi}}, \bibinfo {author} {\bibfnamefont {G.}~\bibnamefont {Parisi}}, \bibinfo {author} {\bibfnamefont {F.}~\bibnamefont {Ricci-Tersenghi}},\ and\ \bibinfo {author} {\bibfnamefont {J.~J.}\ \bibnamefont {Ruiz-Lorenzo}},\ }\bibfield  {title} {\bibinfo {title} {Dilute one-dimensional spin glasses with power-law decaying interactions},\ }\href {https://doi.org/10.1103/PhysRevLett.101.107203} {\bibfield  {journal} {\bibinfo  {journal} {Phys. Rev. Lett.}\ }\textbf {\bibinfo {volume} {101}},\ \bibinfo {pages} {107203} (\bibinfo {year} {2008})}\BibitemShut {NoStop}%
\bibitem [{\citenamefont {Moore}(2010)}]{Moore10}%
  \BibitemOpen
  \bibfield  {author} {\bibinfo {author} {\bibfnamefont {M.~A.}\ \bibnamefont {Moore}},\ }\bibfield  {title} {\bibinfo {title} {Ordered phase of the one-dimensional {Ising} spin glass with long-range interactions},\ }\href {https://doi.org/10.1103/PhysRevB.82.014417} {\bibfield  {journal} {\bibinfo  {journal} {Phys. Rev. B}\ }\textbf {\bibinfo {volume} {82}},\ \bibinfo {pages} {014417} (\bibinfo {year} {2010})}\BibitemShut {NoStop}%
\bibitem [{\citenamefont {Monthus}(2014)}]{Monthus14}%
  \BibitemOpen
  \bibfield  {author} {\bibinfo {author} {\bibfnamefont {C.}~\bibnamefont {Monthus}},\ }\bibfield  {title} {\bibinfo {title} {One-dimensional {Ising} spin-glass with power-law interaction: real-space renormalization at zero temperature},\ }\href {https://doi.org/10.1088/1742-5468/2014/14/P06015} {\bibfield  {journal} {\bibinfo  {journal} {J. Stat. Mech.: Theory Exp.}\ }\textbf {\bibinfo {volume} {2014}}\bibinfo  {number} { (6)},\ \bibinfo {pages} {P06015}}\BibitemShut {NoStop}%
\bibitem [{\citenamefont {Defenu}\ \emph {et~al.}(2023)\citenamefont {Defenu}, \citenamefont {Donner}, \citenamefont {Macr{\`\i}}, \citenamefont {Pagano}, \citenamefont {Ruffo},\ and\ \citenamefont {Trombettoni}}]{DefenuRMP23}%
  \BibitemOpen
\bibfield  {number} {  }\bibfield  {author} {\bibinfo {author} {\bibfnamefont {N.}~\bibnamefont {Defenu}}, \bibinfo {author} {\bibfnamefont {T.}~\bibnamefont {Donner}}, \bibinfo {author} {\bibfnamefont {T.}~\bibnamefont {Macr{\`\i}}}, \bibinfo {author} {\bibfnamefont {G.}~\bibnamefont {Pagano}}, \bibinfo {author} {\bibfnamefont {S.}~\bibnamefont {Ruffo}},\ and\ \bibinfo {author} {\bibfnamefont {A.}~\bibnamefont {Trombettoni}},\ }\bibfield  {title} {\bibinfo {title} {Long-range interacting quantum systems},\ }\href {https://doi.org/10.1103/RevModPhys.95.035002} {\bibfield  {journal} {\bibinfo  {journal} {Rev. Mod. Phys.}\ }\textbf {\bibinfo {volume} {95}},\ \bibinfo {pages} {035002} (\bibinfo {year} {2023})}\BibitemShut {NoStop}%
\bibitem [{\citenamefont {Smith}\ \emph {et~al.}(2016)\citenamefont {Smith}, \citenamefont {Lee}, \citenamefont {Richerme}, \citenamefont {Neyenhuis}, \citenamefont {Hess}, \citenamefont {Hauke}, \citenamefont {Heyl}, \citenamefont {Huse},\ and\ \citenamefont {Monroe}}]{smith2016}%
  \BibitemOpen
  \bibfield  {author} {\bibinfo {author} {\bibfnamefont {J.}~\bibnamefont {Smith}}, \bibinfo {author} {\bibfnamefont {A.}~\bibnamefont {Lee}}, \bibinfo {author} {\bibfnamefont {P.}~\bibnamefont {Richerme}}, \bibinfo {author} {\bibfnamefont {B.}~\bibnamefont {Neyenhuis}}, \bibinfo {author} {\bibfnamefont {P.~W.}\ \bibnamefont {Hess}}, \bibinfo {author} {\bibfnamefont {P.}~\bibnamefont {Hauke}}, \bibinfo {author} {\bibfnamefont {M.}~\bibnamefont {Heyl}}, \bibinfo {author} {\bibfnamefont {D.~A.}\ \bibnamefont {Huse}},\ and\ \bibinfo {author} {\bibfnamefont {C.}~\bibnamefont {Monroe}},\ }\bibfield  {title} {\bibinfo {title} {Many-body localization in a quantum simulator with programmable random disorder},\ }\href {https://doi.org/10.1038/nphys3783} {\bibfield  {journal} {\bibinfo  {journal} {Nat. Phys.}\ }\textbf {\bibinfo {volume} {12}},\ \bibinfo {pages} {907} (\bibinfo {year} {2016})}\BibitemShut {NoStop}%
\bibitem [{\citenamefont {Korenblit}\ \emph {et~al.}(2012)\citenamefont {Korenblit}, \citenamefont {Kafri}, \citenamefont {Campbell}, \citenamefont {Islam}, \citenamefont {Edwards}, \citenamefont {Gong}, \citenamefont {Lin}, \citenamefont {Duan}, \citenamefont {Kim}, \citenamefont {Kim},\ and\ \citenamefont {Monroe}}]{Korenblit2012}%
  \BibitemOpen
  \bibfield  {author} {\bibinfo {author} {\bibfnamefont {S.}~\bibnamefont {Korenblit}}, \bibinfo {author} {\bibfnamefont {D.}~\bibnamefont {Kafri}}, \bibinfo {author} {\bibfnamefont {W.~C.}\ \bibnamefont {Campbell}}, \bibinfo {author} {\bibfnamefont {R.}~\bibnamefont {Islam}}, \bibinfo {author} {\bibfnamefont {E.~E.}\ \bibnamefont {Edwards}}, \bibinfo {author} {\bibfnamefont {Z.-X.}\ \bibnamefont {Gong}}, \bibinfo {author} {\bibfnamefont {G.-D.}\ \bibnamefont {Lin}}, \bibinfo {author} {\bibfnamefont {L.-M.}\ \bibnamefont {Duan}}, \bibinfo {author} {\bibfnamefont {J.}~\bibnamefont {Kim}}, \bibinfo {author} {\bibfnamefont {K.}~\bibnamefont {Kim}},\ and\ \bibinfo {author} {\bibfnamefont {C.}~\bibnamefont {Monroe}},\ }\bibfield  {title} {\bibinfo {title} {Quantum simulation of spin models on an arbitrary lattice with trapped ions},\ }\href {https://doi.org/10.1088/1367-2630/14/9/095024} {\bibfield  {journal} {\bibinfo  {journal} {New J. Phys.}\ }\textbf {\bibinfo {volume} {14}},\ \bibinfo {pages} {095024}
  (\bibinfo {year} {2012})}\BibitemShut {NoStop}%
\bibitem [{\citenamefont {Laumann}\ \emph {et~al.}(2014)\citenamefont {Laumann}, \citenamefont {Pal},\ and\ \citenamefont {Scardicchio}}]{Laumann2014}%
  \BibitemOpen
  \bibfield  {author} {\bibinfo {author} {\bibfnamefont {C.~R.}\ \bibnamefont {Laumann}}, \bibinfo {author} {\bibfnamefont {A.}~\bibnamefont {Pal}},\ and\ \bibinfo {author} {\bibfnamefont {A.}~\bibnamefont {Scardicchio}},\ }\bibfield  {title} {\bibinfo {title} {Many-body mobility edge in a mean-field quantum spin glass},\ }\href {https://doi.org/10.1103/PhysRevLett.113.200405} {\bibfield  {journal} {\bibinfo  {journal} {Phys. Rev. Lett.}\ }\textbf {\bibinfo {volume} {113}},\ \bibinfo {pages} {200405} (\bibinfo {year} {2014})}\BibitemShut {NoStop}%
\bibitem [{\citenamefont {Smelyanskiy}\ \emph {et~al.}(2020)\citenamefont {Smelyanskiy}, \citenamefont {Kechedzhi}, \citenamefont {Boixo}, \citenamefont {Isakov}, \citenamefont {Neven},\ and\ \citenamefont {Altshuler}}]{Smelyanskiy2020}%
  \BibitemOpen
  \bibfield  {author} {\bibinfo {author} {\bibfnamefont {V.~N.}\ \bibnamefont {Smelyanskiy}}, \bibinfo {author} {\bibfnamefont {K.}~\bibnamefont {Kechedzhi}}, \bibinfo {author} {\bibfnamefont {S.}~\bibnamefont {Boixo}}, \bibinfo {author} {\bibfnamefont {S.~V.}\ \bibnamefont {Isakov}}, \bibinfo {author} {\bibfnamefont {H.}~\bibnamefont {Neven}},\ and\ \bibinfo {author} {\bibfnamefont {B.}~\bibnamefont {Altshuler}},\ }\bibfield  {title} {\bibinfo {title} {Nonergodic delocalized states for efficient population transfer within a narrow band of the energy landscape},\ }\href {https://doi.org/10.1103/PhysRevX.10.011017} {\bibfield  {journal} {\bibinfo  {journal} {Phys. Rev. X}\ }\textbf {\bibinfo {volume} {10}},\ \bibinfo {pages} {011017} (\bibinfo {year} {2020})}\BibitemShut {NoStop}%
\bibitem [{\citenamefont {Javanmard}\ \emph {et~al.}(2019)\citenamefont {Javanmard}, \citenamefont {Bera},\ and\ \citenamefont {Heyl}}]{Javanmard2019}%
  \BibitemOpen
  \bibfield  {author} {\bibinfo {author} {\bibfnamefont {Y.}~\bibnamefont {Javanmard}}, \bibinfo {author} {\bibfnamefont {S.}~\bibnamefont {Bera}},\ and\ \bibinfo {author} {\bibfnamefont {M.}~\bibnamefont {Heyl}},\ }\bibfield  {title} {\bibinfo {title} {Accessing eigenstate spin-glass order from reduced density matrices},\ }\href {https://doi.org/10.1103/PhysRevB.99.144201} {\bibfield  {journal} {\bibinfo  {journal} {Phys. Rev. B}\ }\textbf {\bibinfo {volume} {99}},\ \bibinfo {pages} {144201} (\bibinfo {year} {2019})}\BibitemShut {NoStop}%
\bibitem [{\citenamefont {Baldwin}\ \emph {et~al.}(2017)\citenamefont {Baldwin}, \citenamefont {Laumann}, \citenamefont {Pal},\ and\ \citenamefont {Scardicchio}}]{Baldwin2017}%
  \BibitemOpen
  \bibfield  {author} {\bibinfo {author} {\bibfnamefont {C.~L.}\ \bibnamefont {Baldwin}}, \bibinfo {author} {\bibfnamefont {C.~R.}\ \bibnamefont {Laumann}}, \bibinfo {author} {\bibfnamefont {A.}~\bibnamefont {Pal}},\ and\ \bibinfo {author} {\bibfnamefont {A.}~\bibnamefont {Scardicchio}},\ }\bibfield  {title} {\bibinfo {title} {Clustering of nonergodic eigenstates in quantum spin glasses},\ }\href {https://doi.org/10.1103/PhysRevLett.118.127201} {\bibfield  {journal} {\bibinfo  {journal} {Phys. Rev. Lett.}\ }\textbf {\bibinfo {volume} {118}},\ \bibinfo {pages} {127201} (\bibinfo {year} {2017})}\BibitemShut {NoStop}%
\bibitem [{\citenamefont {Mossi}\ and\ \citenamefont {Scardicchio}(2017)}]{Mossi17}%
  \BibitemOpen
  \bibfield  {author} {\bibinfo {author} {\bibfnamefont {G.}~\bibnamefont {Mossi}}\ and\ \bibinfo {author} {\bibfnamefont {A.}~\bibnamefont {Scardicchio}},\ }\bibfield  {title} {\bibinfo {title} {Ergodic and localized regions in quantum spin glasses on the {Bethe} lattice},\ }\href {https://doi.org/10.1098/rsta.2016.0424} {\bibfield  {journal} {\bibinfo  {journal} {Philos. Trans. R. Soc. A}\ }\textbf {\bibinfo {volume} {375}},\ \bibinfo {pages} {20160424} (\bibinfo {year} {2017})}\BibitemShut {NoStop}%
\bibitem [{\citenamefont {Rademaker}\ and\ \citenamefont {Abanin}(2020)}]{rademaker_abanin_2020}%
  \BibitemOpen
  \bibfield  {author} {\bibinfo {author} {\bibfnamefont {L.}~\bibnamefont {Rademaker}}\ and\ \bibinfo {author} {\bibfnamefont {D.~A.}\ \bibnamefont {Abanin}},\ }\bibfield  {title} {\bibinfo {title} {Slow nonthermalizing dynamics in a quantum spin glass},\ }\href {https://doi.org/10.1103/physrevlett.125.260405} {\bibfield  {journal} {\bibinfo  {journal} {Phys. Rev. Lett.}\ }\textbf {\bibinfo {volume} {125}},\ \bibinfo {pages} {260405} (\bibinfo {year} {2020})}\BibitemShut {NoStop}%
\bibitem [{\citenamefont {Correale}\ \emph {et~al.}(2023)\citenamefont {Correale}, \citenamefont {Polkovnikov}, \citenamefont {Schir{\`o}},\ and\ \citenamefont {Silva}}]{Correale23}%
  \BibitemOpen
  \bibfield  {author} {\bibinfo {author} {\bibfnamefont {L.}~\bibnamefont {Correale}}, \bibinfo {author} {\bibfnamefont {A.}~\bibnamefont {Polkovnikov}}, \bibinfo {author} {\bibfnamefont {M.}~\bibnamefont {Schir{\`o}}},\ and\ \bibinfo {author} {\bibfnamefont {A.}~\bibnamefont {Silva}},\ }\bibfield  {title} {\bibinfo {title} {{Probing chaos in the spherical p-spin glass model}},\ }\href {https://doi.org/10.21468/SciPostPhys.15.5.190} {\bibfield  {journal} {\bibinfo  {journal} {SciPost Phys.}\ }\textbf {\bibinfo {volume} {15}},\ \bibinfo {pages} {190} (\bibinfo {year} {2023})}\BibitemShut {NoStop}%
\bibitem [{\citenamefont {Pappalardi}\ \emph {et~al.}(2020)\citenamefont {Pappalardi}, \citenamefont {Polkovnikov},\ and\ \citenamefont {Silva}}]{Pappalardi20}%
  \BibitemOpen
  \bibfield  {author} {\bibinfo {author} {\bibfnamefont {S.}~\bibnamefont {Pappalardi}}, \bibinfo {author} {\bibfnamefont {A.}~\bibnamefont {Polkovnikov}},\ and\ \bibinfo {author} {\bibfnamefont {A.}~\bibnamefont {Silva}},\ }\bibfield  {title} {\bibinfo {title} {{Quantum echo dynamics in the Sherrington-Kirkpatrick model}},\ }\href {https://doi.org/10.21468/SciPostPhys.9.2.021} {\bibfield  {journal} {\bibinfo  {journal} {SciPost Phys.}\ }\textbf {\bibinfo {volume} {9}},\ \bibinfo {pages} {021} (\bibinfo {year} {2020})}\BibitemShut {NoStop}%
\bibitem [{\citenamefont {Thomson}\ \emph {et~al.}(2020)\citenamefont {Thomson}, \citenamefont {Urbani},\ and\ \citenamefont {Schir\`o}}]{Thomson20}%
  \BibitemOpen
  \bibfield  {author} {\bibinfo {author} {\bibfnamefont {S.~J.}\ \bibnamefont {Thomson}}, \bibinfo {author} {\bibfnamefont {P.}~\bibnamefont {Urbani}},\ and\ \bibinfo {author} {\bibfnamefont {M.}~\bibnamefont {Schir\`o}},\ }\bibfield  {title} {\bibinfo {title} {Quantum quenches in isolated quantum glasses out of equilibrium},\ }\href {https://doi.org/10.1103/PhysRevLett.125.120602} {\bibfield  {journal} {\bibinfo  {journal} {Phys. Rev. Lett.}\ }\textbf {\bibinfo {volume} {125}},\ \bibinfo {pages} {120602} (\bibinfo {year} {2020})}\BibitemShut {NoStop}%
\bibitem [{\citenamefont {Hosseinabadi}\ \emph {et~al.}(2024)\citenamefont {Hosseinabadi}, \citenamefont {Chang},\ and\ \citenamefont {Marino}}]{hosseinabadi24}%
  \BibitemOpen
  \bibfield  {author} {\bibinfo {author} {\bibfnamefont {H.}~\bibnamefont {Hosseinabadi}}, \bibinfo {author} {\bibfnamefont {D.~E.}\ \bibnamefont {Chang}},\ and\ \bibinfo {author} {\bibfnamefont {J.}~\bibnamefont {Marino}},\ }\bibfield  {title} {\bibinfo {title} {Quantum-to-classical crossover in the spin glass dynamics of cavity {QED} simulators},\ }\href {https://doi.org/10.1103/physrevresearch.6.043313} {\bibfield  {journal} {\bibinfo  {journal} {Phys. Rev. Res.}\ }\textbf {\bibinfo {volume} {6}},\ \bibinfo {pages} {043313} (\bibinfo {year} {2024})}\BibitemShut {NoStop}%
\bibitem [{\citenamefont {Rigol}\ \emph {et~al.}(2008)\citenamefont {Rigol}, \citenamefont {Dunjko},\ and\ \citenamefont {Olshanii}}]{RigolNature08}%
  \BibitemOpen
  \bibfield  {author} {\bibinfo {author} {\bibfnamefont {M.}~\bibnamefont {Rigol}}, \bibinfo {author} {\bibfnamefont {V.}~\bibnamefont {Dunjko}},\ and\ \bibinfo {author} {\bibfnamefont {M.}~\bibnamefont {Olshanii}},\ }\bibfield  {title} {\bibinfo {title} {Thermalization and its mechanism for generic isolated quantum systems},\ }\href {https://doi.org/10.1038/nature06838} {\bibfield  {journal} {\bibinfo  {journal} {Nature}\ }\textbf {\bibinfo {volume} {452}},\ \bibinfo {pages} {854} (\bibinfo {year} {2008})}\BibitemShut {NoStop}%
\bibitem [{SM()}]{SM}%
  \BibitemOpen
  \href@noop {} {}\bibinfo {note} {See the Supplemental Material appended to this Letter for the finite-size analysis of the diagonal-ensemble response, the isolated-resonance theory of the cold-shell overshoot, the adiabatic-gauge-potential formulation of the dephased response, the multibranch pair-pattern network, the dynamical and spectral diagnostics, and the weak-noise kinetic limit of longitudinal dephasing, which includes Refs.~\cite{Atas2013,Cotler2017,HaakeBook,Leviandier1986,MehtaBook,Niedda2025RG}}\BibitemShut {NoStop}%
\bibitem [{\citenamefont {Parisi}(1983)}]{Parisi83}%
  \BibitemOpen
  \bibfield  {author} {\bibinfo {author} {\bibfnamefont {G.}~\bibnamefont {Parisi}},\ }\bibfield  {title} {\bibinfo {title} {Order parameter for spin-glasses},\ }\href {https://doi.org/10.1103/PhysRevLett.50.1946} {\bibfield  {journal} {\bibinfo  {journal} {Phys. Rev. Lett.}\ }\textbf {\bibinfo {volume} {50}},\ \bibinfo {pages} {1946} (\bibinfo {year} {1983})}\BibitemShut {NoStop}%
\bibitem [{\citenamefont {M{\'e}zard}\ \emph {et~al.}(1987)\citenamefont {M{\'e}zard}, \citenamefont {Parisi},\ and\ \citenamefont {Virasoro}}]{MezardParisiVirasoro1987}%
  \BibitemOpen
  \bibfield  {author} {\bibinfo {author} {\bibfnamefont {M.}~\bibnamefont {M{\'e}zard}}, \bibinfo {author} {\bibfnamefont {G.}~\bibnamefont {Parisi}},\ and\ \bibinfo {author} {\bibfnamefont {M.~A.}\ \bibnamefont {Virasoro}},\ }\href@noop {} {\emph {\bibinfo {title} {Spin Glass Theory and Beyond}}}\ (\bibinfo  {publisher} {World Scientific},\ \bibinfo {address} {Singapore},\ \bibinfo {year} {1987})\BibitemShut {NoStop}%
\bibitem [{\citenamefont {Kim}\ \emph {et~al.}(2014)\citenamefont {Kim}, \citenamefont {Ikeda},\ and\ \citenamefont {Huse}}]{KimIkedaHuse2014}%
  \BibitemOpen
  \bibfield  {author} {\bibinfo {author} {\bibfnamefont {H.}~\bibnamefont {Kim}}, \bibinfo {author} {\bibfnamefont {T.~N.}\ \bibnamefont {Ikeda}},\ and\ \bibinfo {author} {\bibfnamefont {D.~A.}\ \bibnamefont {Huse}},\ }\bibfield  {title} {\bibinfo {title} {Testing whether all eigenstates obey the eigenstate thermalization hypothesis},\ }\href {https://doi.org/10.1103/PhysRevE.90.052105} {\bibfield  {journal} {\bibinfo  {journal} {Phys. Rev. E}\ }\textbf {\bibinfo {volume} {90}},\ \bibinfo {pages} {052105} (\bibinfo {year} {2014})}\BibitemShut {NoStop}%
\bibitem [{\citenamefont {Bruzewicz}\ \emph {et~al.}(2019)\citenamefont {Bruzewicz}, \citenamefont {Chiaverini}, \citenamefont {McConnell},\ and\ \citenamefont {Sage}}]{Bruzewicz2019}%
  \BibitemOpen
  \bibfield  {author} {\bibinfo {author} {\bibfnamefont {C.~D.}\ \bibnamefont {Bruzewicz}}, \bibinfo {author} {\bibfnamefont {J.}~\bibnamefont {Chiaverini}}, \bibinfo {author} {\bibfnamefont {R.}~\bibnamefont {McConnell}},\ and\ \bibinfo {author} {\bibfnamefont {J.~M.}\ \bibnamefont {Sage}},\ }\bibfield  {title} {\bibinfo {title} {Trapped-ion quantum computing: Progress and challenges},\ }\href {https://doi.org/10.1063/1.5088164} {\bibfield  {journal} {\bibinfo  {journal} {Appl. Phys. Rev.}\ }\textbf {\bibinfo {volume} {6}},\ \bibinfo {pages} {021314} (\bibinfo {year} {2019})}\BibitemShut {NoStop}%
\bibitem [{\citenamefont {Britton}\ \emph {et~al.}(2012)\citenamefont {Britton}, \citenamefont {Sawyer}, \citenamefont {Keith}, \citenamefont {Wang}, \citenamefont {Freericks}, \citenamefont {Uys}, \citenamefont {Biercuk},\ and\ \citenamefont {Bollinger}}]{Britton2012}%
  \BibitemOpen
  \bibfield  {author} {\bibinfo {author} {\bibfnamefont {J.~W.}\ \bibnamefont {Britton}}, \bibinfo {author} {\bibfnamefont {B.~C.}\ \bibnamefont {Sawyer}}, \bibinfo {author} {\bibfnamefont {A.~C.}\ \bibnamefont {Keith}}, \bibinfo {author} {\bibfnamefont {C.-C.~J.}\ \bibnamefont {Wang}}, \bibinfo {author} {\bibfnamefont {J.~K.}\ \bibnamefont {Freericks}}, \bibinfo {author} {\bibfnamefont {H.}~\bibnamefont {Uys}}, \bibinfo {author} {\bibfnamefont {M.~J.}\ \bibnamefont {Biercuk}},\ and\ \bibinfo {author} {\bibfnamefont {J.~J.}\ \bibnamefont {Bollinger}},\ }\bibfield  {title} {\bibinfo {title} {Engineered two-dimensional {Ising} interactions in a trapped-ion quantum simulator with hundreds of spins},\ }\href {https://doi.org/10.1038/nature10981} {\bibfield  {journal} {\bibinfo  {journal} {Nature}\ }\textbf {\bibinfo {volume} {484}},\ \bibinfo {pages} {489} (\bibinfo {year} {2012})}\BibitemShut {NoStop}%
\bibitem [{\citenamefont {G{\"a}rttner}\ \emph {et~al.}(2017)\citenamefont {G{\"a}rttner}, \citenamefont {Bohnet}, \citenamefont {Safavi-Naini}, \citenamefont {Wall}, \citenamefont {Bollinger},\ and\ \citenamefont {Rey}}]{Garttner2017}%
  \BibitemOpen
  \bibfield  {author} {\bibinfo {author} {\bibfnamefont {M.}~\bibnamefont {G{\"a}rttner}}, \bibinfo {author} {\bibfnamefont {J.~G.}\ \bibnamefont {Bohnet}}, \bibinfo {author} {\bibfnamefont {A.}~\bibnamefont {Safavi-Naini}}, \bibinfo {author} {\bibfnamefont {M.~L.}\ \bibnamefont {Wall}}, \bibinfo {author} {\bibfnamefont {J.~J.}\ \bibnamefont {Bollinger}},\ and\ \bibinfo {author} {\bibfnamefont {A.~M.}\ \bibnamefont {Rey}},\ }\bibfield  {title} {\bibinfo {title} {Measuring out-of-time-order correlations and multiple quantum spectra in a trapped-ion quantum magnet},\ }\href {https://doi.org/10.1038/nphys4119} {\bibfield  {journal} {\bibinfo  {journal} {Nat. Phys.}\ }\textbf {\bibinfo {volume} {13}},\ \bibinfo {pages} {781} (\bibinfo {year} {2017})}\BibitemShut {NoStop}%
\bibitem [{\citenamefont {Atas}\ \emph {et~al.}(2013)\citenamefont {Atas}, \citenamefont {Bogomolny}, \citenamefont {Giraud},\ and\ \citenamefont {Roux}}]{Atas2013}%
  \BibitemOpen
  \bibfield  {author} {\bibinfo {author} {\bibfnamefont {Y.~Y.}\ \bibnamefont {Atas}}, \bibinfo {author} {\bibfnamefont {E.}~\bibnamefont {Bogomolny}}, \bibinfo {author} {\bibfnamefont {O.}~\bibnamefont {Giraud}},\ and\ \bibinfo {author} {\bibfnamefont {G.}~\bibnamefont {Roux}},\ }\bibfield  {title} {\bibinfo {title} {Distribution of the ratio of consecutive level spacings in random matrix ensembles},\ }\href {https://doi.org/10.1103/PhysRevLett.110.084101} {\bibfield  {journal} {\bibinfo  {journal} {Phys. Rev. Lett.}\ }\textbf {\bibinfo {volume} {110}},\ \bibinfo {pages} {084101} (\bibinfo {year} {2013})}\BibitemShut {NoStop}%
\bibitem [{\citenamefont {Cotler}\ \emph {et~al.}(2017)\citenamefont {Cotler}, \citenamefont {Gur-Ari}, \citenamefont {Hanada}, \citenamefont {Polchinski}, \citenamefont {Saad}, \citenamefont {Shenker}, \citenamefont {Stanford}, \citenamefont {Streicher},\ and\ \citenamefont {Tezuka}}]{Cotler2017}%
  \BibitemOpen
  \bibfield  {author} {\bibinfo {author} {\bibfnamefont {J.~S.}\ \bibnamefont {Cotler}}, \bibinfo {author} {\bibfnamefont {G.}~\bibnamefont {Gur-Ari}}, \bibinfo {author} {\bibfnamefont {M.}~\bibnamefont {Hanada}}, \bibinfo {author} {\bibfnamefont {J.}~\bibnamefont {Polchinski}}, \bibinfo {author} {\bibfnamefont {P.}~\bibnamefont {Saad}}, \bibinfo {author} {\bibfnamefont {S.~H.}\ \bibnamefont {Shenker}}, \bibinfo {author} {\bibfnamefont {D.}~\bibnamefont {Stanford}}, \bibinfo {author} {\bibfnamefont {A.}~\bibnamefont {Streicher}},\ and\ \bibinfo {author} {\bibfnamefont {M.}~\bibnamefont {Tezuka}},\ }\bibfield  {title} {\bibinfo {title} {Black holes and random matrices},\ }\href {https://doi.org/10.1007/JHEP05(2017)118} {\bibfield  {journal} {\bibinfo  {journal} {J. High Energy Phys.}\ }\textbf {\bibinfo {volume} {2017}}\bibinfo  {number} { (5)},\ \bibinfo {pages} {118}}\BibitemShut {NoStop}%
\bibitem [{\citenamefont {Haake}(2010)}]{HaakeBook}%
  \BibitemOpen
\bibfield  {number} {  }\bibfield  {author} {\bibinfo {author} {\bibfnamefont {F.}~\bibnamefont {Haake}},\ }\href {https://doi.org/10.1007/978-3-642-05428-0} {\emph {\bibinfo {title} {Quantum Signatures of Chaos}}},\ \bibinfo {edition} {3rd}\ ed.\ (\bibinfo  {publisher} {Springer},\ \bibinfo {address} {Berlin},\ \bibinfo {year} {2010})\BibitemShut {NoStop}%
\bibitem [{\citenamefont {Leviandier}\ \emph {et~al.}(1986)\citenamefont {Leviandier}, \citenamefont {Lombardi}, \citenamefont {Jost},\ and\ \citenamefont {Pique}}]{Leviandier1986}%
  \BibitemOpen
  \bibfield  {author} {\bibinfo {author} {\bibfnamefont {L.}~\bibnamefont {Leviandier}}, \bibinfo {author} {\bibfnamefont {M.}~\bibnamefont {Lombardi}}, \bibinfo {author} {\bibfnamefont {R.}~\bibnamefont {Jost}},\ and\ \bibinfo {author} {\bibfnamefont {J.-P.}\ \bibnamefont {Pique}},\ }\bibfield  {title} {\bibinfo {title} {Fourier transform: A tool to measure statistical level properties in very complex spectra},\ }\href {https://doi.org/10.1103/PhysRevLett.56.2449} {\bibfield  {journal} {\bibinfo  {journal} {Phys. Rev. Lett.}\ }\textbf {\bibinfo {volume} {56}},\ \bibinfo {pages} {2449} (\bibinfo {year} {1986})}\BibitemShut {NoStop}%
\bibitem [{\citenamefont {Mehta}(2004)}]{MehtaBook}%
  \BibitemOpen
  \bibfield  {author} {\bibinfo {author} {\bibfnamefont {M.~L.}\ \bibnamefont {Mehta}},\ }\href@noop {} {\emph {\bibinfo {title} {Random Matrices}}},\ \bibinfo {edition} {3rd}\ ed.\ (\bibinfo  {publisher} {Elsevier/Academic Press},\ \bibinfo {address} {Amsterdam},\ \bibinfo {year} {2004})\BibitemShut {NoStop}%
\bibitem [{\citenamefont {Niedda}\ \emph {et~al.}(2025)\citenamefont {Niedda}, \citenamefont {Bracci~Testasecca}, \citenamefont {Magnifico}, \citenamefont {Balducci}, \citenamefont {Vanoni},\ and\ \citenamefont {Scardicchio}}]{Niedda2025RG}%
  \BibitemOpen
  \bibfield  {author} {\bibinfo {author} {\bibfnamefont {J.}~\bibnamefont {Niedda}}, \bibinfo {author} {\bibfnamefont {G.}~\bibnamefont {Bracci~Testasecca}}, \bibinfo {author} {\bibfnamefont {G.}~\bibnamefont {Magnifico}}, \bibinfo {author} {\bibfnamefont {F.}~\bibnamefont {Balducci}}, \bibinfo {author} {\bibfnamefont {C.}~\bibnamefont {Vanoni}},\ and\ \bibinfo {author} {\bibfnamefont {A.}~\bibnamefont {Scardicchio}},\ }\bibfield  {title} {\bibinfo {title} {Renormalization group analysis of the many-body localization transition in the random-field {XXZ} chain},\ }\href {https://doi.org/10.1103/gcwf-jdlr} {\bibfield  {journal} {\bibinfo  {journal} {Phys. Rev. B}\ }\textbf {\bibinfo {volume} {112}},\ \bibinfo {pages} {144201} (\bibinfo {year} {2025})}\BibitemShut {NoStop}%
\end{thebibliography}%

\makeatletter
\let\SMsave@hangfrom@section\@hangfrom@section
\let\SMsave@sectioncntformat\@sectioncntformat
\makeatother
\appendix
\setcounter{secnumdepth}{3}
\section*{End Matter}

\section{Derivation of the order parameter  \texorpdfstring{$Q_{\mathrm{DE}}$}{QDE}}
\label{app:derivation}

Let the system be prepared in a normalized eigenstate of $H_I$,
\begin{equation}
H_I \ket{n_I}=E_n^I \ket{n_I},
\qquad
\braket{n_I|n_I}=1.
\end{equation}
At $t=0$ the Hamiltonian is switched to $H_F$. We are interested in the pair operator
\begin{equation}
O_{ij}:=\sigma_i^z\sigma_j^z,
\qquad
i\neq j.
\end{equation}

Let us write the most general spectral decomposition of $H_F$ as
\begin{equation}
H_F=\sum_{\alpha} E_{\alpha} P_{\alpha},
\end{equation}
where $\alpha$ labels distinct eigenvalues and $P_\alpha$ is the orthogonal projector onto the eigenspace associated with $E_\alpha$. Then
\begin{equation}
P_\alpha P_\beta=\delta_{\alpha\beta}P_\alpha,
\qquad
\sum_\alpha P_\alpha=\mathbb I.
\end{equation}
This formulation keeps exact degeneracies explicit.

The evolved state is
\begin{equation}
\ket{\psi(t)}=\ee^{-\ii H_F t}\ket{n_I}
=\sum_\alpha \ee^{-\ii E_\alpha t}P_\alpha\ket{n_I}.
\end{equation}
Hence the density matrix is
\begin{equation}
\rho(t)=\ket{\psi(t)}\bra{\psi(t)}
=
\sum_{\alpha,\beta}
\ee^{-\ii(E_\alpha-E_\beta)t}
P_\alpha\ket{n_I}\bra{n_I}P_\beta.
\end{equation}
The exact time-dependent correlator is therefore
\begin{equation}
C_{ij}(t):=\Tr[\rho(t)O_{ij}]
=
\sum_{\alpha,\beta}
\ee^{-\ii(E_\alpha-E_\beta)t}
\bra{n_I}P_\beta O_{ij}P_\alpha\ket{n_I}.
\label{eq:Cijt_exact_projectors_app}
\end{equation}

Let us define the infinite-time  average
\begin{equation}
\overline{C_{ij}}
:=
\lim_{T\to\infty}\frac1T\int_0^T C_{ij}(t)\,\dd t.
\end{equation}
Substituting Eq.~\eqref{eq:Cijt_exact_projectors_app} and using the elementary identity
\begin{equation}
\lim_{T\to\infty}\frac1T\int_0^T \ee^{-\ii\omega t}\,\dd t
=
\begin{cases}
1,& \omega=0,\\
0,& \omega\neq0,
\end{cases}
\end{equation}
one finds
\begin{equation}
\overline{C_{ij}}=
\sum_\alpha \bra{n_I}P_\alpha O_{ij}P_\alpha\ket{n_I}.
\label{eq:Cij_bar_projectors_app}
\end{equation}
This motivates the dephased density matrix
\begin{equation}
\rho_{\mathrm{DE}}
:=
\sum_\alpha P_\alpha\ket{n_I}\bra{n_I}P_\alpha.
\label{eq:rhoDE_projectors_app}
\end{equation}
It is positive, Hermitian, and normalized:
\begin{align}
\Tr \rho_{\mathrm{DE}}
&=
\sum_\alpha
\Tr\!\left(P_\alpha\ket{n_I}\bra{n_I}P_\alpha\right)
=
\sum_\alpha \bra{n_I}P_\alpha\ket{n_I}
=
1.
\end{align}
Equation~\eqref{eq:Cij_bar_projectors_app} becomes
\begin{equation}
C_{ij}^{\mathrm{DE}}
:=
\overline{C_{ij}}
=
\Tr\!\left(\rho_{\mathrm{DE}}O_{ij}\right).
\label{eq:CijDE_trace_general_app}
\end{equation}
Thus the diagonal ensemble is precisely the infinite-time dephased density matrix, and $C_{ij}^{\mathrm{DE}}$ is the long-time dephased expectation value of $\sigma_i^z\sigma_j^z$.

In the nondegenerate case, $P_m=\ket{m_F}\bra{m_F}$ and
\begin{equation}
\ket{n_I}=\sum_m c_m\ket{m_F},
\quad
c_m:=\braket{m_F|n_I},
\quad
\sum_m |c_m|^2=1.
\end{equation}
Then
\begin{equation}
\rho_{\mathrm{DE}}=
\sum_m |c_m|^2 \ket{m_F}\bra{m_F},
\label{eq:rhoDE_nondeg_app}
\end{equation}
and hence
\begin{equation}
C_{ij}^{\mathrm{DE}}=
\sum_m |\braket{m_F|n_I}|^2
\mel{m_F}{\sigma_i^z\sigma_j^z}{m_F}.
\label{eq:CijDE_nondeg_app}
\end{equation}

The equal-time expectation value in the initial state,
\begin{equation}
\mel{n_I}{\sigma_i^z\sigma_j^z}{n_I}
=
\sum_{m,m'}c_m^\ast c_{m'}
\mel{m_F}{\sigma_i^z\sigma_j^z}{m'_F},
\end{equation}
is not equal in general to Eq.~\eqref{eq:CijDE_nondeg_app}, because the former contains off-diagonal coherences and the latter does not. The physical distinction is that the former refers to $t=0$, while the latter refers to the fully dephased asymptotic state.

Finally, define
\begin{equation}
q_{\rm DE}^{(n)}:=
\frac{2}{L(L-1)}\sum_{i<j}\left(C_{ij}^{\mathrm{DE}}\right)^2.
\label{eq:QDE_definition_app}
\end{equation}
Using Eq.~\eqref{eq:CijDE_trace_general_app}, one obtains the exact projector form
\begin{equation}
q_{\rm DE}^{(n)}=
\frac{2}{L(L-1)}\sum_{i<j}
\left(
\sum_\alpha
\bra{n_I}P_\alpha \sigma_i^z\sigma_j^z P_\alpha\ket{n_I}
\right)^2,
\label{eq:QDE_projectors_app}
\end{equation}
while in the nondegenerate case one recovers
\begin{equation}
q_{\rm DE}^{(n)}=
\frac{2}{L(L-1)}\sum_{i<j}
\left(
\sum_m |\braket{m_F|n_I}|^2
\mel{m_F}{\sigma_i^z\sigma_j^z}{m_F}
\right)^2\!,
\label{eq:QDE_nondeg_app}
\end{equation}
 which is Eq.~\eqref{eq:qDE} in the main text.

\subsection{Exact \texorpdfstring{$\lambda\to0$}{lambda to 0} limit of the pair diagonal ensemble}
\label{app:lambda0}

We prove here that the diagonal-ensemble pair correlations approach their initial values as $\lambda\to0$, despite the degeneracy of the final Hamiltonian at $\lambda=0$. For fixed $h_I>0$,
\begin{equation}
H_F(\lambda)=V-\lambda h_I X .
\end{equation}
Let
\begin{equation}
V=\sum_a v_a P_a^V
\end{equation}
be the spectral decomposition of $V$ into distinct eigenvalues. Since $H_F(\lambda)\to V$ in operator norm as $\lambda\to0$, standard finite-dimensional spectral perturbation theory implies that the total
spectral projector $\Pi_a(\lambda)$ associated with the levels converging to $v_a$ satisfies
\begin{equation}
\Pi_a(\lambda)=\sum_{\alpha\in a}P_\alpha(\lambda)
\xrightarrow[\lambda\to0]{}P_a^V
\end{equation}
in operator norm, where $P_\alpha(\lambda)$ are the exact spectral projectors of $H_F(\lambda)$.

For the Gaussian couplings considered here, accidental degeneracies of $V$ have probability zero: if two Ising configurations $\mathbf s'\neq\pm\mathbf s$ had the same energy, this would impose a
nontrivial linear constraint on the independent continuous random variables $\{\xi_{ij}\}$. Thus, almost surely, each eigenspace of $V$ is the global-spin-flip doublet $\operatorname{span}\{\ket{\mathbf s},\ket{-\mathbf s}\}$. For the pair operator
\begin{equation}
O_{ij}=\sigma_i^z\sigma_j^z
\end{equation}
both states of such a doublet have the same eigenvalue. Therefore
\begin{equation}
P_a^V O_{ij}P_a^V
=
\omega_{ij}^{(a)}P_a^V,
\qquad
\omega_{ij}^{(a)}=\pm1 .
\label{eq:pair_scalar_lambda0}
\end{equation}

For the initial state
$\rho_0^{(n)}=\ket{n_I}\bra{n_I}$, the exact diagonal ensemble at finite
$\lambda$ is
\begin{equation}
\rho_{\mathrm{DE}}^{(n)}(\lambda)
=
\sum_\alpha
P_\alpha(\lambda)\rho_0^{(n)}P_\alpha(\lambda).
\end{equation}
The possible splitting and rotation of states within a doublet do not affect the pair correlator. Indeed, using $P_\alpha(\lambda)\Pi_a(\lambda)=P_\alpha(\lambda)$ and the contractivity of projective pinching in operator norm,
\begin{align}
&
\left\|
\sum_{\alpha\in a}
P_\alpha(\lambda)O_{ij}P_\alpha(\lambda)
-
\omega_{ij}^{(a)}\Pi_a(\lambda)
\right\|
\nonumber\\
&\qquad\le
\left\|
\Pi_a(\lambda)
\left(O_{ij}-\omega_{ij}^{(a)}\right)
\Pi_a(\lambda)
\right\|
\xrightarrow[\lambda\to0]{}0 ,
\end{align}
where Eq.~\eqref{eq:pair_scalar_lambda0} was used in the last step. Hence
\begin{align}
\lim_{\lambda\to0}
\Tr\!\left[
\rho_{\mathrm{DE}}^{(n)}(\lambda)O_{ij}
\right]
&=
\sum_a
\omega_{ij}^{(a)}
\Tr\!\left(\rho_0^{(n)}P_a^V\right)
\nonumber\\
&=
\Tr\!\left(\rho_0^{(n)}O_{ij}\right)
=
\mel{n_I}{\sigma_i^z\sigma_j^z}{n_I}.
\label{eq:lambda0_pair_limit}
\end{align}
Consequently,
\begin{equation}
\lim_{\lambda\to0}q_{\mathrm{DE}}^{(n)}(\lambda)
=
\frac{1}{\Np}\sum_{i<j}
\mel{n_I}{\sigma_i^z\sigma_j^z}{n_I}^{\,2}
=
q_{\mathrm{DE}}^{(n)}(1).
\label{eq:lambda0_q_limit}
\end{equation}
Since $0\le q_{\mathrm{DE}}^{(n)}\le1$, the same relation survives the energy-shell and disorder averages, yielding
\begin{equation}
\lim_{\lambda\to0}\QDE(\lambda,\epsilon)
=
\QDE(1,\epsilon)
\equiv Q_I(\epsilon).
\end{equation}

At $\lambda=0$ itself the conservation of every $O_{ij}$ follows directly from $[O_{ij},V]=0$. The argument above establishes the stronger statement needed in the main text: the exact diagonal ensemble obtained at finite $\lambda$ approaches the same value continuously as $\lambda\to0$, despite the spin-flip degeneracy of $V$.

\clearpage
\makeatletter
\let\@hangfrom@section\SMsave@hangfrom@section
\let\@sectioncntformat\SMsave@sectioncntformat
\@removefromreset{equation}{section}
\makeatother
\onecolumngrid
\begin{center}
{\large\bfseries Supplemental Material for\\[2pt]
``Nonequilibrium Spin-Glass Order in a Long-Range Quantum Ising Glass and Its Sensitivity to Noise''}\\[8pt]
Giuseppe D'Auria and Alessandro Silva\\[2pt]
{\small\itshape International School for Advanced Studies (SISSA), Via Bonomea 265, 34136 Trieste, Italy}
\end{center}
\vspace{6pt}
\twocolumngrid

\setcounter{equation}{0}\setcounter{figure}{0}\setcounter{table}{0}\setcounter{section}{0}
\renewcommand{\thesection}{S\arabic{section}}
\renewcommand{\thesubsection}{\thesection.\arabic{subsection}}
\renewcommand{\theequation}{S\arabic{equation}}
\renewcommand{\thefigure}{S\arabic{figure}}
\renewcommand{\thetable}{S\arabic{table}}
\makeatletter\renewcommand{\p@subsection}{}\makeatother
\renewcommand{\theHequation}{S\arabic{equation}}
\renewcommand{\theHfigure}{S\arabic{figure}}
\renewcommand{\theHtable}{S\arabic{table}}
\renewcommand{\theHsection}{S\arabic{section}}
\renewcommand{\theHsubsection}{S\arabic{section}.\arabic{subsection}}

\begin{figure*}[t]
\centering
\includegraphics[width=0.9\linewidth]
{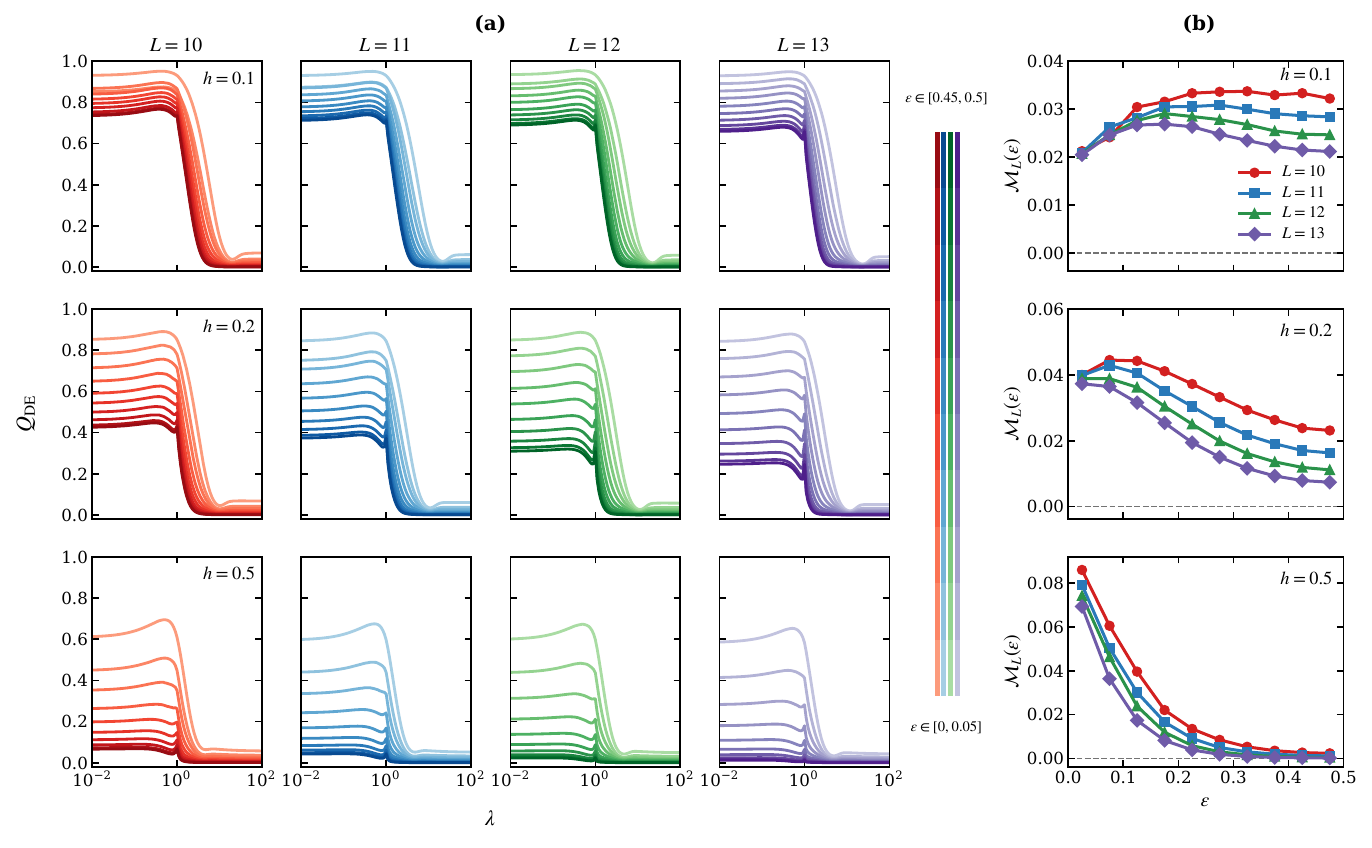}
\caption{
Finite-size diagonal-ensemble response for $\alpha=0.85$.
(a) Shell-resolved $\QDE(\lambda,\epsilon;L)$ for
$h_I=0.1,0.2,0.5$, from top to bottom, and
$L=10,11,12,13$, from left to right. Each panel shows ten energy-density
shells spanning the lower half of the spectrum, $\epsilon\in[0,0.5]$,
with colors ranging from the lightest shell, $\epsilon\in[0,0.05]$, to
the darkest, $\epsilon\in[0.45,0.5]$.  The endpoints $\lambda\to0$ and
$\lambda=1$ coincide at $Q_I(\epsilon;L)$. Cold shells develop an
overshoot at intermediate $0<\lambda<1$, whereas hotter shells
progressively develop an undershoot near $\lambda=1^-$.
(b) Corresponding overshoot amplitude
$\mathcal{M}_L(\epsilon)$ for the four system sizes. Positive values indicate that, at fixed shell and size, $\QDE$ exceeds the common endpoint value for at
least one $0<\lambda<1$.
}
\label{fig:shell_curves}
\end{figure*}

\section{Finite-size diagonal-ensemble response}
\label{sec:fss_nonmonotonic}

Figure~\ref{fig:shell_curves}(a) summarizes the shell-resolved
diagonal-ensemble response for the system sizes accessible to exact
diagonalization, $L=10,11,12,13$, and for
$h_I=0.1,0.2,0.5$. The natural reference is the exact common endpoint
\begin{equation}
Q_I(\epsilon;L)
\equiv
\QDE(1,\epsilon;L)
=
\lim_{\lambda\to0}\QDE(\lambda,\epsilon;L),
\label{eq:QI_finite_size}
\end{equation}
where $\lambda=1$ is the no-quench point and $\lambda\to0$ the
interaction-dominated limit. For $0<\lambda<1$, cold shells develop a
positive excursion above $Q_I$. With increasing energy density this
enhancement weakens, while a negative excursion appears near
$\lambda=1$. For $\lambda>1$, the stronger final transverse field rapidly
suppresses the pair-glass response.

For fixed $(\epsilon,h_I)$, we define the response relative to the common
endpoint as
\begin{equation}
\Delta Q_L(\lambda)
:=
\QDE(\lambda,\epsilon;L)-Q_I(\epsilon;L),
\label{eq:deltaQ_fss}
\end{equation}
with
\begin{equation}
\Delta Q_L(1)=0,
\qquad
\lim_{\lambda\to0}\Delta Q_L(\lambda)=0.
\label{eq:deltaQ_endpoints}
\end{equation}
Its positive and negative excursions are characterized by
\begin{equation}
\mathcal{M}_L(\epsilon)
:=
\max_{0<\lambda<1}
\left[
\QDE(\lambda,\epsilon;L)-Q_I(\epsilon;L)
\right],
\label{eq:overshoot_def}
\end{equation}
and
\begin{equation}
m_L(\epsilon)
:=
\min_{0<\lambda<1}
\left[
\QDE(\lambda,\epsilon;L)-Q_I(\epsilon;L)
\right].
\label{eq:minimum_def}
\end{equation}
Thus $\mathcal{M}_L>0$ measures the overshoot, while $m_L<0$ signals an
undershoot, as depicted in Figure~\ref{fig:shell_curves}(b). The dependence on $h_I$ is left implicit.

The extrema are determined from the disorder-averaged curve on the
original $\lambda$ grid. Performing the extremization only after disorder
averaging avoids the systematic bias that would arise from maximizing or
minimizing each realization separately. If
$\lambda_{\mathcal M,L}$ and $\lambda_{m,L}$ denote the selected grid
points, the corresponding standard errors are
\begin{align}
\sigma_{\mathcal M,L}
&=
\left[
\frac{
\operatorname{Var}_r
\left(
Q_{r,L}(\lambda_{\mathcal M,L})-Q_{r,L}(1)
\right)
}{
N_L
}
\right]^{1/2},
\\
\sigma_{m,L}
&=
\left[
\frac{
\operatorname{Var}_r
\left(
Q_{r,L}(\lambda_{m,L})-Q_{r,L}(1)
\right)
}{
N_L
}
\right]^{1/2},
\label{eq:extrema_sem}
\end{align}
where $r$ labels disorder realizations. We use ten energy-density windows
of width $\Delta\epsilon=0.05$ and $1000$ disorder
realizations for each $(L,h_I)$ dataset.

Because only four sizes are available, we adopt the minimal empirical
finite-size extrapolations
\begin{align}
\mathcal{M}_L(\epsilon)
&=
\mathcal{M}_\infty(\epsilon)
+
\frac{a_{\mathcal M}(\epsilon)}{L},
\\
m_L(\epsilon)
&=
m_\infty(\epsilon)
+
\frac{a_m(\epsilon)}{L},
\\
Q_I(\epsilon;L)
&=
Q_{I,\infty}(\epsilon)
+
\frac{a_I(\epsilon)}{L}.
\label{eq:fss_ansatz}
\end{align}
These forms are used as finite-size trends rather than as a controlled
scaling theory. The fits are unweighted ordinary least-squares (OLS) fits:
the statistical errors in Eq.~\eqref{eq:extrema_sem} are generally
smaller than the residual deviations from linear behavior in $1/L$, so
weighting by those errors would underestimate the uncertainty associated
with the extrapolation itself. We therefore quote the OLS standard error
of the intercept as the extrapolation uncertainty. We regard the sign of an extrapolated intercept as resolved only when its magnitude exceeds twice its OLS standard error.

Because these extrapolations use only $L=10$--$13$, the above conditions
are finite-size diagnostics rather than statistical confidence bounds or
definitions of critical energies.

\subsection{Overshoot}

\begin{figure}[h]
\centering
\includegraphics[width=0.9\linewidth]{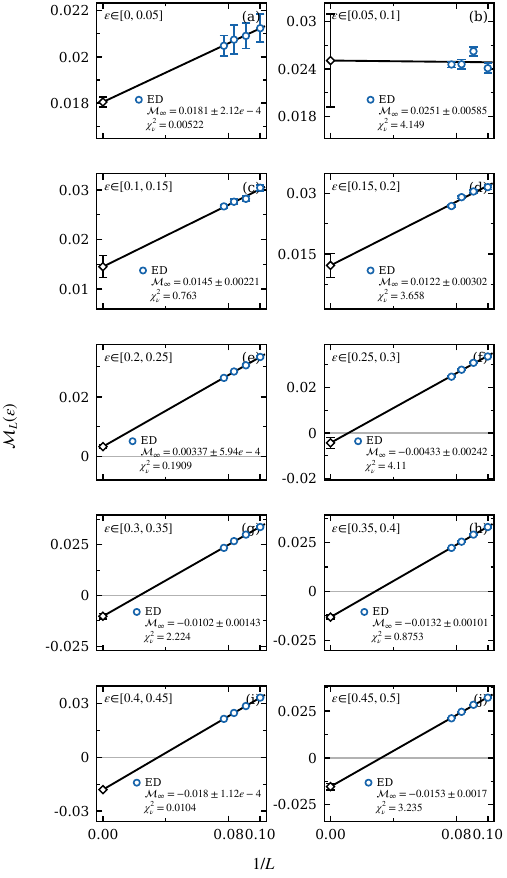}
\caption{
Finite-size scaling of $\mathcal{M}_L(\epsilon)$ for $h_I=0.1$.
Open circles show the finite-size data with their standard errors, the
solid line is the fit
$\mathcal{M}_L=\mathcal{M}_\infty+a_{\mathcal M}/L$, and the open
diamond at $1/L=0$ denotes the extrapolated intercept.
}
\label{fig:M_h01}
\end{figure}

\begin{figure}[h]
\centering
\includegraphics[width=0.9\linewidth]{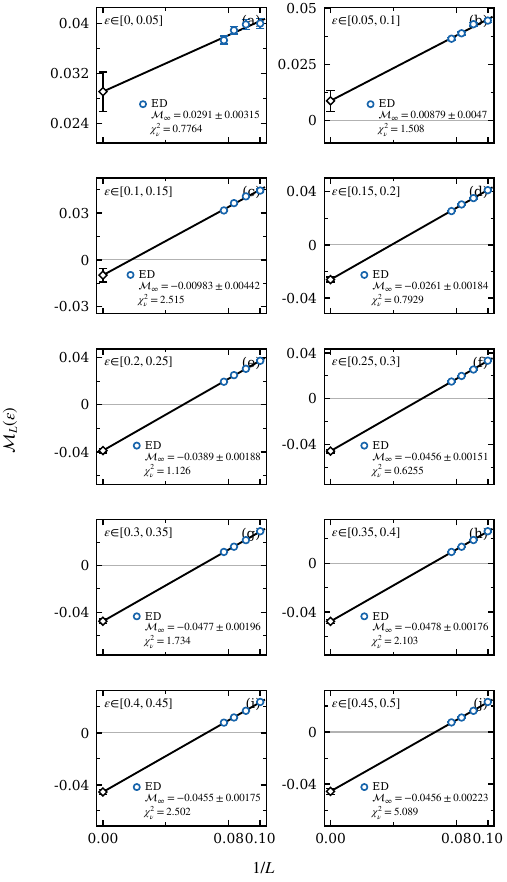}
\caption{
Same as Fig.~\ref{fig:M_h01}, for $h_I=0.2$.
}
\label{fig:M_h02}
\end{figure}

\begin{figure}[h]
\centering
\includegraphics[width=0.9\linewidth]{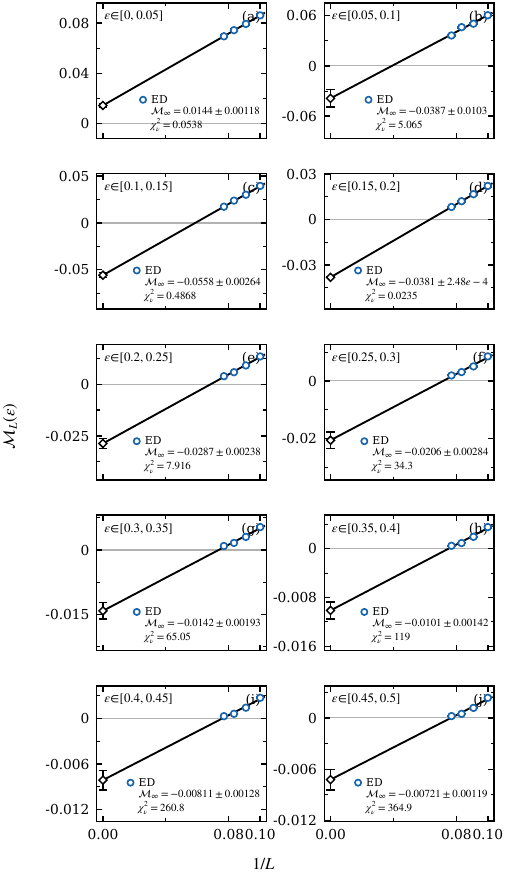}
\caption{
Same as Fig.~\ref{fig:M_h01}, for $h_I=0.5$.
}
\label{fig:M_h05}
\end{figure}

Figures~\ref{fig:M_h01}--\ref{fig:M_h05} show the finite-size evolution of
$\mathcal{M}_L(\epsilon)$ for each energy window.

For $h_I=0.1$, the extrapolation supports a positive overshoot through
$\epsilon\in[0.20,0.25]$. In the next shell,
$\epsilon\in[0.25,0.30]$, the unconstrained linear fit yields a negative
intercept. Since $\mathcal{M}_L\geq0$ by definition, such an intercept is
not physical; rather, it indicates that the observed size dependence is
incompatible with a finite positive limit within the simple $1/L$
extrapolation.

For $h_I=0.2$ and $0.5$, the overshoot survives only in the lowest-energy shell. At higher energies,
$\mathcal{M}_L$ decreases with increasing $L$, while the position of the
maximum shifts toward $\lambda=0$. The accessible sizes therefore do not
support these maxima as finite-$\lambda$ thermodynamic features.

The resulting finite-size trends are
\begin{equation}
\begin{array}{c|c}
h_I & \mathcal{M}_\infty(\epsilon)>0
\\
\hline
0.1 & \epsilon\in[0,0.25]
\\
0.2 & \epsilon\in[0,0.05]
\\
0.5 & \epsilon\in[0,0.05].
\end{array}
\label{eq:overshoot_regions}
\end{equation}
The quoted boundaries are set by the finite shell resolution
$\Delta\epsilon=0.05$ and should not be interpreted as critical energy
densities.

\subsection{Undershoot}

\begin{figure}[h!]
\centering
\includegraphics[width=0.9\linewidth]{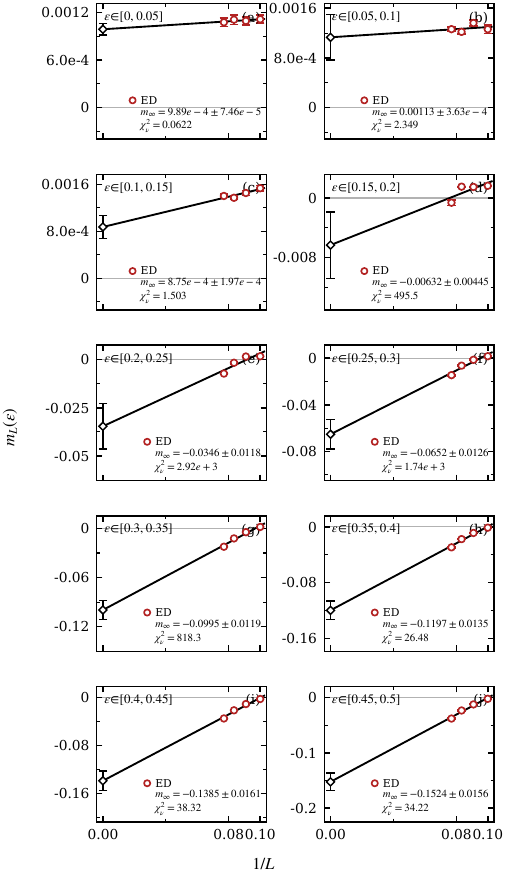}
\caption{
Finite-size scaling of $m_L(\epsilon)$ for $h_I=0.1$.
Negative values indicate an undershoot below $Q_I(\epsilon;L)$.
The solid line is the fit $m_L=m_\infty+a_m/L$, and the open diamond
denotes the extrapolated intercept.
}
\label{fig:m_h01}
\end{figure}

\begin{figure}[h!]
\centering
\includegraphics[width=0.9\linewidth]{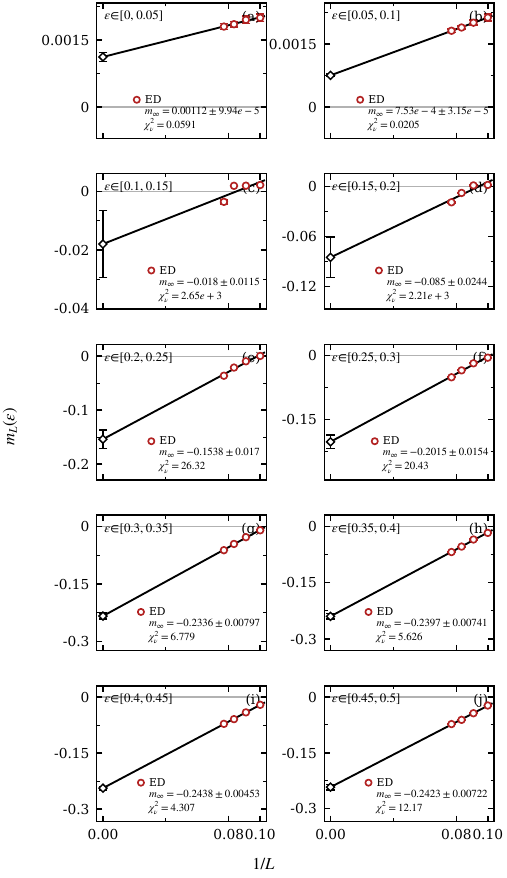}
\caption{
Same as Fig.~\ref{fig:m_h01}, for $h_I=0.2$.
}
\label{fig:m_h02}
\end{figure}

\begin{figure}[h!]
\centering
\includegraphics[width=0.9\linewidth]{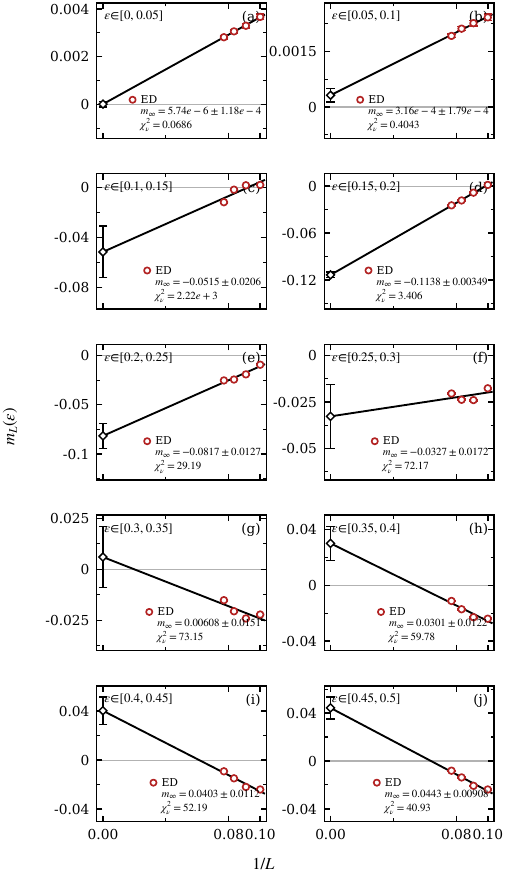}
\caption{
Same as Fig.~\ref{fig:m_h01}, for $h_I=0.5$.
}
\label{fig:m_h05}
\end{figure}

Figures~\ref{fig:m_h01}--\ref{fig:m_h05} show the corresponding signed
minimum $m_L(\epsilon)$. Retaining its sign directly distinguishes an
undershoot, $m_L<0$, from a response that remains above the common
endpoint throughout $0<\lambda<1$.

For $h_I=0.1$, the extrapolated minimum becomes negative in
$\epsilon\in[0.20,0.25]$ and remains negative at larger energy density.
This shell therefore exhibits both a positive overshoot and a negative
minimum in the finite-size extrapolation. For $h_I=0.2$,
$m_\infty<0$ from $\epsilon\in[0.15,0.20]$ upward, whereas for
$h_I=0.5$ a negative intercept is obtained only for
$\epsilon\in[0.10,0.25]$.

A negative $m_\infty$ is not, by itself, sufficient to establish a finite
absolute undershoot. Positivity of the pair-glass order implies the exact
finite-size bound
\begin{equation}
-m_L(\epsilon)
\leq
Q_I(\epsilon;L).
\label{eq:undershoot_bound}
\end{equation}
Hence an absolute undershoot must vanish whenever
$Q_{I,\infty}(\epsilon)=0$. Combining the extrapolations of $m_L$ and
$Q_I$ gives
\begin{equation}
\begin{array}{c|c|c}
h_I
&
m_\infty(\epsilon)<0
&
Q_{I,\infty}(\epsilon)>0\ \text{compatible with data}
\\
\hline
0.1
&
\epsilon\in[0.20,0.50]
&
\epsilon\in[0.20,0.50]
\\
0.2
&
\epsilon\in[0.15,0.50]
&
\epsilon\in[0.15,0.25]
\\
0.5
&
\epsilon\in[0.10,0.25]
&
\text{not resolved}.
\end{array}
\label{eq:undershoot_regions}
\end{equation}
Thus a finite absolute undershoot remains compatible with the available
sizes for $\epsilon\in[0.20,0.50]$ at $h_I=0.1$ and
$\epsilon\in[0.15,0.25]$ at $h_I=0.2$. For $h_I=0.5$, the simultaneous
survival of the reference pair-glass order is not resolved.

Overall, the finite-size data distinguish a cold regime with a robust
positive overshoot, an intermediate regime in which positive and negative
excursions coexist, and a hotter regime where the reference pair-glass
order decreases strongly with system size. The restricted interval
$10\leq L\leq13$ prevents a controlled thermodynamic scaling analysis;
the extrapolations above should therefore be interpreted as finite-size
trends, with the boundary shells providing the natural targets for
larger-size calculations.

\subsection{Results for a larger system size}

\begin{figure*}[t]
\centering
\includegraphics[width=0.65\textwidth]{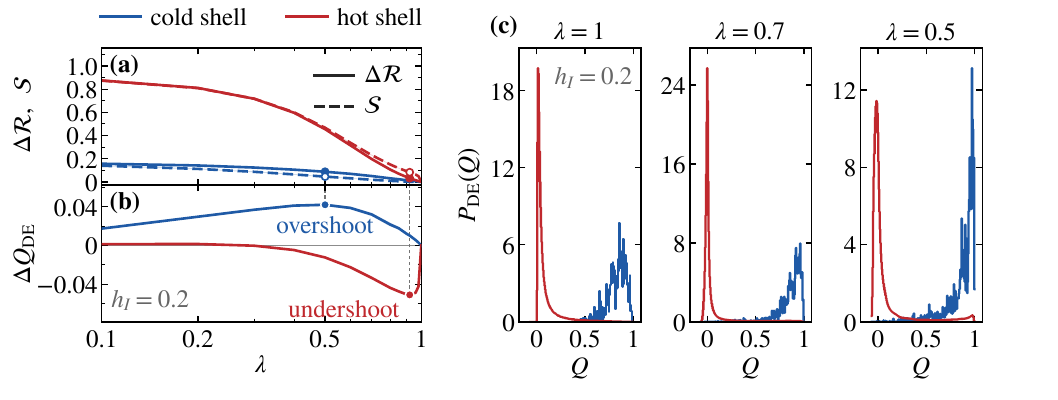}
\caption{Same analysis as in Fig.~\ref{fig:geometry}(d)--(f) of the main text,
but for $L=16$, and $500$ realizations of disorder.}
\label{fig:L16}
\end{figure*}

In Fig.~\ref{fig:L16}, we repeat the analysis presented in Fig.~\ref{fig:geometry}(d)--(f) of the main text for $L=16$, keeping all other parameters fixed. The same qualitative behavior persists at this larger size: the gain in self-overlap dominates in the cold shell, producing an overshoot, whereas the spread of correlations prevails in the hot shell, leading to an undershoot. The corresponding overlap distributions exhibit the same distinct behavior discussed in the main text, providing further support for this interpretation.

\section{Isolated-resonance theory of the cold-shell overshoot}
\label{app:cold_shell_theory}
\label{app:kernel_derivation}
We develop here a microscopic description of the positive excursion of the  diagonal-ensemble pair-glass order $\QDE$ observed in the low-energy, or cold, part of the spectrum. In particular, we show that the overshoot of $\QDE$ in the coldest shells
follows from a simple mechanism: transverse-field resonances between nearly
degenerate classical configurations, each of which increases the
pair-glass order. The starting point is the behavior of individual disorder realizations, illustrated in  Fig.~\ref{fig:resonant_network} of Sec.~\ref{sec:resonant_network} below. In the cold shell, changes in the spectrum and in the longitudinal pair correlations appear to occur mainly through a small number of well-separated avoided crossings. At higher energies these events become more numerous and begin to overlap. This observation suggests modeling the cold-shell response by isolated two-level processes, while neglecting interference between different resonances to leading order. We treat this as an effective description motivated by the finite-size data.

Throughout this section, we consider
the quench protocol explained in the main text, and we restrict to $0\leq\lambda\leq1$.

A computational-basis configuration
$\ket{s}=\ket{s_1,\ldots,s_L}$, with $s_i=\pm1$, has classical interaction
energy
\begin{equation}
v_s
:=
\sum_{i<j}
\frac{\xi_{ij}}{r_{ij}^{\alpha}}
s_i s_j \, ,
\label{eq:classical_energy_cold}
\end{equation}
where we write $J_{ij}:=\xi_{ij}/r_{ij}^{\alpha}$ for brevity.
Consider a rearrangement in which $d$ spins are flipped, and let $\gamma\subset\{1,\ldots,L\}$ denote the cluster of sites involved, with $d=|\gamma|$ and $1\leq d<L$. We denote by $\ket{s^\gamma}$ the configuration obtained from $\ket{s}$ by flipping the spins at these sites. The corresponding classical detuning for such a cluster is
\begin{equation}
\delta_\gamma
:=
v_{s^\gamma}-v_s
=
-2
\sum_{\substack{i\in\gamma\\j\notin\gamma}}
\frac{\xi_{ij}}{r_{ij}^{\alpha}}
s_i s_j .
\label{eq:cluster_detuning_exact}
\end{equation}
Only interactions connecting sites inside the cluster to sites outside it change sign under this rearrangement. When the detuning is comparable to the effective tunnelling amplitude between the two configurations, appreciable hybridization can occur. We now examine this process within an isolated two-level description.

\subsection{Effective two-level Hamiltonian}
\label{sec:ir_block}

Consider a configuration $\ket{s}$ and a nearly degenerate partner $\ket{s^\gamma}$, and assume that all other configurations are far from resonance with both. Eliminating these intermediate configurations through Feshbach--L\"owdin perturbation theory leaves an effective Hamiltonian on the two-dimensional space spanned by $\ket{s}$ and $\ket{s^\gamma}$. We evaluate its off-diagonal element in the forward-scattering approximation to the locator expansion, which retains only the shortest spin-flip paths connecting the two configurations and gives the leading contribution to the effective tunnelling.

Since the two configurations differ by $d$ spin flips, each shortest path involves $d$ applications of the transverse field. The leading tunnelling amplitude is therefore
\begin{equation}
t_\gamma(\lambda)
\simeq
(\lambda h_I)^d\,T_\gamma,
\label{eq:t_gamma_cold}
\end{equation}
where, up to an overall sign that can be absorbed into the relative phase of the two states,
\begin{equation}
T_\gamma
=
\sum_{\pi\in S_d}
\prod_{m=1}^{d-1}
\frac{1}{v_{s_m(\pi)}-v_s}.
\label{eq:Tgamma_cold}
\end{equation}
Here $\pi$ labels the $d!$ shortest paths connecting the two configurations, and $s_m(\pi)$ is the intermediate configuration reached after the first $m$ flips along the path $\pi$. This approximation requires the intermediate configurations to remain sufficiently far from resonance for the perturbative denominators to be well defined. Processes involving additional resonant configurations must instead be described by larger blocks.

Choosing the relative phase of the two states such that $t_\gamma>0$, and subtracting the mean diagonal energy, we obtain the effective two-level Hamiltonian
\begin{equation}
h_\gamma(\lambda)
=
\frac{\delta_\gamma}{2}\,\tau_z
-
t_\gamma(\lambda)\,\tau_x,
\label{eq:h_eff_general}
\end{equation}
with the pseudospin operators
\begin{equation}
\tau_z=\ket{s^\gamma}\!\bra{s^\gamma}-\ket{s}\!\bra{s},
\qquad
\tau_x=\ket{s}\!\bra{s^\gamma}+\ket{s^\gamma}\!\bra{s}.
\label{eq:pseudospin_definitions}
\end{equation}
The same elimination also shifts the diagonal energies. At second order in the transverse field, the detuning becomes
\begin{align}
\delta_\gamma
\;&\to\;
\delta_\gamma
+
(\lambda h_I)^2\left(\Sigma_{s^\gamma}-\Sigma_s\right),
\\
\Sigma_s
&:=
\sum_{u\notin\{s,s^\gamma\}}
\frac{|\bra{u}X\ket{s}|^2}{v_s-v_u},
\label{eq:detuning_self_energy}
\end{align}
with the analogous expression for $\Sigma_{s^\gamma}$.

The relative size of these two effects depends on $d$. For a single-spin flip, $d=1$, there are no intermediate configurations: $T_\gamma=1$ and the tunnelling is simply the direct matrix element, $t_\gamma=\lambda h_I$.
The self-energy shift is then smaller than the tunnelling by a factor of order $h_I/|\delta|$, where $|\delta|$ is a typical detuning of the eliminated configurations, and can be neglected to leading order. For collective rearrangements, $d\geq2$, the shift is of the same order as $t_\gamma$ for $d=2$ and larger for $d\geq3$, so that the effective detuning depends on $\lambda$ already at leading order. Moreover, the resonance window $|\delta_\gamma|\lesssim t_\gamma$ shrinks as $h_I^d$, so that collective resonances are rarer by additional powers of $h_I$. For these reasons, we restrict to single-spin flips, which form the leading cold-shell channel, and discuss the collective ones in Sec.~\ref{sec:ir_validity}.

\subsection{A single resonance}
\label{sec:ir_single}

For small $h_I$, a cold initial eigenstate $\ket{n_I}$ is close to a classical configuration $\ket{s}$, with $s_i=\pm1$. Because of the global spin-flip symmetry, the eigenstates are in fact symmetric or antisymmetric combinations of $\ket{s}$ and $\ket{-s}$. For $L>2$, however, this does not change any of the matrix elements used below, and we can work with $\ket{s}$ directly. The transverse field couples $\ket{s}$ to its $L$ neighbors $\ket{s^k}$, obtained by flipping spin $k$, with matrix element $-\lambda h_I$. The classical energy cost of the flip is
\begin{equation}
\delta_k
=
-2s_k\sum_{j\neq k}J_{kj}s_j .
\label{eq:single_flip_detuning}
\end{equation}
These detunings average to $-4v_s/L$, where $v_s<0$ is the classical energy of $\ket{s}$, so for cold configurations most of them are positive and large. A neighbor with $|\delta_k|\gg h_I$ only weakly dresses the state, whereas a neighbor with $|\delta_k|\lesssim h_I$ is resonant and strongly hybridizes with it. Both cases are described by Eq.~\eqref{eq:h_eff_general} with $\gamma=\{k\}$ and $t_\gamma=\lambda h_I$,
\begin{equation}
h_k(\lambda)
=
\frac{\delta_k}{2}\,\tau_z
-
\lambda h_I\,\tau_x,
\label{eq:h_eff_main}
\end{equation}
where the self-energy shift of $\delta_k$ has been neglected since, as shown in Sec.~\ref{sec:ir_block}, it is of relative order $h_I/|\delta|$.

Flipping spin $k$ reverses $\sigma_i^z\sigma_j^z$ for the $L-1$ pairs that contain $k$, i.e., for a fraction $2/L$ of all pairs, and leaves the other pairs unchanged. Only these affected pairs matter. For each of them, we measure the correlation in units of its classical value $s_is_j$,
\begin{equation}
c
:=
\frac{\langle\sigma_i^z\sigma_j^z\rangle}{s_is_j}
=
-\langle\tau_z\rangle,
\label{eq:normalized_pair_memory}
\end{equation}
where the expectation value is taken in the quantum state under consideration within the two-level subspace: the initial eigenstate for $c_I$ and the post-quench diagonal ensemble for $c_{\mathrm{DE}}$.
Thus, $c=1$ on $\ket{s}$ and $c=-1$ on $\ket{s^k}$.

The eigenstates of $h_k(\lambda)$ are tilted away from the classical axis by an angle $\theta$, with $\tan\theta=2\lambda h_I/|\delta_k|$. In terms of the dimensionless detuning $z:=|\delta_k|/(2h_I)$, the initial and final
tilts are
\begin{equation}
\tan\theta_I=\frac1z,
\qquad
\tan\theta_F=\frac{\lambda}{z},
\label{eq:mixing_angles_cold}
\end{equation}
so that reducing the transverse field reduces the tilt, $\theta_F\leq\theta_I$. The initial state is the eigenstate of $h_k(1)$ connected to $\ket{s}$, with $c_I=\cos\theta_I=z/\sqrt{1+z^2}$. After the quench it is a superposition of the two eigenstates of $h_k(\lambda)$, which carry $c=\pm\cos\theta_F$, with weights $p_\pm=[1\pm\cos(\theta_I-\theta_F)]/2$. Dephasing removes the coherence between them, leaving
\begin{equation}
c_{\mathrm{DE}}
=
(p_+-p_-)\cos\theta_F
=
\cos(\theta_I-\theta_F)\,\cos\theta_F .
\label{eq:cDE_geom}
\end{equation}
The two factors express the competition discussed in the main text. The factor $\cos\theta_F\geq\cos\theta_I$ is the rigidity gain: each final eigenstate is more classical than the initial one. The factor $\cos(\theta_I-\theta_F)\leq1$ is the spread: the initial state is shared between two eigenstates with opposite $c$.

Since $\QDE$ is built from squared correlators, the net effect on an affected pair is
\begin{equation}
K(z,\lambda)
:=
c_{\mathrm{DE}}^2-c_I^2
=
\lambda(1-\lambda)\,
\frac{z^2\left(\lambda^2+\lambda+2z^2\right)}
{\left(z^2+\lambda^2\right)^2\left(1+z^2\right)} .
\label{eq:kernel_main}
\end{equation}
Every factor is non-negative, so that $K>0$ for all $0<\lambda<1$ and $z>0$, and $K$ vanishes both at $\lambda=1$ and as $\lambda\to0$. A single resonance therefore always increases the pair-glass order, and the increase
disappears at both endpoints, as required by Eq.~\eqref{eq:endpoints}. In the notation of the main text, $K=\Delta\Rig-\Spread$ for this two-level system, with $\Delta\Rig=\cos^2\theta_F-\cos^2\theta_I$ and
$\Spread=\cos^2\theta_F\sin^2(\theta_I-\theta_F)$. Near $\lambda=1$ the rigidity gain is linear in $1-\lambda$ and the spread is quadratic, and Eq.~\eqref{eq:kernel_main} shows that the gain wins at every $\lambda$. Since the affected pairs are a fraction $2/L$ of all pairs, the resonance changes $q_{\rm DE}$ of the initial state by $(2/L)\,K$.

\subsection{Average over a cold shell}
\label{sec:ir_average}

When resonances are rare, the $L$ neighbors of $\ket{s}$ act independently and their contributions add,
\begin{equation}
q_{\rm DE}^{(n)}(\lambda)-q_{\rm DE}^{(n)}(1)
\simeq
\frac2L\sum_{k=1}^{L}
K\!\left(\frac{|\delta_k|}{2h_I},\lambda\right).
\label{eq:per_state_response}
\end{equation}
By definition, $\QDE(\lambda)$ is the average of $q_{\rm DE}^{(n)}(\lambda)$ over the $N_\epsilon$ initial states of the shell and over disorder realizations, which we denote by an overline. Applying this average to Eq.~\eqref{eq:per_state_response} gives
\begin{align}
\nonumber
\Delta Q_{\mathrm{cold}}(\lambda)
&:=
\QDE(\lambda)-\QDE(1)
\\
&\simeq
2\,\overline{
\frac{1}{N_\epsilon}\sum_{n}
\frac1L\sum_{k=1}^{L}
K\!\left(\frac{|\delta_k^{(n)}|}{2h_I},\lambda\right)},
\label{eq:cold_average_explicit}
\end{align}
where $\delta_k^{(n)}$ is the detuning of Eq.~\eqref{eq:single_flip_detuning} for the configuration $s$ associated with the initial state $\ket{n_I}$.
The right-hand side depends on the detunings only through their statistics. We therefore introduce the distribution $\rho(\delta)$ of the single-flip detunings, sampled uniformly over the sites $k$, the initial
states of the shell, and the disorder realizations. It is defined by
\begin{equation}
\int_{-\infty}^{\infty}\dd\delta\,\rho(\delta)\,f(\delta)
:=
\overline{
\frac{1}{N_\epsilon}\sum_{n}
\frac1L\sum_{k=1}^{L}
f\big(\delta_k^{(n)}\big)}
\label{eq:rho_definition}
\end{equation}
for any function $f$, and is normalized to unity. With $f(\delta)=K(|\delta|/2h_I,\lambda)$, Eq.~\eqref{eq:cold_average_explicit} becomes
\begin{equation}
\Delta Q_{\mathrm{cold}}(\lambda)
\simeq
2\int_{-\infty}^{\infty}\dd\delta\,
\rho(\delta)\,
K\!\left(\frac{|\delta|}{2h_I},\lambda\right).
\label{eq:cold_channel_integral}
\end{equation}
The disorder average enters only through $\rho$, while the dynamics of each resonance is contained in $K$. Since $K\geq0$, the integral is positive for every $0<\lambda<1$. It receives its main contributions from two regions of $\delta$, which have different $\lambda$ dependence.

\paragraph{Resonant region, $|\delta|\lesssim h_I$.} Here $\rho(\delta)$ is nearly constant on the scale $h_I$ and can be replaced by $\rho(0)$. The remaining integral over $z$ is elementary, 
\begin{equation}
\int_0^\infty K(z,\lambda)\,\dd z
=
\frac{\pi}{2}\,\mathcal H(\lambda),
\qquad
\mathcal H(\lambda)
:=
\frac{\lambda(1-\lambda)(5+3\lambda)}{2(1+\lambda)^2},
\label{eq:H_appendix}
\end{equation}
and gives the contribution $4\pi h_I\rho(0)\,\mathcal H(\lambda)$.

\paragraph{Nonresonant region, $|\delta|\gg h_I$.} Here the kernel reduces to 
\begin{equation}
K\simeq\frac{8h_I^2}{\delta^2}\,G(\lambda),
\qquad
G(\lambda):=\lambda(1-\lambda),
\label{eq:K_hard_expansion}
\end{equation}
which is exactly the result of second-order perturbation theory in the transverse field. Integrated over $\rho$, it gives
$16h_I^2\,\Phi\,G(\lambda)$, where
\begin{equation}
\Phi
=
\lim_{c\to0^+}
\left[
\int_{|\delta|>c}\frac{\rho(\delta)}{\delta^2}\,\dd\delta
-
\frac{2\rho(0)}{c}
\right]
\label{eq:Phi_finite_part}
\end{equation}
is the mean of $1/\delta^2$ with the resonant region subtracted, so that it is not counted twice.

Altogether,
\begin{align}
\Delta Q_{\mathrm{cold}}(\lambda)
&=
A_H\,\mathcal H(\lambda)
+
B_G\,G(\lambda)
+o\!\left(h_I^2\right),
\\
A_H&=4\pi h_I\,\rho(0),
\quad
B_G=16h_I^2\,\Phi .
\label{eq:coldfit_main}
\end{align}
This expansion is exact as $h_I\to0$, provided $\rho$ is smooth at $\delta=0$. It follows by splitting $\rho(\delta)=\rho(0)+[\rho(\delta)-\rho(0)]$ in Eq.~\eqref{eq:cold_channel_integral}: the first term gives the
$\mathcal H$ contribution exactly, and the second gives the $G$ contribution by dominated convergence, since $K\leq 8h_I^2G(\lambda)/\delta^2$ for all $\delta$ and, $K$ being even in $\delta$, only the even part of $\rho(\delta)-\rho(0)$, of order $\delta^2$, contributes.

The two terms have different origins and sizes. The resonant term is of first order in $h_I$ and always positive, since it counts the rare nearly degenerate neighbors. The nonresonant term is of second order and its sign is not fixed: $\Phi>0$ when small detunings are depleted, as expected for cold configurations, but $\Phi<0$ for a distribution peaked at $\delta=0$.
The first-order term is remarkable. The transformation $\prod_i\sigma_i^z$ maps $H(h)$ onto $H(-h)$ and leaves every $\sigma_i^z\sigma_j^z$ invariant, so that $q_{\rm DE}^{(n)}$ of any single sample is even in $h_I$ and starts at order $h_I^2$. The first-order term therefore arises only after the disorder average, from the rare samples with a resonance. At weak field, the overshoot is thus dominated by rare resonances rather than by perturbative dressing.

\subsection{Shape of the overshoot}
\label{sec:ir_endpoints}

The functions $\mathcal H$ and $G$ contain no free parameter and fix the shape of the response. Both vanish at $\lambda=0$ and $\lambda=1$, so Eq.~\eqref{eq:coldfit_main} respects the exact endpoint identity~\eqref{eq:endpoints}. Near the no-quench point,
$\mathcal H(\lambda)\simeq G(\lambda)\simeq1-\lambda$, so that
\begin{equation}
\left.\partial_\lambda\Delta Q_{\mathrm{cold}}\right|_{\lambda=1^-}
=
-\left(A_H+B_G\right)<0 .
\label{eq:cold_initial_slope}
\end{equation}
Since $A_H$ dominates $B_G$ at small $h_I$, $\QDE$ rises immediately when $\lambda$ is lowered below $1$. Near the interaction-dominated endpoint, $\mathcal H(\lambda)\simeq\frac52\lambda$ and $G(\lambda)\simeq\lambda$, so that
\begin{equation}
\Delta Q_{\mathrm{cold}}(\lambda)
\simeq
\left(\tfrac52A_H+B_G\right)\lambda,
\qquad
\lambda\to0,
\label{eq:cold_tail_appendix}
\end{equation}
and $\QDE$ returns linearly and from above to its endpoint value.

In between, both functions are concave. The function $\mathcal H$ has a single maximum at $\lambda_{\mathcal H}=0.3867\ldots$, the root in $(0,1)$ of $3\lambda^3+9\lambda^2+9\lambda-5=0$, and $G$ has a single maximum at $\lambda_G=1/2$. For $A_H>0$ and $B_G\geq0$, the response therefore has a single maximum, which lies between $\lambda_{\mathcal H}$ and $1/2$, close to $\lambda_{\mathcal H}$ when resonances dominate. The overshoot is a broad maximum at intermediate $\lambda$, whose position is set by the universal functions $\mathcal H$ and $G$ rather than by a specific quench amplitude.

\subsection{Range of validity}
\label{sec:ir_validity}

The theory rests on two conditions. First, $h_I$ must be small compared  with the typical detuning, so that cold eigenstates are close to classical configurations. Second, resonances must be rare: the mean number of resonant neighbors per initial state,
\begin{equation}
\nu=4Lh_I\,\rho(0),
\label{eq:nu_def}
\end{equation}
must satisfy $\nu\ll1$. The neglected processes are of three kinds: initial states with two simultaneous resonances, which occur with probability of order $\nu^2$; resonances involving the collective flip of two spins, whose tunnelling amplitude~\eqref{eq:t_gamma_cold} is of order $h_I^2$ and whose detuning~\eqref{eq:detuning_self_energy} depends on $\lambda$ at the same order; and the shifts of $\delta_k$ mentioned above. The first two contribute at order $h_I^2$, like $B_G$, but are suppressed by the small density of resonances; the third contributes only at order $h_I^3$. The resonant term $A_H\mathcal H(\lambda)$ is therefore the controlled leading prediction.
In the comparison with exact diagonalization, we treat $A_H$ and $B_G$ as fit parameters, so that $B_G$ also absorbs these residual corrections.
Finally, at fixed $h_I$ the number of resonances $\nu$ grows with $L$ unless $\rho(0)$ decreases. The isolated-resonance picture is thus a finite-size description of the cold edge of the spectrum, consistent with the finite-size trends of Sec.~\ref{sec:fss_nonmonotonic}.

\begin{figure}[t]
\centering
\includegraphics[width=0.9\linewidth]
{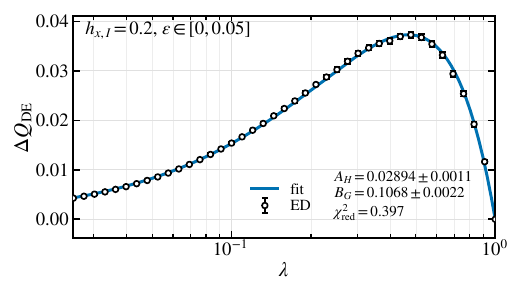}
\caption{
Cold-shell diagonal-ensemble response for $L=13$, $\alpha=0.85$, and $h_I=0.2$, averaged over $10^3$ disorder realizations. The plotted quantity is $\Delta\QDE(\lambda)=\QDE(\lambda)-\QDE(1)$ for the shell $\epsilon\in[0,0.05]$. Exact-diagonalization data are compared with the two-kernel form $A_H\mathcal H(\lambda)+B_GG(\lambda)$ of Eq.~\eqref{eq:coldfit_main}, with $A_H$ and $B_G$ determined by a fit. The two-kernel form captures the positive excursion at intermediate $\lambda$ and the linear return to the exact interaction-dominated endpoint as $\lambda\to0$.
}
\label{fig:microscopic_lambda_flow_hx02}
\end{figure}

Figure~\ref{fig:microscopic_lambda_flow_hx02} summarizes the mechanism.
Lowering $\lambda$ weakens the transverse tunnelling between nearly degenerate configurations, so that the final eigenstates become more classical. The initial state is split between only two of them, and for such a
two-level split the gain in rigidity always outweighs the spread, so that $\QDE$ rises above its no-quench value. As $\lambda\to0$ the tunnelling vanishes and the response returns continuously to the exact endpoint value.

\section{Adiabatic-gauge-potential formulation of the dephased response}
\label{sec:AGP}

The geometric interpretation introduced in the main text can be formulated directly in terms of the adiabatic gauge potential (AGP) associated with the quench coordinate $\lambda$. We use again
\begin{equation}
H(\lambda)=V-\lambda h_I X,
\qquad
\partial_\lambda H=-h_I X,
\end{equation}
with $\lambda=1$ the no-quench point and $\ket{n_I}\equiv\ket{n(1)}$ the fixed initial eigenstate. The correlation patterns $\mathbf c_m$, their weighted center, and the corresponding decomposition $q_{\rm DE}^{(n)}=\mathcal R^{(n)}-\Spread^{(n)}$ are defined as in the main text.

Let
\begin{equation}
H(\lambda)\ket{m(\lambda)}
=
E_m(\lambda)\ket{m(\lambda)}
\end{equation}
and choose the parallel-transport gauge
\begin{equation}
\braket{m(\lambda)|\partial_\lambda m(\lambda)}=0.
\end{equation}
The AGP matrix elements are then
\begin{equation}
\AGP_{m\ell}(\lambda)
:=
\ii\braket{m(\lambda)|\partial_\lambda\ell(\lambda)}
=
-\ii h_I
\frac{X_{m\ell}(\lambda)}
{E_\ell(\lambda)-E_m(\lambda)},
\label{eq:app_AGP_def}
\end{equation}
where $m\neq\ell$, and
\begin{equation}
X_{m\ell}(\lambda)
=
\bra{m(\lambda)}X\ket{\ell(\lambda)},
\qquad
\AGP_{mm}=0.
\end{equation}
The construction can be carried out independently in the two global-$\mathbb Z_2$ parity sectors. For the continuous disorder distribution considered here, accidental degeneracies within a fixed sector occur with probability zero.

It is useful first to make explicit how the AGP acts on the dephasing operation itself. Introducing the instantaneous spectral projectors
\begin{equation}
P_m(\lambda)=\ketbra{m(\lambda)}{m(\lambda)}
\end{equation}
and the dephasing map
\begin{equation}
\mathcal D_\lambda(O)
=
\sum_m P_m(\lambda) O P_m(\lambda),
\end{equation}
one has
\begin{equation}
\partial_\lambda P_m
=
-\ii[\AGP,P_m].
\end{equation}
A direct differentiation therefore gives the exact identity
\begin{equation}
\partial_\lambda\mathcal D_\lambda(O)
=
-\ii[\AGP,\mathcal D_\lambda(O)]
+
\ii\mathcal D_\lambda([\AGP,O]).
\label{eq:app_dD_exact}
\end{equation}
Thus the variation of a dephased observable contains both the AGP rotation of the instantaneous diagonal sector and the projection of the rotated observable back onto that sector.

For $\mathcal O_{ij}=\sigma_i^z\sigma_j^z$, the diagonal-ensemble correlator is
\begin{equation}
c_{ij}^{(n)}(\lambda)
=
\bra{n_I}
\mathcal D_\lambda(\mathcal O_{ij})
\ket{n_I},
\end{equation}
and Eq.~\eqref{eq:app_dD_exact} gives
\begin{align}
\partial_\lambda c_{ij}^{(n)}
=
\ii\bra{n_I}
\Big[
\mathcal D_\lambda([\AGP,\mathcal O_{ij}])
-
[\AGP,\mathcal D_\lambda(\mathcal O_{ij})]
\Big]
\ket{n_I}.
\label{eq:app_dc_exact}
\end{align}
Equation~\eqref{eq:app_dc_exact} is exact for arbitrary quench amplitude and provides the operator form underlying the correlation-space picture discussed in the main text.

The structure becomes particularly transparent close to the no-quench point. At $\lambda=1$, the initial state belongs to the basis defining $\mathcal D_1$, and therefore
\begin{equation}
\left.
\partial_\lambda c_{ij}^{(n)}
\right|_{\lambda=1}
=
\ii
\bra{n_I}
[\AGP(1),\mathcal O_{ij}]
\ket{n_I}.
\label{eq:app_dc_anchor}
\end{equation}
Using Eq.~\eqref{eq:app_AGP_def},
\begin{equation}
\left.
\partial_\lambda c_{ij}^{(n)}
\right|_{\lambda=1}
=
2h_I\,
\Re\sum_{m\neq n}
\frac{
\bra{n_I}\mathcal O_{ij}\ket{m_I}
\bra{m_I}X\ket{n_I}
}{
E_m^I-E_n^I
},
\label{eq:app_dc_anchor_spectral}
\end{equation}
where $\ket{m_I}\equiv\ket{m(1)}$ and $E_m^I\equiv E_m(1)$. Hence the initial deformation of the frozen pair pattern is controlled by low-frequency AGP matrix elements projected onto the same pair sector entering $Q_{\rm DE}$.

The corresponding first derivative of the state-resolved order parameter is
\begin{equation}
\left.
\partial_\lambda q_{\rm DE}^{(n)}
\right|_{\lambda=1}
=
\frac{2}{\Np}
\sum_{i<j}
c_{n,ij}^{I}
\left.
\partial_\lambda c_{ij}^{(n)}
\right|_{\lambda=1},
\label{eq:app_q_slope}
\end{equation}
with $c_{n,ij}^{I}=\bra{n_I}\mathcal O_{ij}\ket{n_I}$. Since the interaction-strengthening branch corresponds to $\lambda<1$, a negative derivative in Eq.~\eqref{eq:app_q_slope} implies an immediate increase of $q_{\rm DE}^{(n)}$ upon reducing the transverse field. Combined with the exact return to the same value as $\lambda\rightarrow0$ discussed in the main text and in Sec.~\ref{sec:fss_nonmonotonic}, such a negative slope is sufficient to produce an intermediate overshoot.

There is, however, an important distinction between this linear deformation and the spreading of the diagonal-ensemble weight. Consider a small perturbation of the quench parameter around the no quench value
\begin{equation}
\lambda=1+\delta\lambda,
\qquad
|\delta\lambda|\ll1.
\end{equation}
For
\begin{equation}
a_m(\lambda)=\braket{m(\lambda)|n_I},
\qquad
p_m(\lambda)=|a_m(\lambda)|^2,
\end{equation}
the AGP gives, for $m\neq n$,
\begin{equation}
a_m(1+\delta\lambda)
=
\ii\delta\lambda\,\AGP_{mn}^{I}
+
O(\delta\lambda^2),
\end{equation}
and therefore
\begin{equation}
p_m(1+\delta\lambda)
=
\delta\lambda^2
|\AGP_{mn}^{I}|^2
+
O(|\delta\lambda|^3),
\qquad
m\neq n.
\label{eq:app_p_pert}
\end{equation}
Normalization similarly gives
\begin{equation}
p_n(1+\delta\lambda)
=
1-
\delta\lambda^2
\sum_{m\neq n}
|\AGP_{mn}^{I}|^2
+
O(|\delta\lambda|^3).
\end{equation}

Defining the distance between the unperturbed correlation patterns,
\begin{equation}
d_{mn}^2
=
\|\mathbf c_m(1)-\mathbf c_n(1)\|^2,
\end{equation}
the variance form of $\Spread$ immediately yields
\begin{equation}
\Spread^{(n)}(1+\delta\lambda)
=
\frac{\delta\lambda^2}{\Np}
\sum_{m\neq n}
|\AGP_{mn}^{I}|^2
d_{mn}^2
+
O(|\delta\lambda|^3),
\label{eq:app_spread_pert_AGP}
\end{equation}
or, equivalently,
\begin{equation}
\Spread^{(n)}(1+\delta\lambda)
=
\frac{h_I^2\delta\lambda^2}{\Np}
\sum_{m\neq n}
\frac{|X_{mn}^{I}|^2}
{(E_m^I-E_n^I)^2}
d_{mn}^2
+
O(|\delta\lambda|^3).
\label{eq:app_spread_pert}
\end{equation}
This is the perturbative expression quoted in the main text. It identifies $\Spread$ as a pattern-weighted fidelity susceptibility: small level spacings enhance the response through the usual AGP denominator, but only transitions between eigenstates carrying different frozen pair patterns contribute appreciably. In particular,
\begin{equation}
\left.
\partial_\lambda\Spread^{(n)}
\right|_{\lambda=1}
=0.
\label{eq:app_spread_zero_slope}
\end{equation}
Thus the rigidity of the adiabatically connected pattern can change already at linear order, whereas cancellation between distinct populated patterns enters only at quadratic order. This separation explains why a weak interaction-strengthening quench can initially increase $Q_{\rm DE}$ even though the quench eventually populates several final eigenstates. At larger amplitudes this perturbative separation ceases to be sufficient, and the full competition between $\mathcal R$ and $\Spread$ must be followed along the quench trajectory.

\subsection{Autonomous AGP flow and pattern transport}
\label{app:agp_flow}

The finite-amplitude evolution can be reconstructed without repeated exact diagonalizations by integrating an autonomous system generated by the same AGP. For the fixed initial state $\ket{n_I}$, define
\begin{equation}
a_m(\lambda)
:=
\braket{m(\lambda)|n_I},
\qquad
p_m(\lambda)
:=
|a_m(\lambda)|^2,
\end{equation}
and
\begin{equation}
\mathcal O_{ij}^{m\ell}(\lambda)
:=
\bra{m(\lambda)}
\sigma_i^z\sigma_j^z
\ket{\ell(\lambda)}.
\end{equation}
The quantities $\{E_m,X_{m\ell},a_m,\mathcal O_{ij}^{m\ell}\}$ satisfy the closed system
\begin{subnumcases}{\label{eq:app_autonomous_system}}
\partial_\lambda E_m=-h_I X_{mm},
\label{eq:app_flow_E}\\
\AGP_{m\ell}
=-\ii h_I\frac{X_{m\ell}}{E_\ell-E_m},
\quad m\neq\ell,
\label{eq:app_flow_A}\\
\partial_\lambda X_{m\ell}
=
\ii\sum_r
\left(
\AGP_{mr}X_{r\ell}
-
X_{mr}\AGP_{r\ell}
\right),
\label{eq:app_flow_X}\\
\partial_\lambda a_m
=
\ii\sum_\ell
\AGP_{m\ell}a_\ell,
\label{eq:app_flow_a}\\
\partial_\lambda\mathcal O_{ij}^{m\ell}
=
\ii\sum_r
\left(
\AGP_{mr}\mathcal O_{ij}^{r\ell}
-
\mathcal O_{ij}^{mr}\AGP_{r\ell}
\right).
\label{eq:app_flow_O}
\end{subnumcases}
A single exact diagonalization at $\lambda=1$ supplies
\begin{equation}
E_m(1),
\qquad
X_{m\ell}(1),
\qquad
a_m(1)=\delta_{mn},
\qquad
\mathcal O_{ij}^{m\ell}(1),
\end{equation}
after which Eq.~\eqref{eq:app_autonomous_system} generates the complete $\lambda$ dependence.

Numerically, it is convenient to propagate the unitary matrix $\mathcal U(\lambda)$ whose columns are the instantaneous eigenvectors in the computational basis,
\begin{equation}
\partial_\lambda\mathcal U
=
-\ii\,\mathcal U\AGP.
\label{eq:app_S_flow}
\end{equation}
Since $\sigma_i^z\sigma_j^z$ is diagonal in that basis,
\begin{equation}
c_{m,ij}(\lambda)
=
\sum_{\mathbf s}
|\mathcal U_{\mathbf s,m}(\lambda)|^2s_i s_j,
\label{eq:app_c_from_S}
\end{equation}
while the diagonal-ensemble pair correlator is
\begin{equation}
c_{ij}(\lambda)
=
\sum_m p_m(\lambda)c_{m,ij}(\lambda).
\end{equation}

 Given that the spread is the pair-averaged variance of the populated patterns, which is the eigenstate-to-eigenstate fluctuation mentioned in the main text, i.e. 
\begin{equation}
\Spread^{(n)}= \frac{1}{\Np}\sum_{i<j}\sum_m p_m\left(c_{m,ij}-c_{ij}\right)^2 ,
\label{eq:app_spread_variance}
\end{equation}
this expression is useful because its derivative separates naturally into two contributions,
\begin{align}
\partial_\lambda\Spread
&=
\frac{1}{\Np}
\sum_{i<j}\sum_m
(\partial_\lambda p_m)
\left[c_{m,ij}-c_{ij}\right]^2
\nonumber\\
&\quad+
\frac{2}{\Np}
\sum_{i<j}\sum_m
p_m
\left[c_{m,ij}-c_{ij}\right]
\partial_\lambda c_{m,ij}.
\label{eq:app_dDpat_dlambda}
\end{align}
The first term, $S_{\mathrm{sel}}^{(\lambda)}$, describes the redistribution of weight among instantaneous eigenstates, whereas the second, $S_{\mathrm{geom}}^{(\lambda)}$, describes the deformation of the pair patterns themselves. Terms proportional to $\partial_\lambda c_{ij}$ cancel because
\begin{equation}
\sum_m p_m(c_{m,ij}-c_{ij})=0.
\end{equation}
Using Eq.~\eqref{eq:app_autonomous_system},
\begin{align}
\partial_\lambda p_m
&=
-2\sum_\ell
\Im\left[
a_m^\ast\AGP_{m\ell}a_\ell
\right],
\label{eq:app_dp_dlambda}\\
\partial_\lambda c_{m,ij}
&=
-2\sum_{r\neq m}
\Im\left[
\AGP_{mr}\mathcal O_{ij}^{rm}
\right],
\label{eq:app_dc_dlambda}
\end{align}
where $\Im$ is the imaginary part.
Since $\Spread(1)=0$, the complete spread follows from
\begin{equation}
\Spread^{(\mathrm{AGP})}(\lambda)
=
\int_1^\lambda
\left[
S_{\mathrm{sel}}^{(\lambda)}(x)
+
S_{\mathrm{geom}}^{(\lambda)}(x)
\right]\dd x .
\label{eq:app_Dpat_integral}
\end{equation}
For $0<\lambda<1$ this may equivalently be written as
\begin{equation}
\Spread^{(\mathrm{AGP})}(\lambda)
=
-\int_\lambda^1
\left[
S_{\mathrm{sel}}^{(\lambda)}(x)
+
S_{\mathrm{geom}}^{(\lambda)}(x)
\right]\dd x .
\end{equation}

\begin{figure}[t]
\centering
\includegraphics[width=\linewidth]{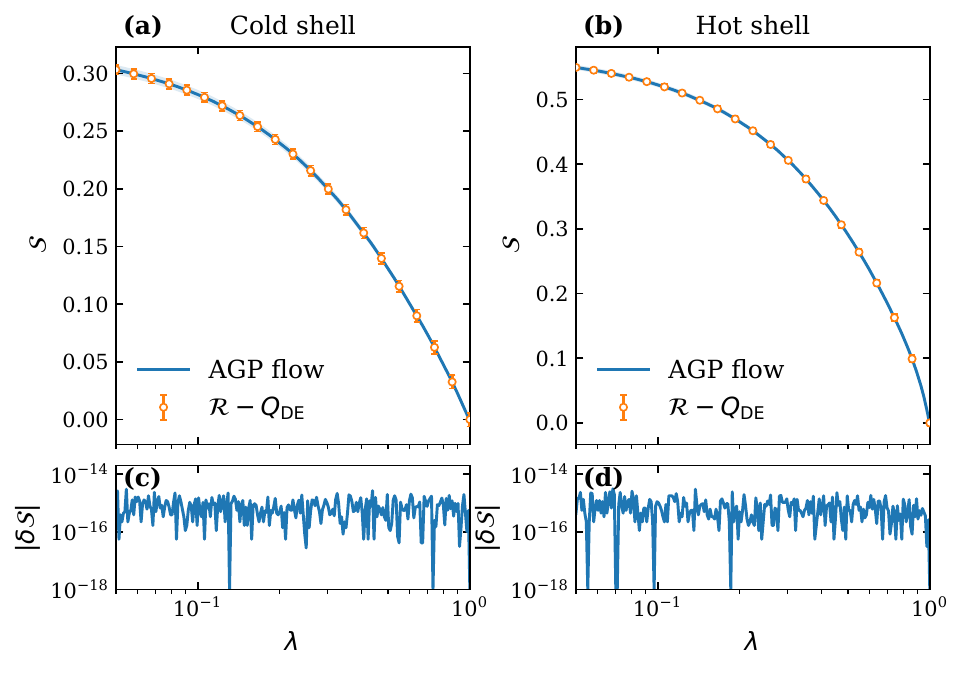}
\caption{
Convergence of the autonomous AGP flow in the $\lambda$ parametrization for $L=10$ and $\alpha=0.85$. The spread $\Spread$ reconstructed from Eq.~\eqref{eq:app_Dpat_integral} is compared with the algebraic reference $\mathcal R-Q_{\mathrm{DE}}$ obtained by exact diagonalization. The residual discrepancy remains at numerical precision, $|\delta\Spread|\sim10^{-15}$.
}
\label{fig:agp_convergence}
\end{figure}

Figure~\ref{fig:agp_convergence} verifies this construction directly. The spread obtained by integrating the AGP flow is indistinguishable, up to numerical precision, from the algebraic value $\mathcal R-Q_{\rm DE}$ computed by exact diagonalization over the full range of $\lambda$.

The same flow also resolves the qualitative difference between cold and hot shells. A useful diagnostic is the inverse participation ratio of the initial state in the instantaneous basis,
\begin{equation}
P_2(\lambda)
=
\sum_m p_m^2(\lambda).
\label{eq:app_IPR}
\end{equation}
For an isolated two-level avoided crossing with populations $p$ and $1-p$,
\begin{equation}
P_2
=
p^2+(1-p)^2
\geq\frac12.
\label{eq:app_two_level_bound}
\end{equation}
Hence $P_2<1/2$ necessarily signals multibranch fragmentation rather than a single isolated resonance.

\begin{figure}[t]
\centering
\includegraphics[width=\linewidth]{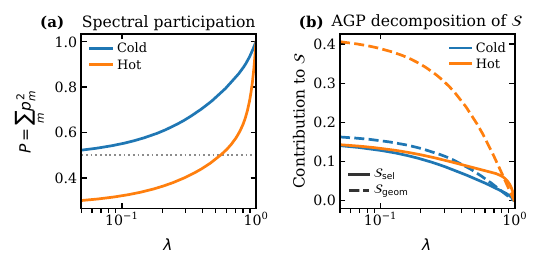}
\caption{
Microscopic diagnostics of the autonomous AGP flow in the $\lambda$ parametrization for $L=10$ and $\alpha=0.85$. (a) Inverse participation ratio $P_2(\lambda)=\sum_m p_m^2(\lambda)$ for representative cold and hot shells. The cold shell remains close to the two-level bound $P_2\geq1/2$, whereas $P_2<1/2$ in the hot shell signals multibranch fragmentation. (b) Cumulative selection and geometric contributions to the pattern spread obtained by integrating the two terms in Eq.~\eqref{eq:app_dDpat_dlambda}. The geometric contribution is strongly enhanced in the hot shell.
}
\label{fig:agp_microscopic}
\end{figure}

This distinction is visible in Fig.~\ref{fig:agp_microscopic}(a). The cold shell remains close to the two-level bound over the interaction-strengthening branch, consistently with the isolated-resonance description developed in the preceding section. The hot shell instead reaches $P_2<1/2$, showing that several spectral branches are populated simultaneously. Figure~\ref{fig:agp_microscopic}(b) shows the corresponding decomposition of $\Spread$. While both redistribution and pattern deformation contribute, the geometric term is strongly enhanced in the hot shell. The resulting suppression of $Q_{\rm DE}$ is therefore not caused merely by spreading over many eigenstates, but by spreading over eigenstates carrying mutually incompatible frozen pair patterns.

The perturbative and finite-amplitude descriptions therefore fit together naturally. Close to $\lambda=1$, pattern incompatibility starts only at second order through the susceptibility in Eq.~\eqref{eq:app_spread_pert}; farther from the anchor, the autonomous flow follows its nonperturbative buildup and separates weight redistribution from genuine deformation in correlation space. Cold shells remain close to a dilute sequence of few-level resonances, allowing the rigidity gain discussed in the main text to dominate. In hotter shells, multibranch fragmentation and the associated geometric spread become sufficiently strong to overcome that gain and generate the undershoot.

\section{From isolated resonances to a multibranch pair-pattern network}
\label{sec:resonant_network}

The isolated-resonance construction of Sec.~\ref{app:cold_shell_theory} describes the cold-shell response in terms of well separated few-level hybridizations.  The state-resolved data show directly how this picture breaks down upon moving toward denser parts of the spectrum.  For a representative disorder realization, the cold shell displays a small number of sharply resolved avoided crossings, each accompanied by a localized rearrangement of the pair correlators and of the diagonal-ensemble response.  At larger energy density these events proliferate and overlap, until no single avoided crossing controls the response.  The resulting evolution is instead a
multibranch resonant network, consistent with the decrease of the inverse participation ratio and the enhanced geometric contribution to $\Spread$ discussed in Sec.~\ref{app:agp_flow}.

\begin{figure}[t]
\includegraphics[width=\linewidth]{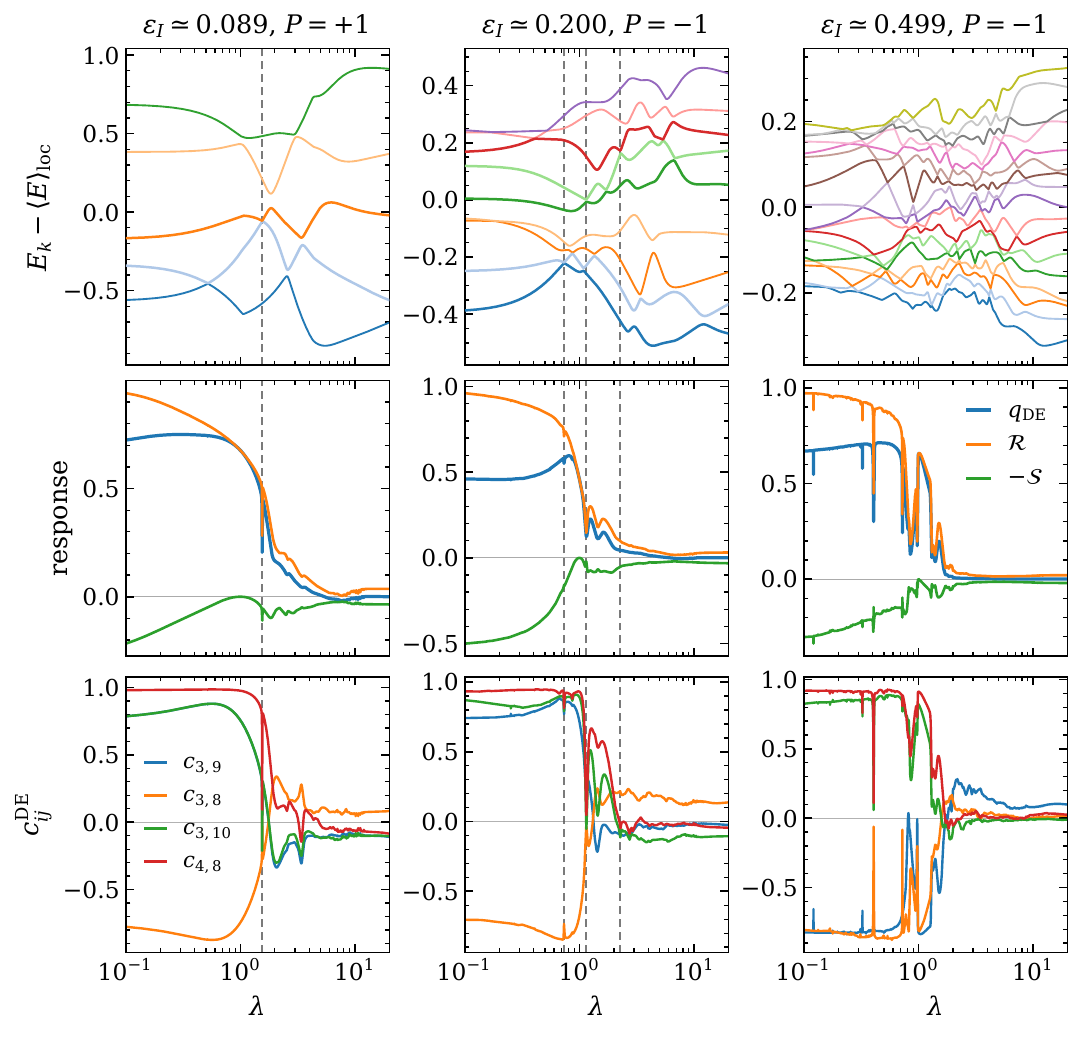}
\caption{
Microscopic evolution from isolated to overlapping resonances for a representative disorder realization at $L=10$, $\alpha=0.85$, and $h_I=0.5$. The three columns correspond to cold, intermediate, and hot initial eigenstates, respectively. The upper row shows the instantaneous energy levels $E_k(\lambda)-\langle E(\lambda)\rangle_{\mathrm{loc}}$, measured relative to the mean energy of the displayed levels at each $\lambda=h_F/h_I$. The middle row displays the diagonal-ensemble order $q_{\mathrm{DE}}$, rigidity $\mathcal{R}$, and negative pattern spread $-\mathcal{S}$, while the lower row shows representative pair correlators $c_{ij}^{\mathrm{DE}}$. Vertical dashed lines mark selected avoided crossings. In the cold shell, individual avoided crossings can be associated with distinct rearrangements of the pair pattern. With increasing energy density, the resonances become denser and overlap, producing a collective multibranch response. The spectra are resolved within fixed global-parity sectors so that exact crossings between different symmetry sectors are excluded.
}
\label{fig:resonant_network}
\end{figure}

Figure~\ref{fig:resonant_network} also makes clear that a small gap alone is not sufficient to produce a large change in $\QDE$.  The response is strong only when the resonant branches also carry appreciably different pair patterns.  This is the finite-amplitude counterpart of the pattern-weighted AGP susceptibility~\eqref{eq:app_spread_pert}: the spectral denominator selects near-resonant branches, whereas their distance in correlation space controls how efficiently the resonance generates pattern spreading.

\begin{figure}[t]
\includegraphics[width=\linewidth]{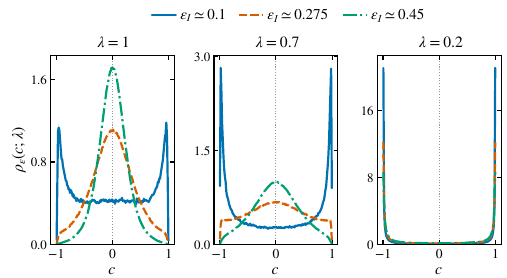}
\caption{
Distribution of diagonal pair correlators of the populated final eigenstates.  Cold shells retain substantial weight at large $|c|$, whereas hotter shells develop a broad distribution centered near $c=0$.  Reducing $\lambda$ shifts weight back toward large $|c|$, reflecting the increasing longitudinal rigidity of the individual final eigenstates.
}
\label{fig:pair_corr_statistics}
\end{figure}

A complementary view is obtained from the distribution of the diagonal pair correlators carried by the populated final eigenstates, as depicted in Fig.~\ref{fig:pair_corr_statistics}.  Using the weights and pair patterns defined in the main text, we introduce only the scalar marginal
\begin{equation}
\rho_{\epsilon,\lambda}(c)
=
\overline{
\frac{1}{|\mathcal W_\epsilon|}
\sum_{n\in\mathcal W_\epsilon}
\frac{1}{\Np}
\sum_{i<j}\sum_m
p_m^{(n)}(\lambda)\,
\delta\!\left[c-c_{m,ij}(\lambda)\right]
},
\label{eq:pair_corr_distribution}
\end{equation}
where the overline denotes the disorder average, and $\mathcal W_\epsilon$ are energy shells.  Its second moment is precisely the rigidity already introduced in the main text,
\begin{equation}
\int dc\,c^2\rho_{\epsilon,\lambda}(c)
=
\Rig(\lambda,\epsilon).
\label{eq:rho_second_moment}
\end{equation}
Thus the distribution provides a direct microscopic visualization of the same quantity entering $\QDE=\Rig-\Spread$.

The two diagnostics therefore expose complementary aspects of the same crossover.  The cold-shell overshoot is associated with a dilute sequence of resonances between a few mutually compatible pair patterns.  In hotter shells, the proliferation of overlapping avoided crossings populates many branches at once; although the individual eigenstates become increasingly longitudinal as $\lambda$ is reduced, their pair patterns explore more
independent directions and cancel more efficiently in the diagonal ensemble.  This microscopic progression connects the isolated-resonance picture of Sec.~\ref{app:cold_shell_theory}  with the nonperturbative AGP transport and correlation-space spreading analyzed in Sec.~\ref{sec:AGP}.

\section{Dynamical spreading and spectral diagnostics}
\label{sec:spectral}

A large value of $\QDE$ does not by itself distinguish a dynamically confined state from one that spreads over many eigenstates of the post-quench Hamiltonian. We therefore complement the diagonal-ensemble analysis with dynamical and spectral diagnostics. 

\subsection{Return-probability dynamics}
\label{supp:return_probability}

For an initial eigenstate $\ket{n_I}$ of $H_I$, we define the return probability
\begin{equation}
R_n(t)
:=
\left|
\bra{n_I}
\ee^{-\ii H_F t}
\ket{n_I}
\right|^2
=
\left|
\sum_m
p_m^{(n)}
\ee^{-\ii E_m^F t}
\right|^2,
\label{eq:R_def_supp}
\end{equation}
where $p_m^{(n)}=|\braket{m_F|n_I}|^2$ are the diagonal-ensemble weights.
For a nondegenerate spectrum,
\begin{equation}
\overline{R_n}^{\,\infty}
=
\sum_m
\left[p_m^{(n)}\right]^2,
\end{equation}
so that the infinite-time return probability is the inverse participation ratio of the initial state in the final eigenbasis.

Because $R_n(t)$ develops broad sample-to-sample fluctuations, we use the typical return probability
\begin{equation}
R_{\mathrm{typ}}(t)
=
\exp\!\left[
\overline{\ln R_n(t)}
\right],
\label{eq:Rtyp_supp}
\end{equation}
where the overline denotes the disorder average within the specified initial energy shell. We also characterize the intermediate-time decay by fitting
\begin{equation}
-\ln R_{\mathrm{typ}}(t)
=
\beta_{\mathrm{fit}}\ln t+c,
\qquad
20\leq t\leq300.
\label{eq:beta_fit}
\end{equation}

\begin{figure*}[t]
\centering
\includegraphics[width=0.8\linewidth]
{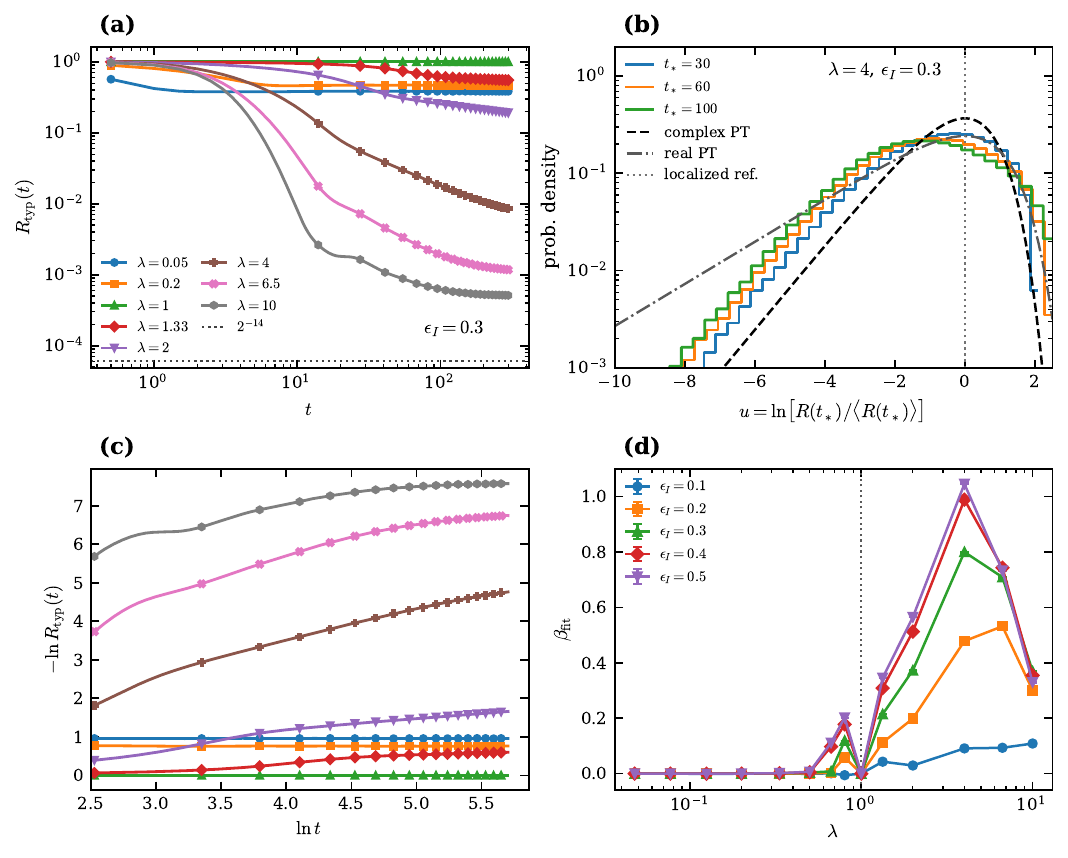}
\caption{
Return-probability diagnostics for $L=14$, $\alpha=0.85$, and
$h_I=0.1$.
(a) Typical return probability $R_{\mathrm{typ}}(t)$ for the initial shell
$\epsilon_I=0.3$ and representative values of
$\lambda=h_F/h_I$. The dotted line marks the Hilbert-space scale
$2^{-L}$. Field-increasing quenches, $\lambda>1$, produce the strongest
decay, whereas the dynamics remains substantially more rigid for
$\lambda<1$; at $\lambda=1$, $R_{\mathrm{typ}}(t)=1$ identically.
(b) Distribution of
$u=\ln[R_n(t_\ast)/\overline{R_n(t_\ast)}]$
for a representative $\lambda>1$ quench and several times $t_\ast$.
Dashed and dash-dotted curves show the complex and real Porter--Thomas
benchmarks. The broad negative-$u$ tail indicates heterogeneous
dephasing, while the deviations from Porter--Thomas statistics exclude
fully random-matrix-like spreading.
(c) $-\ln R_{\mathrm{typ}}(t)$ versus $\ln t$ for the shell in panel (a).
(d) Fitted exponent $\beta_{\mathrm{fit}}$ versus $\lambda$ for several
initial energy shells. The largest decay rates occur for $\lambda>1$,
whereas $\beta_{\mathrm{fit}}$ remains comparatively small throughout
most of the $\lambda<1$ regime in which the diagonal-ensemble overshoot
is observed.
}
\label{fig:return_probability_hx01}
\end{figure*}

Figure~\ref{fig:return_probability_hx01} shows representative results for $L=14$ and $h_I=0.1$. The dynamics is strongly asymmetric around the no-quench point. For $\lambda>1$, increasing the transverse field produces
the strongest decay of $R_{\mathrm{typ}}(t)$ and brings it closest to the
finite Hilbert-space scale $2^{-L}$. In contrast, for $\lambda<1$ the
return probability remains substantially larger over the accessible time
window. Correspondingly, $\beta_{\mathrm{fit}}$ is largest on the
field-increasing side, especially in hotter shells, while remaining small
over most of the regime $0<\lambda<1$ where the pair-glass overshoot
occurs.

A weaker secondary structure in $\beta_{\mathrm{fit}}$ can appear for
$\lambda$ moderately below unity. It marks the onset of mixing among
several final eigenstates, but this mixing alone is insufficient to produce
the overshoot or the undershoot: the populated eigenstates must additionally carry mutually
compatible longitudinal pair-correlation patterns.

To characterize the fluctuations of the return probability, we consider
\begin{equation}
u
=
\ln\!\left[
\frac{R_n(t_\ast)}
{\overline{R_n(t_\ast)}}
\right]
\label{eq:u_def_supp}
\end{equation}
at representative times $t_\ast$. A broad distribution with a long
negative-$u$ tail signals heterogeneous dephasing. As random-matrix
benchmarks we use
\begin{align}
p_{\mathrm{cPT}}(u)
&=
\ee^u\exp[-\ee^u],
\\
p_{\mathrm{rPT}}(u)
&=
\frac{1}{\sqrt{2\pi}}
\ee^{u/2}
\exp[-\ee^u/2],
\label{eq:PT_distributions}
\end{align}
corresponding to complex and real Porter--Thomas statistics,
respectively. The numerical distributions are broad but remain visibly
different from both benchmarks, showing that strong dephasing does not
imply fully random-matrix-like dynamics.

\subsection{Short-range spectral correlations}
\label{supp:gap_ratio}

We next determine whether the post-quench spectrum is Poisson-like or
exhibits random-matrix level repulsion. Because the Hamiltonian has an
exact global $\mathbb Z_2$ symmetry, the analysis must be performed
separately in the two parity sectors.
Within each parity sector, define
\begin{equation}
r_n
=
\frac{
\min(\delta_n,\delta_{n+1})
}{
\max(\delta_n,\delta_{n+1})
},
\qquad
\delta_n=E_{n+1}-E_n.
\label{eq:gap_ratio}
\end{equation}
For an energy shell $\mathcal W_\epsilon$, we compute
\begin{equation}
\overline{r}(\epsilon,\lambda;L)
=
\overline{
\frac{1}{|\mathcal W_\epsilon|}
\sum_{n\in\mathcal W_\epsilon}r_n
}^{\,(\pm)},
\label{eq:r_bar}
\end{equation}
where the overline denotes disorder averaging and the superscript
$(\pm)$ denotes the average over the two parity sectors. The dependence
on the fixed initial field $h_I$ is left implicit. The standard reference
values are
\begin{equation}
 r_{\mathrm P}
=
2\ln2-1
\simeq0.3863,
\qquad
 r_{\mathrm{GOE}}
\simeq0.5307.
\label{eq:r_benchmarks}
\end{equation}
They correspond, respectively, to uncorrelated Poisson levels and to the
Gaussian orthogonal ensemble (GOE), appropriate here because the Hamiltonian
can be represented by a real matrix.

\begin{figure}[t]
\centering
\includegraphics[width=0.95\linewidth]{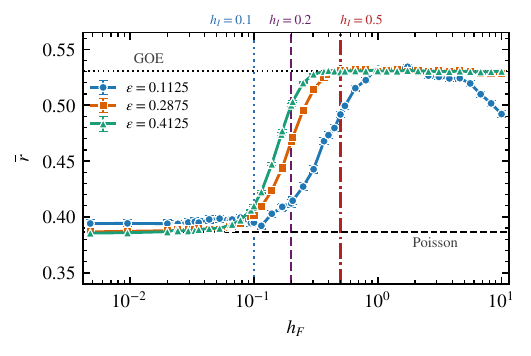}
\caption{
Parity-resolved adjacent-gap ratio $\overline r(\epsilon,h_F;L)$ for
$L=15$, averaged over $10^3$ disorder realizations in three energy shells.
Horizontal dashed and dotted lines denote the Poisson and GOE benchmarks
$ r_{\mathrm P}$ and $ r_{\mathrm{GOE}}$, respectively.
The vertical lines mark the no-quench fields
$h_F=h_I=0.1,0.2,0.5$. The interaction-dominated limit is Poisson-like,
whereas increasing $h_F$ produces an energy-dependent crossover to a broad
GOE-like window.
}
\label{fig:gap_ratio_lambda}
\end{figure}

Figure~\ref{fig:gap_ratio_lambda} shows the shell-resolved ratio for $L=15$
as a function of the final transverse field $h_F$. The three vertical lines
mark the no-quench values $h_F=h_I$ used elsewhere in the quench protocol.
The limit $h_F\rightarrow0$ is uniformly close to the Poisson benchmark
$ r_{\mathrm P}$, as expected when the transverse hybridization of
classical Ising configurations is switched off. Increasing $h_F$ produces a
smooth but pronounced crossover toward GOE statistics. The crossover is
energy dependent: it is reached first in the hotter part of the spectrum,
whereas the lowest-energy shell requires a larger transverse field. This
ordering is consistent with the enhanced rigidity of the glassy low-energy
states. At large $h_F$ the cold shell eventually departs again from the GOE
value while the higher-energy windows remain close to it over a much broader
interval. Because the cold window lies near a spectral edge and contains
substantially fewer levels, we do not interpret this high-field downturn as
evidence for a second universal Poisson regime.

The position of the no-quench Hamiltonian within this crossover sharpens the
spectral interpretation of the dynamical results. For $h_I=0.1$, the finite
systems lie on the Poisson-proximate side of the crossover, so reducing
$\lambda=h_F/h_I$ further suppresses transverse hybridization. For $h_I=0.2$,
the no-quench point falls inside the crossover region. For $h_I=0.5$, by
contrast, the no-quench spectrum is already close to GOE over the
intermediate and hot shells. Thus the pronounced undershoot found for the
larger initial field is naturally associated with quenches out of a much more
strongly hybridized spectral regime.

\begin{figure*}[t]
\centering
\includegraphics[width=\linewidth]{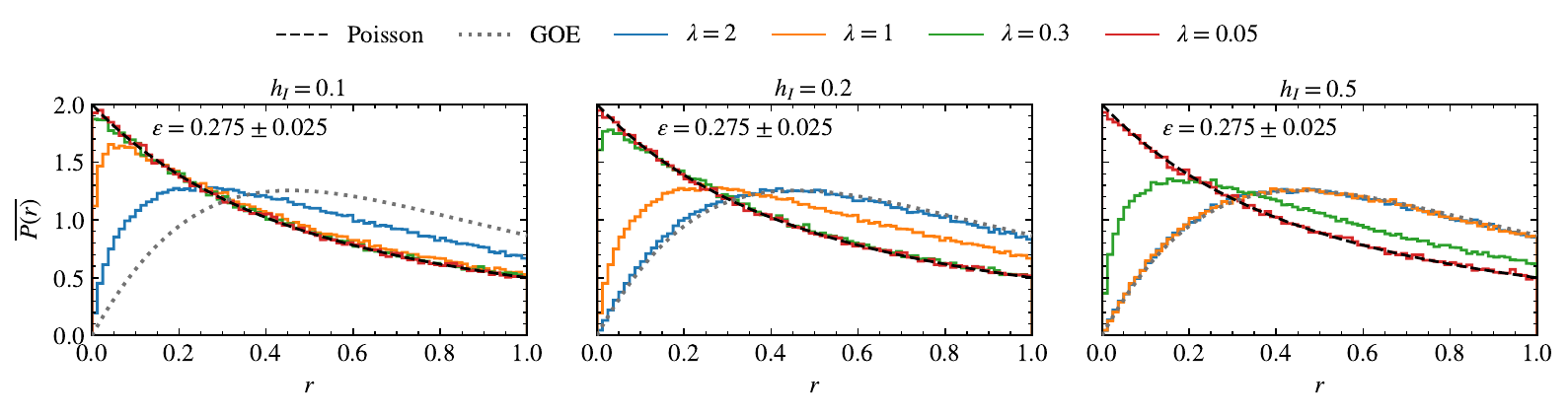}
\caption{
Probability density $\overline{P(r)}$ of the parity-resolved adjacent-gap
ratio for $L=15$ and $\epsilon=0.275\pm0.025$, averaged over $10^3$ disorder
realizations. Columns correspond to $h_I=0.1,0.2,0.5$; colored curves show
representative values of $\lambda=h_F/h_I$. Dashed and dotted black curves
are the Poisson and GOE predictions in Eq.~\eqref{eq:Pr_benchmarks}. The
progressive loss of weight near $r=0$ provides a distribution-level measure
of the crossover from weakly repelling, Poisson-like levels to GOE-like
spectral correlations.
}
\label{fig:gap_ratio_distribution}
\end{figure*}

We also study the full gap-ratio probability density
$\overline{P(r)}$. For the folded ratio $0\leq r\leq1$, the reference
distributions are
\begin{equation}
P_{\mathrm P}(r)=\frac{2}{(1+r)^2},
\qquad
P_{\mathrm{GOE}}(r)
=
\frac{27}{4}\,
\frac{r+r^2}{(1+r+r^2)^{5/2}},
\label{eq:Pr_benchmarks}
\end{equation}
where the second expression is the GOE Wigner-like
surmise~\cite{Atas2013}. Figure~\ref{fig:gap_ratio_distribution} displays
$\overline{P(r)}$ in the intermediate-energy window
$\epsilon=0.275\pm0.025$ for several values of $h_I$ and
$\lambda=h_F/h_I$. When $h_F$ is small, the distribution retains the large
weight near $r=0$ characteristic of weak level repulsion and follows the
Poisson curve closely. Increasing the transverse field progressively depletes
this small-$r$ weight and generates the broad maximum characteristic of GOE
statistics. In particular, for $h_I=0.5$ the distributions at $\lambda=1$
and $\lambda=2$ are nearly indistinguishable from the GOE reference, whereas
small $\lambda$ restores a Poisson-like distribution. The evolution of
$\overline{P(r)}$ therefore confirms that the crossover seen in
$\overline r$ reflects a reorganization of the complete local level
statistics, not merely of their first moment.

The preceding results describe the spectrum at a fixed system size. To test
whether the apparent Poisson--GOE crossover is reinforced or reversed as the
system grows, we analyze the finite-size flow of the gap ratio. We follow the
RG construction introduced for many-body spectral statistics by Niedda,
Bracci Testasecca, Magnifico, Balducci, Vanoni, and Scardicchio
~\cite{Niedda2025RG}. Their useful observation is that the normalized gap
ratio
\begin{equation}
\phi
=
\frac{
\overline r- r_{\mathrm P}
}{
 r_{\mathrm{GOE}}- r_{\mathrm P}
}
\label{eq:phi_gap_flow}
\end{equation}
provides a natural running spectral variable: $\phi=0$ is Poisson-like and
$\phi=1$ is GOE-like. The corresponding beta function is defined as
\begin{equation}
\beta(\phi)
=
\frac{\dd\ln\phi}{\dd\ln N}.
\label{eq:beta_gap_flow}
\end{equation}
Here the level statistics are computed separately in a parity block of
Hilbert-space dimension $N=2^{L-1}$. Consequently,
\begin{equation}
\beta(\phi)
=
\frac{1}{\ln2}\frac{\dd\ln\phi}{\dd L}.
\label{eq:beta_gap_flow_L}
\end{equation}
The multiplicative factor relating $2^{L-1}$ to $2^L$ is immaterial for this
logarithmic derivative.

\begin{figure}[t]
\centering
\includegraphics[width=\linewidth]{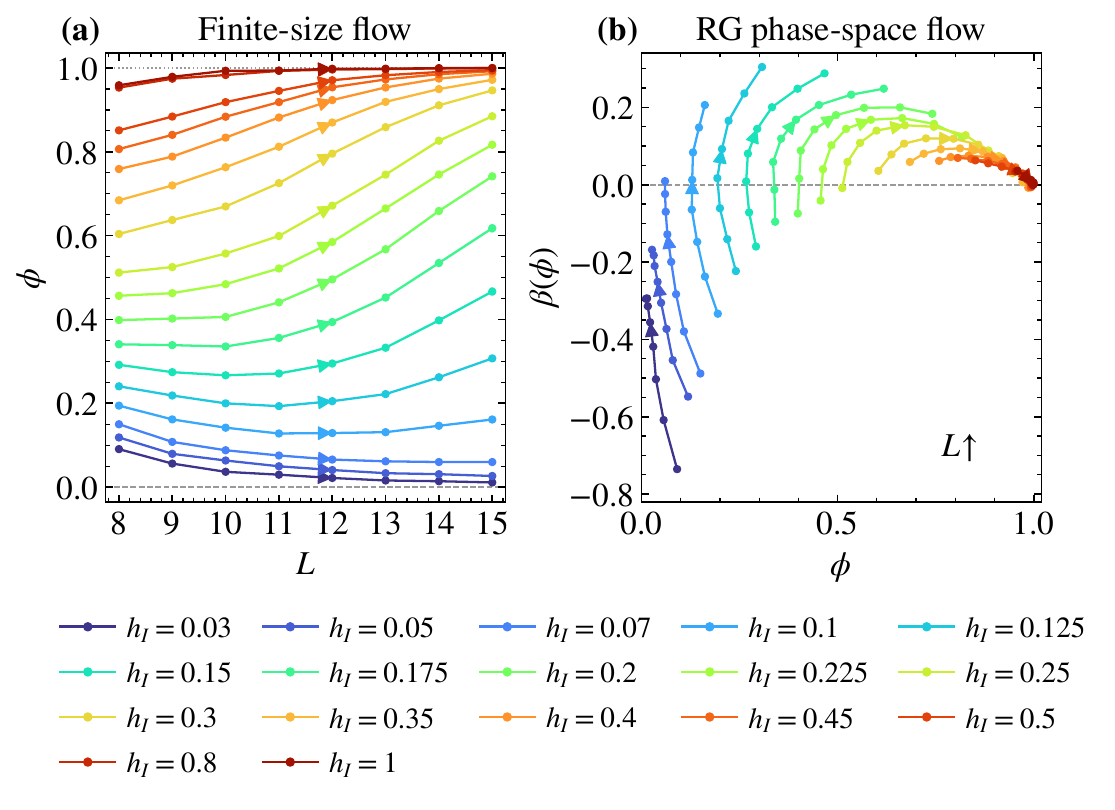}
\caption{
Finite-size RG flow of the full-spectrum disorder-averaged adjacent-gap
ratio for $L=8,\ldots,15$.
(a) Normalized spectral variable
$\phi=(\overline r- r_{\mathrm P})/
( r_{\mathrm{GOE}}- r_{\mathrm P})$
for fixed fields along the no-quench family $h_F=h_I$. Small fields flow
toward the Poisson boundary, large fields flow toward GOE, and intermediate
fields display a nonmonotonic size dependence.
(b) The same trajectories in the $(\phi,\beta)$ plane, with
$\beta=\dd\ln\phi/\dd\ln N$ and $N=2^{L-1}$. The label $L\uparrow$ indicates
increasing system size. The sign change of $\beta$ at intermediate $h_I$
exposes a turning flow that is hidden in a single-size analysis. The lack of
collapse onto a unique $\beta(\phi)$ curve indicates appreciable corrections
to simple one-parameter scaling over the accessible sizes.
}
\label{fig:gap_ratio_rg_flow}
\end{figure}

For this finite-size analysis we use the full-spectrum, parity-resolved
disorder-averaged ratio $\overline r$. This avoids taking a size derivative
across slightly different microcanonical windows and allows us to use all
sizes $L=8,\ldots,15$ on equal footing. The RG trajectories refer to the
no-quench Hamiltonian family, for which $h_F=h_I\equiv h$; we retain the
notation $h_I$ in Fig.~\ref{fig:gap_ratio_rg_flow} to connect directly with
the initial Hamiltonians used in the quench protocol. At each fixed $h_I$ we
interpolate the numerical $\overline r(h;L)$ to a common field value and fit
$\ln\phi(L)$ by a cubic polynomial over the available sizes
~\cite{Niedda2025RG}. The derivative of this smooth interpolation gives
Eq.~\eqref{eq:beta_gap_flow_L}.

Figure~\ref{fig:gap_ratio_rg_flow}(a) shows a clear field-dependent
finite-size flow. For $h_I\lesssim0.07$, $\phi$ decreases with $L$,
corresponding to $\beta<0$ and a flow toward Poisson statistics, whereas at
large fields $\phi$ increases toward the GOE limit. At intermediate fields,
$h_I\simeq0.1$--$0.2$, the flow is nonmonotonic and develops a turning point,
which appears in panel~(b) as a crossing of $\beta=0$.

This structure is closely analogous, at the level of the RG diagnostic, to
the turning trajectories emphasized by Niedda \textit{et al.}
~\cite{Niedda2025RG}. A strict one-parameter scaling hypothesis would require
all microscopic Hamiltonians to lie on a unique curve $\beta=\beta(\phi)$.
Our trajectories do not collapse onto such a single curve over the accessible
sizes: different fields can reach comparable values of $\phi$ with visibly
different $\beta$, and the intermediate-field flows possess a genuine
turning point. The data therefore indicate that a second scaling direction is relevant at these sizes. A more detailed characterization of the associated  two-parameter RG flow would require a dedicated finite-size scaling analysis and lies beyond the scope of the present work.
\subsection{Long-range spectral rigidity}
\label{supp:sff}

The adjacent-gap ratio probes  local level repulsion. To test
long-range correlations, we compute the shell-resolved spectral form
factor
\begin{equation}
K(\tau)
=
\frac{
\overline{|Z(\tau)|^2}
}{
\overline{N_{\mathrm{eff}}}
},
\qquad
K_{\mathrm c}(\tau)
=
\frac{
\overline{|Z(\tau)|^2}
-
|\overline{Z(\tau)}|^2
}{
\overline{N_{\mathrm{eff}}}
},
\label{eq:SFF_supp}
\end{equation}
with
\begin{equation}
Z(\tau)
=
\sum_{n\in\mathcal W_\epsilon}
\ee^{-\ii\tau E_n}.
\label{eq:Z_def_supp}
\end{equation}
Here $N_{\mathrm{eff}}$ is the number of levels in the shell, and the
average includes disorder and symmetry sectors. The connected subtraction
removes the smooth contribution associated with the finite spectral window
and isolates the correlations responsible for the random-matrix
ramp~\cite{Leviandier1986,MehtaBook,HaakeBook,Cotler2017}.

The spectral form factor confirms the gap-ratio analysis. At intermediate
$\lambda$, the connected component develops a pronounced growth toward
the plateau and follows the GOE trend over a finite time interval,
particularly away from the lowest-energy shell. In contrast, as
$\lambda\to0$ the connected form factor lacks a clear GOE-like ramp,
consistent with weak long-range spectral rigidity in the
interaction-dominated glass. The cold shell exhibits larger finite-size
deviations throughout, reflecting the persistence of glassy structure near
the spectral edge.

\begin{figure*}[t]
\centering
\includegraphics[width=0.9\linewidth]{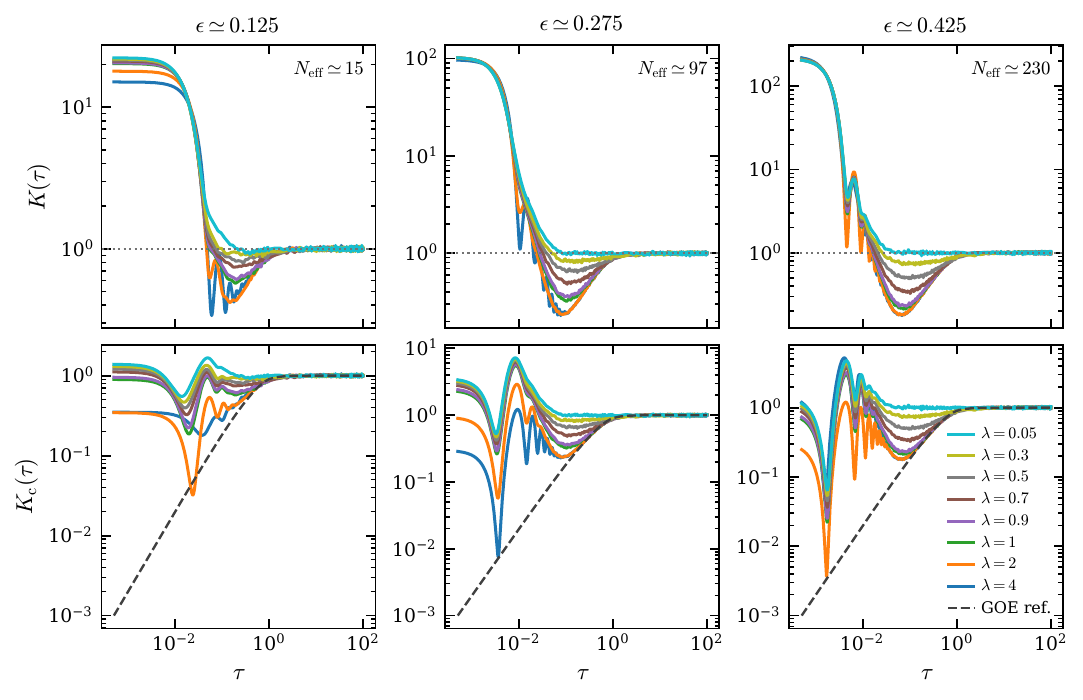}
\caption{
Shell-resolved spectral form factor for $L=12$, averaged over
$10^3$ disorder realizations, for representative values of $\lambda$ and
energy density. Top row: normalized total form factor $K(\tau)$.
Bottom row: connected component $K_{\mathrm c}(\tau)$. Columns correspond
to cold, intermediate, and hot shells. The dashed line in the lower panels
denotes the GOE reference. At intermediate $\lambda$, the connected form
factor develops a GOE-like ramp over a finite time interval, particularly
away from the spectral edge. In the interaction-dominated
$\lambda\to0$ regime, this ramp is strongly suppressed, consistently with
the Poisson-like adjacent-gap statistics.
}
\label{fig:sff_total_connected_supp}
\end{figure*}

Taken together, the return probability and spectral diagnostics separate
the pair-glass overshoot from ordinary many-body spreading. Increasing the
transverse field, $\lambda>1$, produces the strongest decay of the initial
state and the most pronounced random-matrix-like behavior. The overshoot,
by contrast, occurs for $\lambda<1$, where the dynamics is comparatively
rigid and the spectrum evolves toward the interaction-dominated
Poisson-like regime. Its origin is therefore not enhanced ergodicity, but
the constructive reorganization of the diagonal-ensemble weight among
final eigenstates whose longitudinal pair-correlation patterns remain
mutually compatible.

\section{Weak-noise kinetic limit of longitudinal dephasing}
\label{sec:weak_noise_gksl}

The main text studies the finite-time evolution generated by the full GKSL equation starting from the initial eigenstate $\ket{n_I}$. Here we derive the complementary weak-noise limit, which separates the fast Hamiltonian dephasing from the slow bath-induced redistribution of diagonal-ensemble weights. This formulation provides an analytic description of the noise sensitivity observed in the full-GKSL simulations and identifies the microscopic quantity controlling its onset.

We start from the same generator as in the main text,
\begin{equation}
\frac{d\rho}{dt}
=
\mathcal L_0\rho+\gamma\mathcal D_z\rho,
\qquad
\mathcal L_0\rho=-\ii[H_F,\rho],
\label{eq:gksl_supp}
\end{equation}
with
\begin{equation}
\mathcal D_z\rho
=
\sum_{i=1}^{L}
\left(
\sigma_i^z\rho\sigma_i^z-\rho
\right).
\label{eq:Dz_supp}
\end{equation}
For $\gamma$ much smaller than the relevant Hamiltonian frequencies, the coherent dynamics occurs on a parametrically shorter time scale than the dissipative redistribution. The natural slow variable is therefore
\begin{equation}
s=\gamma t,
\end{equation}
and the kinetic regime is obtained in the scaling limit
\begin{equation}
\gamma\rightarrow0,
\qquad
t\rightarrow\infty,
\qquad
s=\gamma t
\ \ {\rm fixed}.
\label{eq:van_hove_limit}
\end{equation}

Let
\begin{equation}
\mathcal P A
=
\sum_E\Pi_E A\Pi_E
\label{eq:secular_projector}
\end{equation}
denote the projection onto the kernel of the Hamiltonian Liouvillian, where $\Pi_E$ projects onto an eigenspace of $H_F$. Averaging the rapidly oscillating components generated by $\mathcal L_0$ gives the secular slow dynamics
\begin{equation}
\frac{d\widetilde\rho}{ds}
=
\mathcal P\mathcal D_z\mathcal P\,\widetilde\rho.
\label{eq:secular_generator}
\end{equation}
The initial condition of this slow evolution is not an additional preparation protocol: it is the Hamiltonian-dephased component of the original state,
\begin{equation}
\widetilde\rho(0)
=
\mathcal P
\ketbra{n_I}{n_I}
=
\rho_{\rm DE}.
\label{eq:slow_initial_condition}
\end{equation}
Thus Eq.~\eqref{eq:secular_generator} is the weak-noise kinetic limit of the full evolution considered in the main text.

For a nondegenerate finite-size spectrum,
\begin{equation}
\widetilde\rho(s)
=
\sum_m p_m(s)
\ketbra{m_F}{m_F},
\end{equation}
and Eq.~\eqref{eq:secular_generator} reduces to the Pauli master equation
\begin{equation}
\frac{dp_m}{ds}
=
\sum_n K_{mn}p_n,
\label{eq:pauli_master_supp}
\end{equation}
with
\begin{equation}
K_{mn}
=
\sum_{i=1}^{L}
\left|
\bra{m_F}\sigma_i^z\ket{n_F}
\right|^2,
\qquad
m\neq n,
\label{eq:Kmn_supp}
\end{equation}
and
\begin{equation}
K_{nn}
=
-\sum_{m\neq n}K_{mn}.
\label{eq:Knn_supp}
\end{equation}
Exact degeneracies are retained as secular blocks in Eq.~\eqref{eq:secular_generator}. Since $\sigma_i^z$ is Hermitian,
\begin{equation}
K_{mn}=K_{nm},
\label{eq:K_symmetric}
\end{equation}
so the kinetic generator satisfies detailed balance with respect to the uniform distribution. 

The pair-glass response we consider along the slow dynamics is
\begin{equation}
Q(\lambda,s)
=
\frac{1}{\Np}
\sum_{i<j}
\left[
\Tr\!\left(
\widetilde\rho(s)\sigma_i^z\sigma_j^z
\right)
\right]^2,
\label{eq:Q_slow}
\end{equation}
with
\begin{equation}
Q(\lambda,0)
=
Q_{\rm DE}(\lambda).
\label{eq:Qslow_initial}
\end{equation}
Within the kinetic regime, $\gamma$ and $t$ therefore enter only through their product $s=\gamma t$. The same accumulated dephasing can be realized by different rate--time pairs, provided the scale separation underlying Eq.~\eqref{eq:van_hove_limit} remains valid.

The interaction-dominated endpoint is protected more strongly than the secular approximation alone would imply.
Moreover, the adjoint dissipator satisfies
\begin{align}
\mathcal D_z^\dagger(O_{ij})
&=
\sum_k
\left(
\sigma_k^z O_{ij}\sigma_k^z-O_{ij}
\right)=0.
\end{align}
Hence the full adjoint GKSL generator obeys
\begin{equation}
\mathcal L^\dagger(O_{ij})=0
\qquad
(\lambda=0),
\end{equation}
independently of the state. Every longitudinal pair correlator is therefore conserved exactly,
\begin{equation}
Q(0,\gamma;t)=Q_I,
\qquad
\forall\,\gamma\geq0,
\quad
\forall\,t\geq0.
\label{eq:protected_supp}
\end{equation}
The protection of the $\lambda=0$ reference is thus a property of the full GKSL dynamics and not an artifact of the secular reduction.

The finite-$\lambda$ behavior is qualitatively different. When the rate graph generated by Eq.~\eqref{eq:Kmn_supp} is connected, Eq.~\eqref{eq:pauli_master_supp} approaches the uniform distribution. Since
\begin{equation}
\Tr\!\left(
\sigma_i^z\sigma_j^z
\right)=0,
\qquad
i\neq j,
\end{equation}
the asymptotic pair pattern vanishes and
\begin{equation}
\lim_{s\rightarrow\infty}
Q(\lambda,s)=0,
\qquad
\lambda>0,
\label{eq:Q_asymptotic}
\end{equation}
whereas Eq.~\eqref{eq:protected_supp} holds at $\lambda=0$. Longitudinal dephasing therefore separates an exactly pinned reference from a finite-$\lambda$ many-body structure that is progressively erased.

\begin{figure*}[t]
\centering
\includegraphics[width=0.7\linewidth]{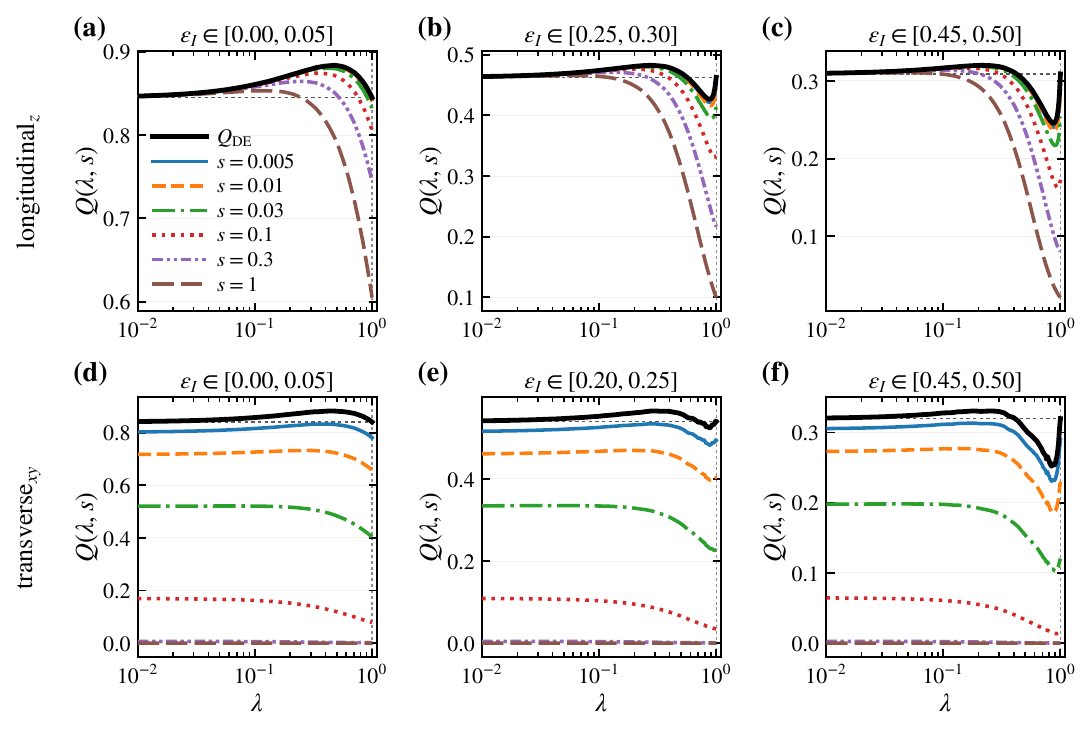}
\caption{
Weak-noise evolution of the pair-glass response $Q(\lambda,s)$
for $L=12$, $\alpha=0.85$, and $h_I=0.2$.
Panels (a)--(c) show longitudinal dephasing
[$s=\gamma t$], while (d)--(f) show transverse dephasing
[$s=(\gamma_x+\gamma_y)t$, with $\gamma_x=\gamma_y$].
The initial energy shells are indicated above each panel.
Black curves denote the diagonal-ensemble reference
$Q_{\rm DE}(\lambda)$; colored curves correspond to increasing
slow times $s$.
Longitudinal noise preserves the small-$\lambda$ response
while suppressing the finite-$\lambda$ signal, whereas
transverse noise induces relaxation across the entire
$\lambda$ range, with a marked dependence on the initial energy shell.
}
\label{fig:transverse_noise_slow}
\end{figure*}

This separation is visible directly in Fig.~\ref{fig:transverse_noise_slow}(a)-(c).
Starting from $Q_{\rm DE}(\lambda)$ at $s=0$, accumulated dephasing progressively suppresses the interior response while leaving the $\lambda\rightarrow0$ endpoint unchanged. The effect is strongly shell dependent. In the cold shell the overshoot survives over an extended range of slow times before being suppressed, whereas the intermediate and hot shells lose their finite-$\lambda$ structure substantially more rapidly.

The onset of this degradation can be obtained in closed form. Denote the pair-correlation pattern of the final eigenstate $\ket{m_F}$ by
\begin{equation}
\mathbf c_m
=
\left(
\bra{m_F}
\sigma_i^z\sigma_j^z
\ket{m_F}
\right)_{i<j},
\end{equation}
and define the instantaneous center
\begin{equation}
\overline{\mathbf c}(s)
=
\sum_m p_m(s)\mathbf c_m.
\end{equation}
Then
\begin{equation}
Q(\lambda,s)
=
\frac{
\left\|
\overline{\mathbf c}(s)
\right\|^2
}{\Np}.
\label{eq:Q_cbar_noise}
\end{equation}
At $s=0$, let
\begin{equation}
u_m
:=
\overline{\mathbf c}(0)\cdot\mathbf c_m.
\label{eq:um_noise}
\end{equation}
Using Eq.~\eqref{eq:pauli_master_supp} and the symmetry of $K_{mn}$ gives
\begin{align}
\nonumber
R(\lambda)
&:=
-
\left.
\frac{\partial Q}{\partial s}
\right|_{s=0}
\\
&=
-\frac{1}{\Np}
\sum_{m\neq n}
K_{mn}
\left(
p_n-p_m
\right)
\left(
u_m-u_n
\right),
\label{eq:initial_noise_slope}
\end{align}
where $p_m\equiv p_m(0)$ are the diagonal-ensemble populations.

Equation~\eqref{eq:initial_noise_slope} connects the open-system sensitivity directly to the correlation-space geometry of the closed-system problem. A transition contributes only if the bath connects states that are both populated unequally and distinguished by their projection onto the frozen diagonal-ensemble pattern. In particular, the rate vanishes if the populations are uniform or if the connected eigenstates carry identical pair patterns. Thus the same incompatibility among frozen patterns that limits $Q_{\rm DE}$ after the quench also controls its sensitivity to longitudinal dephasing.

The protected endpoint follows from the same expression. As $\lambda\rightarrow0$, the Hamiltonian eigenstates approach longitudinal spin configurations and the $\sigma_i^z$ jumps no longer generate transitions between distinct pair patterns. Consequently,
\begin{equation}
R(\lambda)\rightarrow0
\qquad
{\rm as}
\qquad
\lambda\rightarrow0,
\end{equation}
consistently with the exact result Eq.~\eqref{eq:protected_supp}.

\begin{figure}[t]
\centering
\includegraphics[width=0.8\linewidth]{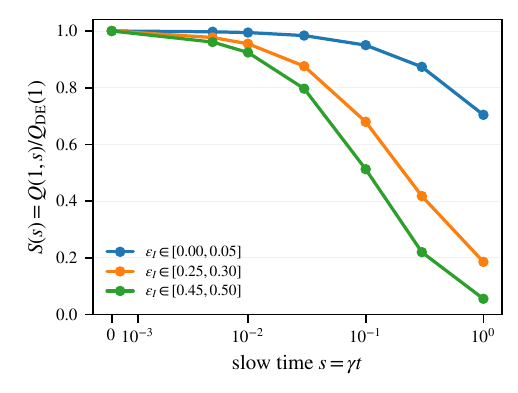}
\caption{
Shell-resolved survival of the no-quench pair-glass signal in the weak-noise kinetic regime for $L=12$, $\alpha=0.85$, and $h_I=0.2$. The normalized quantity $S(s)=Q(1,s)/Q_{\rm DE}(1)$ is shown as a function of the slow time $s=\gamma t$ for the cold, intermediate, and hot shells $\epsilon_I\in[0,0.05]$, $[0.25,0.30]$, and $[0.45,0.50]$. The markedly faster decay of the hotter shells demonstrates the strong energy dependence of the susceptibility to longitudinal dephasing.
}
\label{fig:local_z_survival}
\end{figure}

A complementary view is obtained at the no-quench point $\lambda=1$. Figure~\ref{fig:local_z_survival} shows the normalized survival
\begin{equation}
S(s)
=
\frac{Q(1,s)}{Q_{\rm DE}(1)}
\label{eq:survival_noise}
\end{equation}
for three representative energy shells. Although all curves start from unity, their subsequent decay differs markedly across the spectrum. The cold shell remains close to its closed-system value over substantially longer slow times, while the hotter shells decay much more rapidly. This confirms that the response to a fixed accumulated dephasing is controlled by the spectral and correlation-space structure of the initial shell rather than by the noise strength alone.

The secular description therefore provides the analytic weak-noise counterpart of the finite-time full-GKSL results shown in the main text. The latter retain coherent transients and depend separately on $\gamma$ and $t$ outside the asymptotic kinetic regime, whereas Eq.~\eqref{eq:secular_generator} isolates the slow redistribution that survives after averaging over the fast Hamiltonian motion. Both descriptions lead to the same qualitative structure: the $\lambda=0$ endpoint is exactly protected, while finite-$\lambda$ pair-glass order is noise sensitive in a strongly shell-dependent manner. The initial rate Eq.~\eqref{eq:initial_noise_slope} further shows that this sensitivity is governed by the same correlation-space pattern structure that controls the closed-system nonequilibrium response.

We also consider purely transverse noise,
$\mathcal L=\mathcal L_0+\gamma_x\mathcal D_x+\gamma_y\mathcal D_y$,
where $\mathcal D_a\rho=\sum_i(\sigma_i^a\rho\sigma_i^a-\rho)$.
The secular dynamics is parametrized by the slow time
$s=(\gamma_x+\gamma_y)t$, at fixed relative noise rates.
Unlike longitudinal dephasing, transverse noise does not protect
the zero-field reference. Indeed, the full GKSL equation gives
the exact result
\begin{equation}
Q(0,s)=Q_I e^{-8s}.
\label{eq:transverse_endpoint}
\end{equation}
Figure~\ref{fig:transverse_noise_slow}(d)-(f) show that transverse noise
progressively suppresses both the reference order and the
finite-quench response. The attenuation nevertheless remains
shell dependent, reflecting the different spectral structures
sampled by the dissipative dynamics.
\FloatBarrier

\end{document}